\documentclass{article}       
\usepackage{eme}                
\usepackage{graphics}
\usepackage{epsfig}
\begin{document}
\ifx\href\undefined\else\hypersetup{linktocpage=true}\fi
\branch{C}   
%
\title{Investigation of top mass measurements with the ATLAS detector at LHC}
\author{I. Borjanovi\'c\inst{1}\and I. Efthymiopoulos\inst{2}\and
F. Fassi\inst{3}\and P. Grenier\inst{2,4}\and P. Homola\inst{5}\and
V. Kostioukhine\inst{6}\and R. Leitner\inst{7}\and I. Menda\v s\inst{1}\and
D Pallin\inst{4}\and D. Popovi\'c\inst{1}\and P. Roy\inst{4}
\and V. Simak\inst{5,8}\and L. Simic\inst{1}\and G. \v Skoro\inst{9}\and
J. Valenta\inst{5}}
\institute{
Institute of Physics, P.O. Box 68, 11081 Belgrade, Yugoslavia
\and
CERN, Experimental Physics Division,CH 1211 Geneva 23, Switzerland
\and
IFIC-Departamento de Fisica Atomica, Molecular y Nuclear, \\
Avenida Dr. Moliner 50, E-46100 Burjassot (Valencia), Spain
\and
LPC Clermont-Ferrand, Universite Blaise Pascal, CNRS/IN2P3,\\
F-63177 Aubi\`ere, France
\and
Faculty of Nuclear Sciences and Physical Engineering, Czech Technical University,
Prague
\and
IHEP, PO Box 35, Pobeda street 1, RU-142284 Protvino, Moscow Region, Russia
\and
Faculty of Mathematics and Physics of the Charles University, Prague
\and
Institute of Physics of the Czech Academy of Sciences, Prague.\and
Faculty of Physics, University of Belgrade, P.O. Box 368, 11001 Belgrade, Yugoslavia}
%
%
\maketitle
\begin{abstract}
Several methods for the determination of the mass of the top quark
with the ATLAS detector at the LHC are presented. All dominant
decay channels of the top quark can be explored. The measurements
are in most cases dominated by systematic uncertainties. New
methods have been developed to control those related to the
 detector. The results indicate that a total error on the top mass at the level of 1 GeV should
be achievable.
\end{abstract}

\tableofcontents
\section{Introduction}
A precise measurement of the top quark mass will be a main goal
of top physics at the LHC. The combined top mass value from the Tevatron run I is
$m_t = 174.3 \pm 5.1$ GeV \cite{tevatron}, and the expected accuracy obtained
after run II will be 3 GeV \cite{cdf_run2}.

Motivations for an accurate determination
of the top quark mass are numerous. It is a fundamental parameter of the Standard
Model (SM) and should therefore be measured as precisely as possible. An accurate
value of the top mass would help to provide a rigorous consistency check of the
SM and to constrain some parameters of the model such as the mass of
the Higgs boson.
Furthermore, a high level of accuracy on the top mass value (for example improving
the accuracy down to $\Delta m_t \sim 1$ GeV) is also desirable, both within the
SM and the Minimal Supersymetric Standard Model (MSSM) framework \cite{yellowreport}.
In the SM, such an accuracy would significantly improve the precision on the W boson
mass prediction while in the MSSM, it would put constraints on the parameters of the
scalar top sector and would therefore allow sensitive test of the model by
comparing predictions with direct observations.

Because the top quark, as other quarks, cannot be observed as a
free particle, the  top quark mass is a purely theoretical notion
and depends on the concept adopted for its definition. With
increasingly-precise measurements on the horizon, it is important
to have a firm grasp of exactly what is meant by the top quark
mass. Thus far the top quark mass has been experimentally defined
by the position of the peak in the invariant mass distribution of
the top quark's decay products, a W boson and a b quark jet. This
closely corresponds to the pole mass of the top quark, defined as
the real part of the pole in  the  top quark propagator.

The renormalisation scale dependence is less than 10 MeV for the
range of the scale between 30-150 GeV. The dominant theoretical
uncertainty for the top mass caused by uncertainty on the strong
coupling constant is less than 150 MeV to the binding energy,
which would give uncertainty of 75 MeV in the pole mass.
Corresponding theoretical uncertainty in the $\overline{MS}$ mass
would be about $\pm$12 MeV\cite{tlee}. However, this definition is
still adequate for the analysis of top quark production at LHC
where uncertainty in the top mass measurement will be of order 1
GeV. Because of fragmentation effects it is believed that the top
quark mass determination in an hadronic environment is inherently
uncertain to $\cal O(\Lambda_{QCD})$
\cite{yellowreport}\cite{smithandwill}.

At the LHC, the top quark will be produced mainly in pairs through the hard
process $gg \rightarrow t \bar{t}$ ($90 \%$ of the total $t \bar{t}$ cross-section)
and $q \bar{q} \rightarrow t \bar{t}$ (remaining $10 \%$ of the cross-section).
The next-to-leading order cross-section prediction for $t \bar{t}$ production is
$\sigma(t \bar{t}) = 833$ pb \cite{crosssection}. Thus the LHC will be a top factory
as more than 8 million $t \bar{t}$ pairs will be produced per year at low luminosity
(corresponding to an integrated luminosity of $10 \ \rm fb^{-1}$).
The electroweak single top production processes, whose cross-sections are in total
approximately one third of those of $t \bar{t}$ production, have not been
investigated for the determination of the top mass.

Within the SM, the top quark decays almost exclusively into a W boson and
a b-quark ($t \rightarrow W b$). Depending on the decay mode of the W bosons
the $t \bar{t}$ events can be classified into three channels: the lepton
plus jets channel, the dilepton channel and the all jets channel.
In the lepton plus jets channel, one of the W boson decays leptonically
($W \rightarrow l \nu$) and the other one hadronically ($W \rightarrow jj$).
Considering electrons and muons, the branching ratio is
$\rm BR = 2 \times 2/9 \times 6/9 \simeq 30 \%$. The final state topology is
$gg \rightarrow t \bar{t} \rightarrow (jjb)(l \nu b)$.
In the dilepton channel, both W bosons decay leptonically with
$\rm BR = 2/9 \times 2/9 \simeq 5 \%$. In the all jets channel,
both W bosons decay hadronically with $\rm BR = 6/9 \times 6/9 \simeq 44 \%$.

This paper summarizes studies of the top mass measurement, including updates
of studies presented in the ATLAS Technical Design Report \cite{atlastdr} as
well as several new analysis.

Unless otherwise indicated, all analysis were performed with events generated
with Pythia \cite{pythia} and passed through the ATLAS fast simulation package
Atlfast \cite{atlfast} for particles and jets reconstruction and momenta smearing.
Jets are defined as massless objects by summing the momenta of clusters of energy deposited in the calorimeters.
Clusters are associated to form jets using a cone algorithm with $\delta(R)<0.4$.
A tagging efficiency of 60\% for b-jets was assumed. Cross-checks of some
results have been made using the detailed GEANT-based simulation of the ATLAS
detector. The top mass is extracted by an analytic fit to the event by event reconstructed invariant mass.


%
\section{Top mass measurement in the lepton plus jets channel}

The lepton plus jets channel is probably the most promising channel for
an accurate determination of the top quark mass. Three methods to measure the
top mass are envisaged. The simplest method consists in extracting the top mass
from the three jets invariant mass of the hadronic top decay. In the second method,
the entire $t\bar t$ system is fully exploited to determine the top quark mass
from a kinematic fit. In the last method, the top mass is still determined
from a kinematic fit, but the jets are reconstructed using a continuous
algorithm.

\subsection{Event selection and background rejection}

Taking into account the total $t \bar{t}$ cross-section and the branching ratio,
one can expect 2.5 millions $t \bar{t}$ pairs with this topology to be
produced per year assuming an integrated luminosity $10fb^{-1}$.

The signal final state $t \bar{t} \rightarrow Wb \ Wb \rightarrow jjb \ l \nu b $
(with $l = e,\mu$) is characterized by one high transverse momentum
lepton, large transverse missing energy $E_T^{miss}$, and high jet multiplicity.
The following background processes were considered:
$b \bar{b} \rightarrow l \nu + jets$, $W + jets \rightarrow l \nu + jets$,
$Z + jets \rightarrow l^+ l^- + jets$, $WW \rightarrow l \nu + jets$,
$WZ \rightarrow l \nu + jets$, and $ZZ \rightarrow l^+ l^- + jets$. At production
level, the signal over background ratio is very unfavorable (S/B $\sim 10^{-5}$).

\begin{table}[h]
\begin{center}
\begin{tabular}{lcc} \hline
Process             & Cross-section &  Total efficiency \\
                    &    (pb)     &      ($\%$)         \\
\hline
$t \bar{t}$ signal                    & 250             & 3.5               \\
$b \bar{b} \rightarrow l \nu + jets$  &$2.2 \times 10^6$&$3 \times 10^{-8}$ \\
$W + jets \rightarrow l \nu + jets$   &$7.8 \times 10^3$&$2 \times 10^{-4}$ \\
$Z + jets \rightarrow l^+ l^- + jets$ &$1.2 \times 10^3$&$6 \times 10^{-5}$ \\
$WW \rightarrow l \nu + jets$         & 17.1            &$7 \times 10^{-3}$ \\
$WZ \rightarrow l \nu + jets$         & 3.4             &$1 \times 10^{-2}$ \\
$ZZ \rightarrow l^+ l^- + jets$       & 9.2             &$3 \times 10^{-3}$ \\
\hline
\end{tabular}
\caption{\label{lepton_efficiency} {\it Cross-section and
selection efficiency for signal and background processes. For the
background, the hard scattering processes are generated with a cut
on the transverse momentum at 20 GeV}}
\end{center}
\end{table}

A high level of rejection was obtained using the following requirements: one
isolated lepton with $p_T > 20$ GeV, $E_T^{miss} > 20$ GeV, and at least four jets
reconstructed with a cone size of $\Delta R = 0.4$ with $p_T > 40$ GeV and
$| \eta | < 2.5$, of which at least two are tagged as b-jets. The efficiency of
the selection for the various background processes is shown in table
\ref{lepton_efficiency}. After selection cuts, the signal over background
ratio is extremely good (S/B $\sim 78$), and the remaining number of signal
events is approximately 87000 (for an integrated luminosity of $10^{-1}$fb).

\begin{figure}
\begin{center}
\epsfig{file=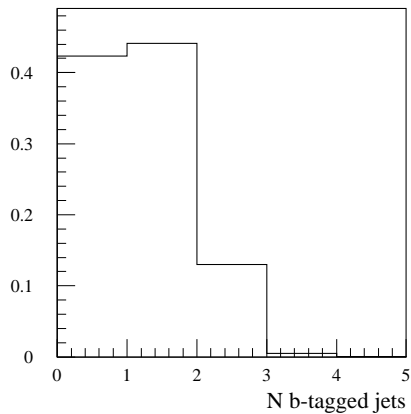,width=6.0cm}
\caption{\label{lepton_nbtagged} {\it $t\bar t$ sample as a function
 of the number of b-tagged jets in the event.}}
\end{center}
\end{figure}

The requirement of having at least two b-tagged jets in the final state helps in
rejecting a large part of the physical background, but also reduces considerably
the signal sample. The fraction of signal events with at least two b-tagged jets
is three times smaller than the fraction with at least one  b-tagged jet (see
figure \ref{lepton_nbtagged}). Requiring only one b-tagged jet would decrease the
signal over background ratio from 78 to 28, which would still be acceptable.
This extended sample can be used for the top mass measurement using the hadronic top
decay, requiring the tagged b-jet to be the one belonging to the hadronic top final
state.

\subsection{Top mass measurement using the hadronic top decay}

In the following method, the top mass will be determined from the invariant mass of the
three jets arising from the hadronic top decay ($t \rightarrow Wb \rightarrow jjb$).
To accurately reconstruct the decay, one should: i) identify the jets associated to
the hadronic top decay among all other jets, ii) precisely calibrate the jet
energies and directions.

\subsubsection{Jet association}

\begin{figure}
\begin{center}
\includegraphics[scale=0.75]{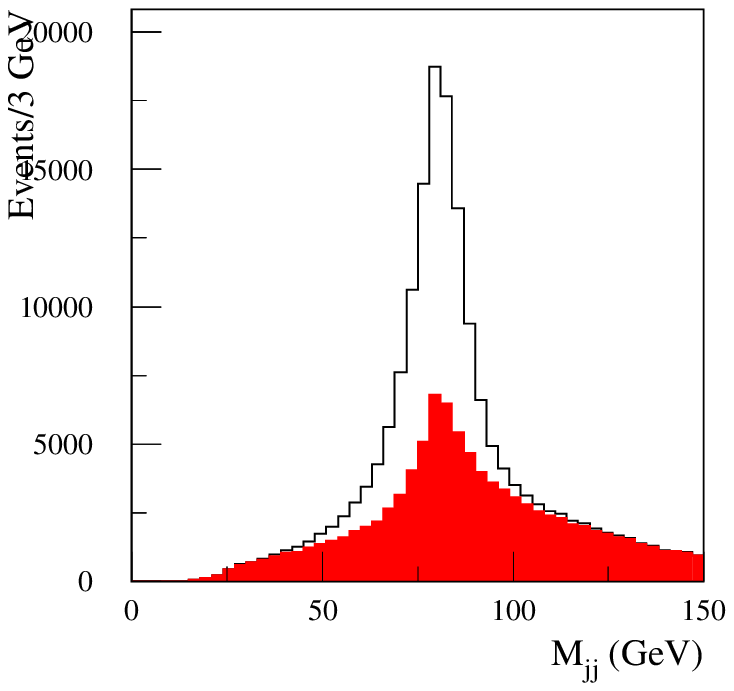}
\includegraphics[scale=0.75]{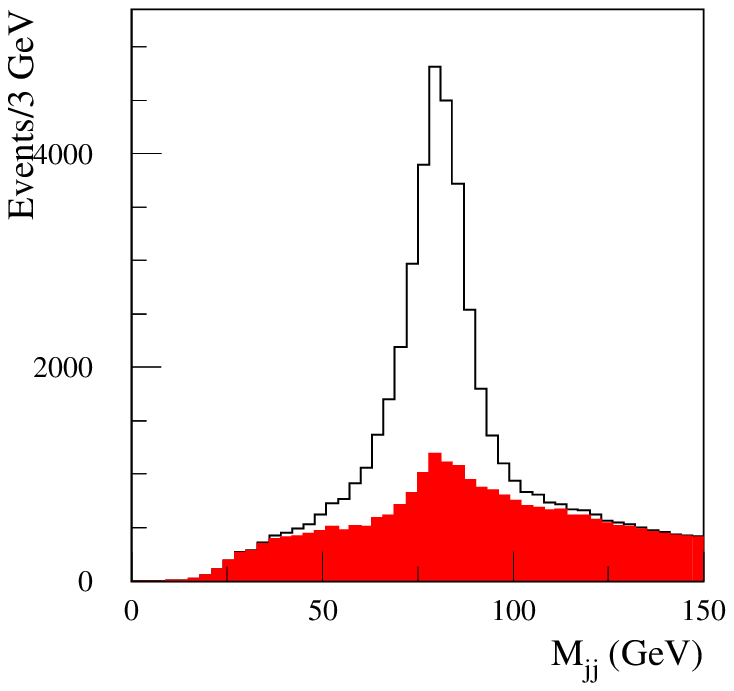}
\caption{\label{lepton_mw} \it{ Dijet invariant mass distributions for events with at least 1
b-tagged jet (left plot), and at least 2 b-tagged jets (right plot). The shaded
area represents the combinatorial background. Both plots are for 10 fb$^{-1}$.}}
\end{center}
\end{figure}

At least four jets are expected in the event: two from the hadronic W decay and
two b-jets. Additional jets will be produced by initial state radiation (ISR) and
final state radiation (FSR) effects. The association of jets to the original partons
is done as follows. The hadronic W decay is first reconstructed: all the
non b-tagged jets are paired together and the jet pair with an invariant mass
closest to the W mass is taken as the right combination for the W.
The di-jet invariant mass distributions for events selected by requiring two
b-tagged jets or at least one are shown on figure \ref{lepton_mw}.
When the two associated jets are reconstructed, 80 $\%$ of the true W decays
are selected, which is realized in 45 $\%$ of the cases. This leads, in a mass
window $|M_{jj}-M_W^{PDG}|<20$ GeV, to a purity of 55 $\%$ for events selected
with at least one b-tagged jet and 66 $\%$ for events selected with two b-tagged
jets. The width of the mass distribution is 7.4 GeV for both cases.

\begin{figure}
\begin{center}
\mbox{
\includegraphics[scale=0.75]{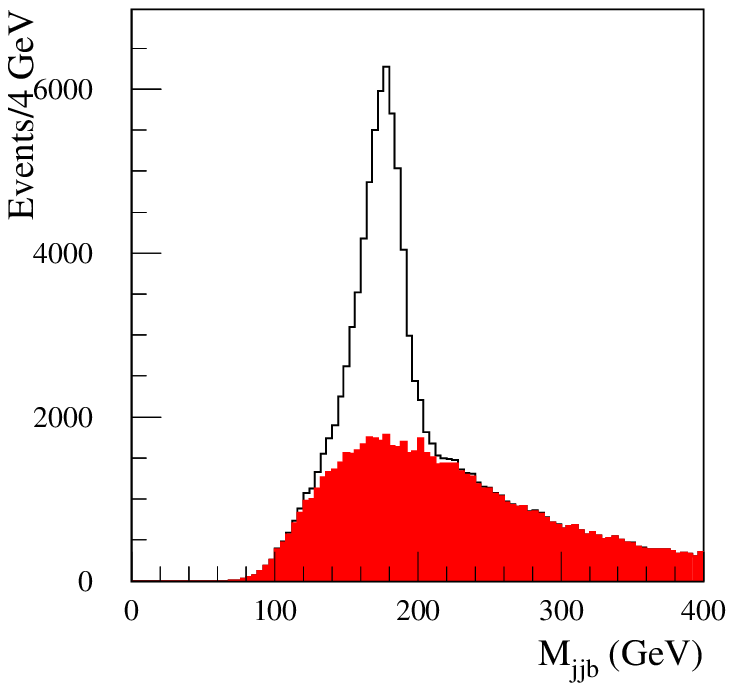}
\includegraphics[scale=0.75]{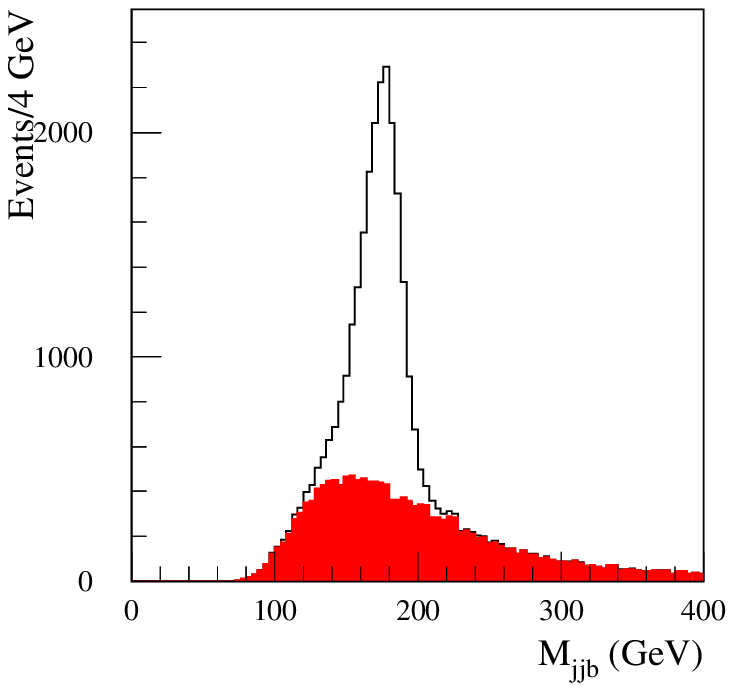}
}
\caption{\label{lepton_top1bet2b} \it{Top mass distributions for the 1 b-tagged
jet sample (left plot) and the 2 b-tagged jets sample (right plot). The shaded
area represents the combinatorial background. Both plots are for 10 fb$^{-1}$.}}
\end{center}
\end{figure}

The next step is to associate a b-tagged jet to the reconstructed W. When events are
selected with only one b-tagged jet, the association is performed if the b-tagged
jet is closer to the reconstructed W than to the isolated lepton
($\Delta R(b,W)<\Delta R(b,lepton)$). The efficiency of this association criteria
is $82 \%$. In the presence of two b-tagged jets, the chosen b-tagged
jet is the one giving the highest $p_T$ for the reconstructed top, giving an
association efficiency of $81 \%$. The reconstructed three jets mass distributions
are presented on figure \ref{lepton_top1bet2b} for the two event selection
criteria. The peak width is 12 GeV in both cases.

The overall association purity and efficiency within a top mass window of
$\pm$35 GeV around the top mass peak are summarized in table \ref{lepton_njball}.
The top mass determination will not be limited by the statistics even if the
analysis is restricted to the two b-tagged jets sample. Nevertheless, the large
one b-tagged jet sample allows further dedicated cuts with negligible impact
on the statistical precision of the top mass determination.
In the sample with two b-tagged jets, the overall reconstruction
efficiency is $1.2 \%$, leading to 30000 events for one year of running at
low luminosity (per 10 fb$^{-1}$).
For simplicity, only this case will considered in the following.

\begin{table}[h]
\begin{center}
\begin{tabular}{lcc} \cline{2-3}
 & 1 b-tagged jet sample & 2 b-tagged jets sample \\ \hline
Top purity ($\%$) & 65 & 69 \\
Total efficiency ($\%$) & 2.5 & 1.2 \\ \hline
\end{tabular}
\caption{\label{lepton_njball} \it{Summary table for the two samples considered.
Purity and total efficiency are related to a top mass window of $\pm$35 GeV around
the generated $m_t$ value.}}
\end{center}
\end{table}

\subsubsection{In situ jet energy and direction calibration}

As the top mass is determined from the invariant mass of a three-jet system,
the accuracy of the measurement depends on how well the jet energies and
directions are reconstructed. A mis-measurement of $1 \%$ of the jet energies
induces a top mass shift of 1.6 GeV. Similarly, a mis-measurement of $1 \%$ of
the cosinus of opening angle between the two jets from the W and between the reconstructed
W and the b-jet induces a top mass shift of 1.2 GeV. Therefore an excellent
absolute energy scale and angle measurement are required to precisely determine
the top quark mass.

Numerous effects have to be taken into account to determine the initial parton
energy from the energies deposited in the ATLAS calorimeters. Prior to data taking,
the accumulated knowledge on the detector performances and characteristics,
on physics effects like initial and final state radiations, underlying or minimum
bias events plus the impact of the jet finding algorithms will allow to reach a
5-10  $\%$ on the absolute energy scale \cite{atlastdr}.
In-situ calibrations will fix the absolute energy scale through the study of known
processes, taking into account in a global way the remaining inaccuracies on the
knowledge on the various effects described previously.

It has been shown that an accurate absolute energy calibration of light quark jets and b-jets
can be extracted from Z+jet events \cite{atlastdr,lepton_zjet1}, within an expected
precision of about 1 $\%$. However, this calibration applied to the W mass
reconstruction \cite{lepton_zjet2,lepton_zjet3}, leads to a shifted W mass. Due
to the energy sharing between jets, the opening angles are systematically
underestimated leading to a less precise mass measurement \cite{lepton_pallin1}.

Here, to avoid any dependency from external inputs, it is proposed to perform an
situ calibration in which both the absolute energy
and direction calibration are extracted from the $ W \rightarrow jj$ channel itself.
For this purpose, a cleaner sample of W
candidates has been selected from the $t \bar{t}$ events. Initially, the jets
are not calibrated but corrected for cell energy sharing effects. In addition
to the preselection cuts, the di-jet invariant mass is required to fall within
a mass window of $\pm$ 20 GeV around the peak value and the three-jet invariant
mass to fall within a mass window of $\pm$ 15 GeV around the peak value.

The in-situ calibration is performed through a $ \chi ^2 $ minimization
procedure in which the dijet mass is constrained to the known W mass.

Non-calibrated jet energies are shifted from the initial parton
energies due mainly to the jet cone algorithm procedure and FSR effects. However
the induced shift is in general smaller than the energy resolution of the jets.
This allows to fix the $\sigma _{E_i}$ term to the intrinsic calorimeter
energy resolution. The same approach is employed for the jet directions.

\begin{figure}[ht!]
\begin{minipage}{0.49\textwidth}
\begin{center}
\epsfig{file=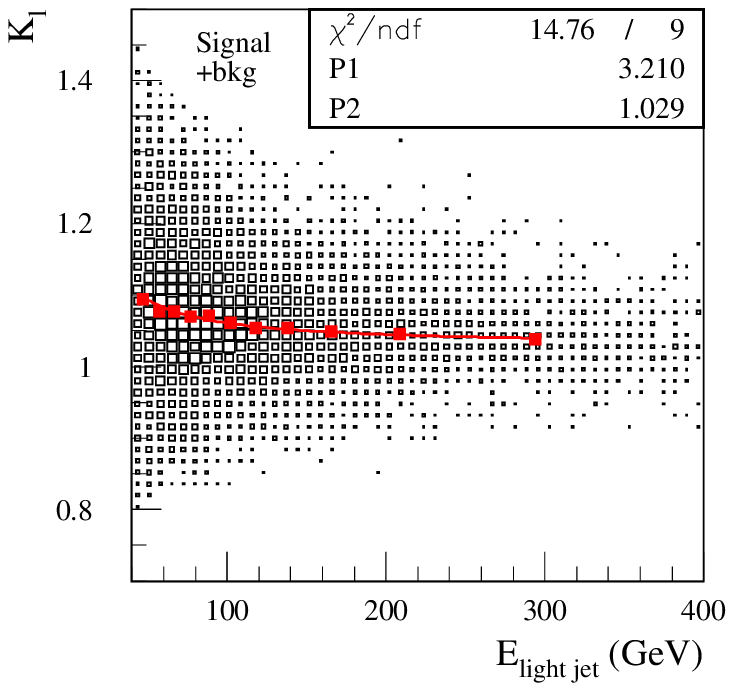,width=6.0cm}
\caption{\label{lepton_figcalk1} {\it Calibration factors obtained event by event,
and parametrization after calibration fit.}}
\end{center}
\end{minipage}
\hfill
\begin{minipage}{0.49\textwidth}
\begin{center}
\epsfig{file=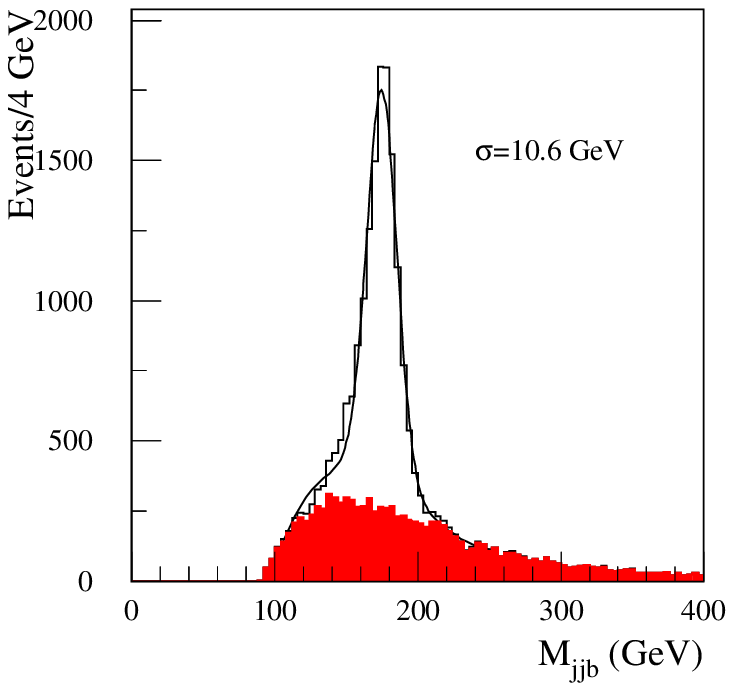,width=6.0cm}
\caption{\label{lepton_figtophad} {\it Final jjb invariant mass distribution.}}
\end{center}
\end{minipage}
\end{figure}

\begin{figure}
\hbox{
\includegraphics[scale=0.75]{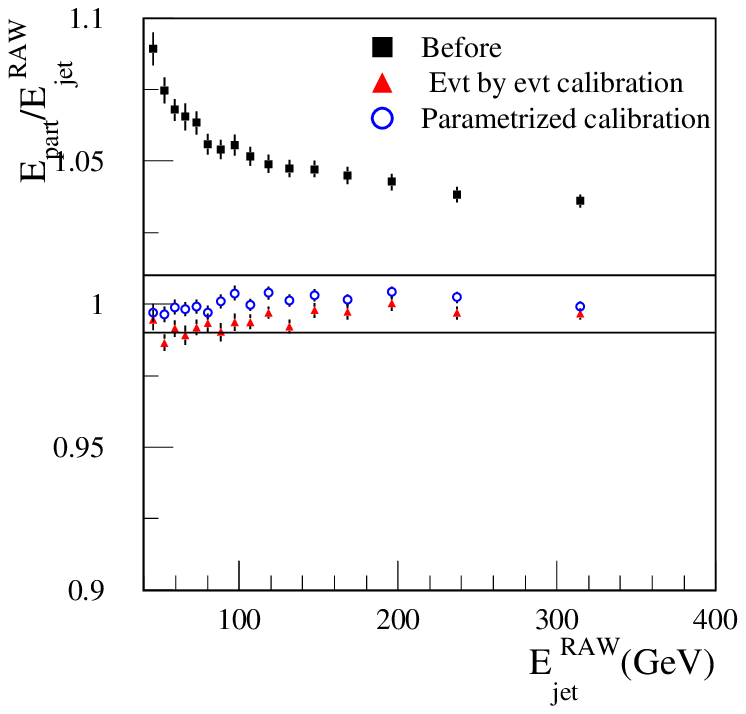}
\includegraphics[scale=0.75]{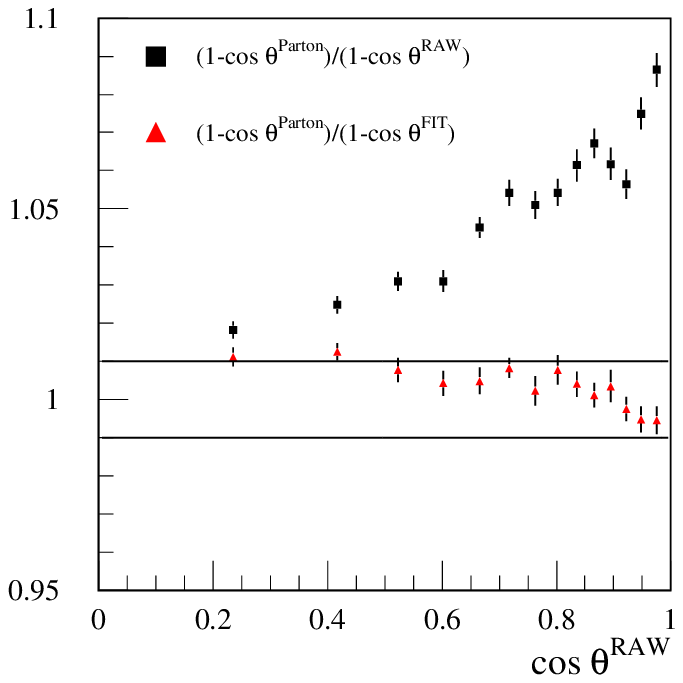}
}
\caption{\label{lepton_figcal1} \it{Results of the calibration fit on jet energy
and direction. The two solid lines show the $ \pm 1 \% $ precision level.}}
\end{figure}

An energy correction factor K is obtained at the end of the fitting procedure,
for each jet and for each event. The distribution as a function of the raw initial
jet energy is shown in figure \ref{lepton_figcalk1}. Finally, the function
$\rm K=P2+\frac{P1}{E^{RAW}}$ is fitted to the distribution (the resulting
parameters are shown on the figure) leading to a calibration function,
without an a priori knowledge on the initial calibration function shape.
In figure \ref{lepton_figcal1} the comparison between the initial parton
energy and the calibrated jet energy at various steps of the procedure is
represented. One can see the impressive effect of the in-situ calibration
procedure as the ratio of the parton energy to the reconstructed jet energy
remains well below the level of $1 \%$. The improvement brought by this
procedure is also clear on the reconstruction of the opening angle between
the two jets, as seen on figure \ref{lepton_figcal1}. It should be noted
that the combinatorial background does not introduce a sizable bias on
the calibration factors.

\subsubsection{Top mass reconstruction}

The selected three jets invariant mass distribution is shown in figure
\ref{lepton_figtophad}. The light quark jets were calibrated as described
above and the b-quark jets were calibrated using Z+b events \cite{atlastdr}.
The mass peak is in agreement within 100 MeV of the generated value.
The peak width is around 11 GeV, leading to a statistical error on the top mass
of the order of 100 MeV for one year of running at low luminosity (per 10 fb$^{-1}$).

\subsubsection{Full simulation results}

The analysis presented in section 2.2 of the determination of the top quark mass
using the hadronic top decay has been repeated using fully simulated events. For
this purpose, 30000 $t \bar{t}$ events were processed through the GEANT-based
ATLAS detector simulation package.

Events were generated under restrictive conditions
at generator level. These conditions include, for example, cuts on the
transverse momentum of the $t \bar{t}$ decay products. Therefore any direct comparison
with the results presented in section 2.2 should be avoided. The comparison is made using
the same generated events which have been passed through both the fast and full
simulation package.

\begin{figure}
\begin{center}
\begin{tabular}{cc}
\mbox{\epsfig{file=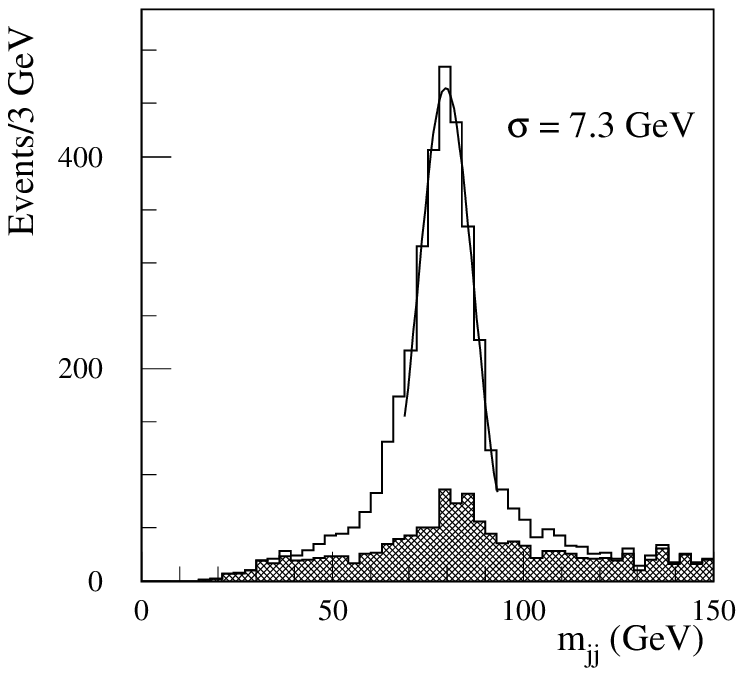,width=5.0cm}} &
\mbox{\epsfig{file=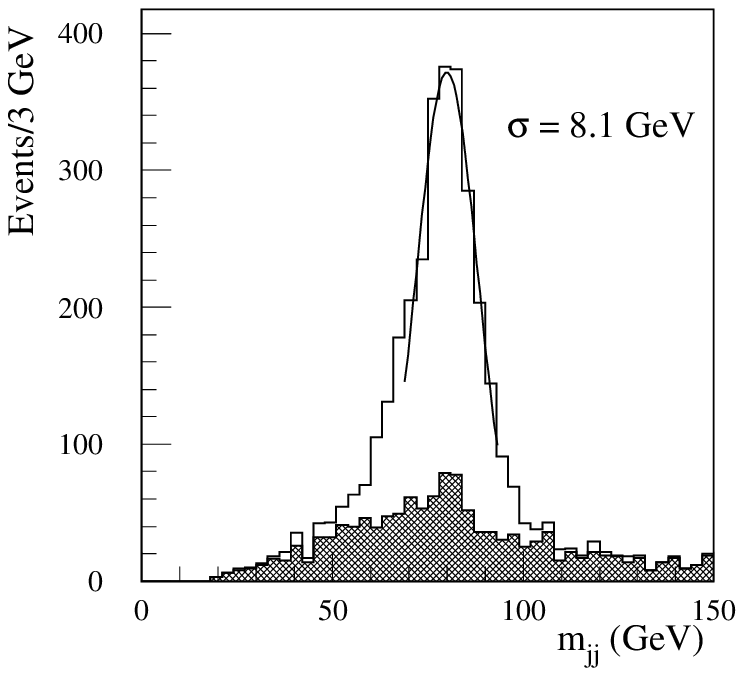,width=5.0cm}} \\
\mbox{\epsfig{file=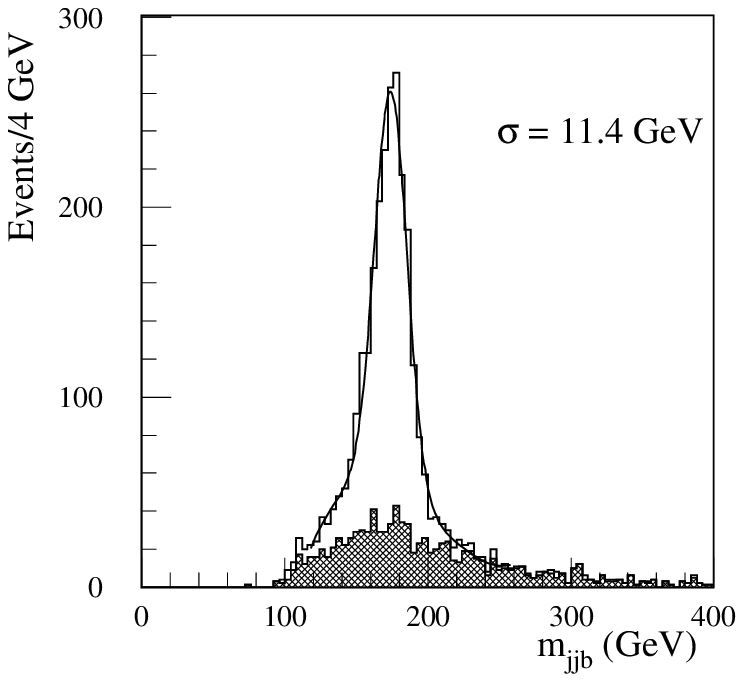,width=5.0cm}} &
\mbox{\epsfig{file=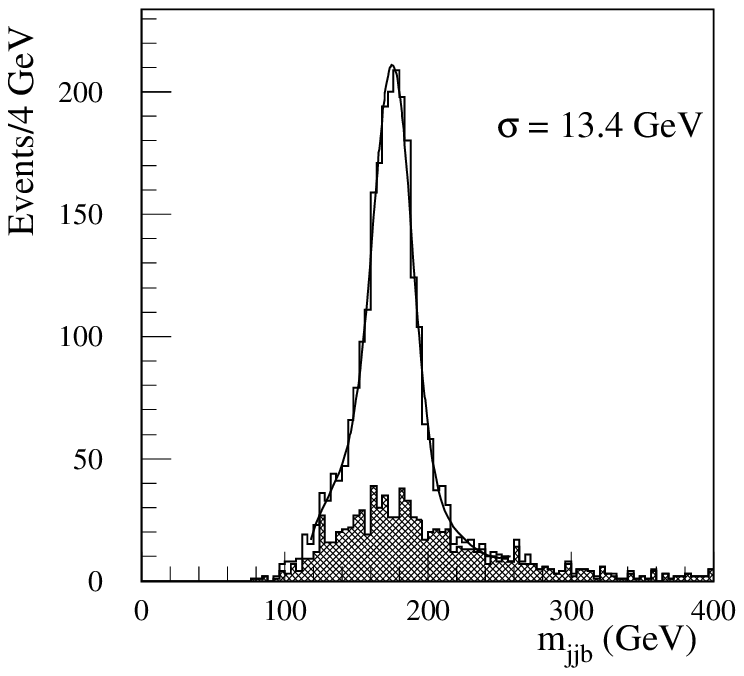,width=5.0cm}} \\
\end{tabular}
\caption{ \label{lepton_fullsimfig} {\it Top mass determination using the hadronic
top decay: $m_{jj}$ and $m_{jjb}$ invariant mass
distributions. The left-handed plots represent the distributions obtained
from fast simulation and the right-handed plots represent the distribution
obtained from full simulation.}}
\end{center}
\end{figure}

Figure \ref{lepton_fullsimfig} represents the $m_{jj}$ and $m_{jjb}$ distributions for
fast and full simulation. The $m_{jj}$ invariant mass resolution is $7.3 \ \rm GeV$
for fast simulation and $8.1 \ \rm GeV$ for full simulation. The $m_{jjb}$ invariant
mass resolution is $11.4 \ \rm GeV$ for fast simulation and $13.4 \ \rm GeV$ for
full simulation. In the top mass window $175 \pm 35 \ \rm GeV$, the signal purity
and overall efficiency are $P = (79 \pm 2) \%$ and $E = (6.4 \pm 0.5) \%$
for fast simulation and $P = (78 \pm 2) \%$ and $E = (5.7 \pm 0.5) \%$ for
full simulation. These results are summarized in table \ref{lepton_fullsimtab}.

\begin{table}[h]
\begin{center}
\begin{tabular}{lcc}          \hline
Quantity & Fast simulation & Full simulation \\ \hline
$m_{jj}$ resolution (GeV) & 7.3 & 8.1   \\
$m_{jjb}$ resolution (GeV) & 11.4 & 13.4 \\
Signal purity ($\%$) in $175 \pm 35 \ \rm GeV$ & 79 & 78 \\
Signal efficiency ($\%$) in $175 \pm 35 \ \rm GeV$ &
6.4 & 5.7 \\ \hline
\end{tabular}
\caption{ \label{lepton_fullsimtab} {\it Top mass determination using the hadronic
top decay: comparison between fast and full simulation.}}
\end{center}
\end{table}

The results obtained for the signal purity and overall efficiency
as well for the $m_{jj}$ and $m_{jjb}$ invariant mass resolutions are in reasonable
agreement between fast and full simulation, though the resolutions from
GEANT are somewhat worse. In addition, the shape and amount
of the combinatorial background for both the W and top masses reconstruction
are also in good agreement between the two types of simulations.

\subsubsection{Systematic uncertainties}

To estimate the effect of an absolute jet energy scale uncertainty,
different miscalibration coefficients were applied to the reconstructed
jet energy. A top mass shift per percent of miscalibration was obtained.
For light quark jets, the effect is small as the jet are re-calibrated in-situ.
For b-quark jets, a $1 \%$ miscalibration induces a top mass shift of 0.7 GeV.

The presence of initial state radiation of incoming partons (ISR) and final state
radiation from the top decay products (FSR) can impact the
measurement of the top mass. To estimate their effect, a top mass shift due
to ISR was computed as the difference between the value of the top mass determined
with ISR switched on (usual data set) and ISR switched off. The same approach was
employed for FSR. The level of knowledge of ISR and FSR is of order of $10 \%$.
Therefore, as more conservative estimate, the systematic uncertainty on the top
mass was taken to be $20 \%$ of the corresponding mass shifts.

\begin{table}
\begin{center}
\begin{tabular}{l c} \hline
 Systematics  & $\rm \delta m_{t}$ (GeV)\\ \hline
Light jet energy scale & 0.2 \\
b jet energy scale & 0.7$\times x\%$ \\
Initial State Radiation & 0.1 \\
Final State Radiation & $1$ \\
b-quark fragmentation & 0.1 \\
Combinatorial background & 0.1 \\
Statistical error & 0.1 \\ \hline
\end{tabular}
\caption{\label{lepton_systhad} \it{Summary of systematic errors in the
inclusive lepton plus jets sample. x represents the level at which the b jet energy
scale will be known.}}
\end{center}
\end{table}

The b-quark fragmentation was described by the Peterson fragmentation function
\cite{lepton_peterson}. This function is parametrized in terms of one variable
$\varepsilon_b$. The default value was set at $\varepsilon_b = -0.006$, with
an uncertainty of 0.0025 \cite{lepton_aleph}. The top mass was determined
with another sample of events generated with $\varepsilon_b = -0.006 + 0.0025$.
The difference of the top mass value between the usual sample and the latter was
taken to be the systematic uncertainty on the top mass due to the knowledge
of $\varepsilon_b$.

Uncertainties due to the combinatorial background (which is
the main background) were also estimated by varying the assumptions of the
background shape and size in the fitting procedure. Fits of the three-jet invariant
mass distribution were performed using a Gaussian shape for the signal and
either a polynomial or a threshold function for the background. The resulting
systematic error on the top mass was 0.1 GeV.

All the results are summarized in table \ref{lepton_systhad}. The main
contributions are from FSR and b-quark energy scale. Adding in quadrature all
the contribution leads to a total systematic uncertainty on the top mass of the
order of 1.3 GeV, provided the b-quark jets can be calibrated within $1 \%$.

\subsection{Top mass measurement using a kinematic fit}

In the previous section, it was shown that the top quark mass can be measured
in the lepton plus jets channel with an accuracy better than 2 GeV. This error
is totally dominated by systematic effects, in particular the b-quark jet
energy scale and FSR. In order to reduce further the systematic uncertainties,
another method is proposed in the following, where the entire $t \bar{t}$
final state is reconstructed by a kinematic fit. This method is aimed to reduce
the impact of poorly reconstructed jets (due to effects arising from FSR and
semi-leptonic decays of b-quark jets). The final state can divided into two parts:
i) the leptonic part corresponding to the leptonic
top decay ($t(\bar{t}) \rightarrow l \nu b (\bar{b})$) and ii) the hadronic part
corresponding to the hadronic top decay ($\bar{t}(t) \rightarrow j j \bar{b} (b)$).

\begin{figure}
\begin{center}
\includegraphics[scale=0.75]{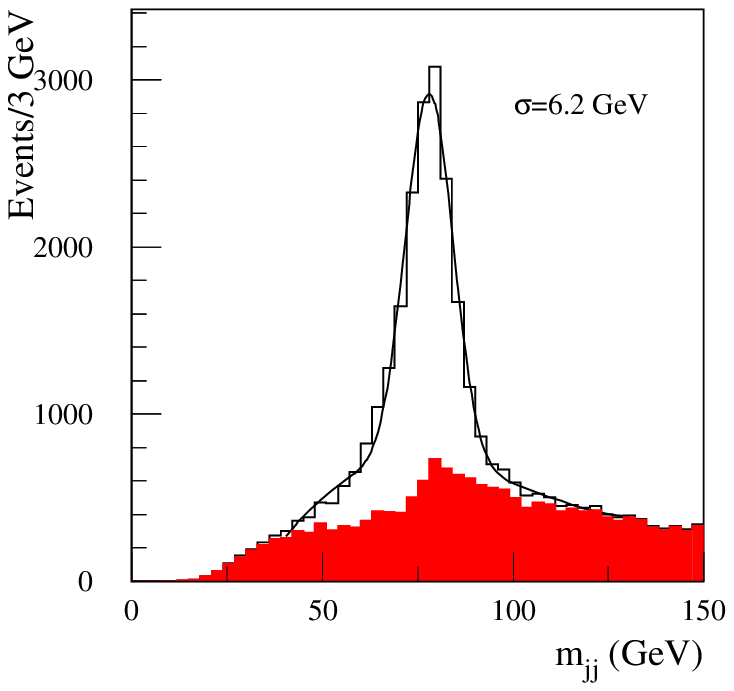}
\includegraphics[scale=0.75]{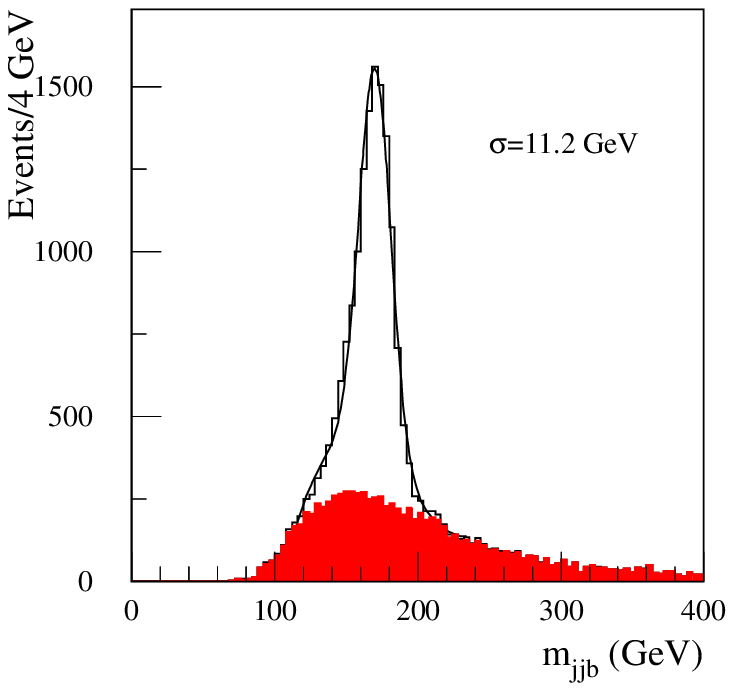}
\caption{\label{lepton_masse} \it{ Hadronic part: reconstructed W mass
(left plot) and reconstructed top mass (right plot). The
combinatorial background contributions are shown (shaded areas)
together with the full invariant mass distributions (full line).}}
\end{center}
\end{figure}

The hadronic part is reconstructed in a similar way to the previous section. For
the light quark jets, the absolute energy scale was taken from the in-situ
calibration described above and for b-quark jets, it was taken from Z+b events
\cite{atlastdr}. The invariant mass distribution of the selected di-jet pairs is
reconstructed with a width on 6.2 GeV (see figure \ref{lepton_masse}). The
three-jet system is reconstructed with a width of 11.2 GeV (see figure
\ref{lepton_masse}). Over the entire mass range, the purity is $51 \%$ and is
increased to $71 \%$ within a mass window of $\pm$ 35 GeV around the peak value.

\begin{figure}[ht!]
\begin{minipage}{0.49\textwidth}
\begin{center}
\epsfig{file=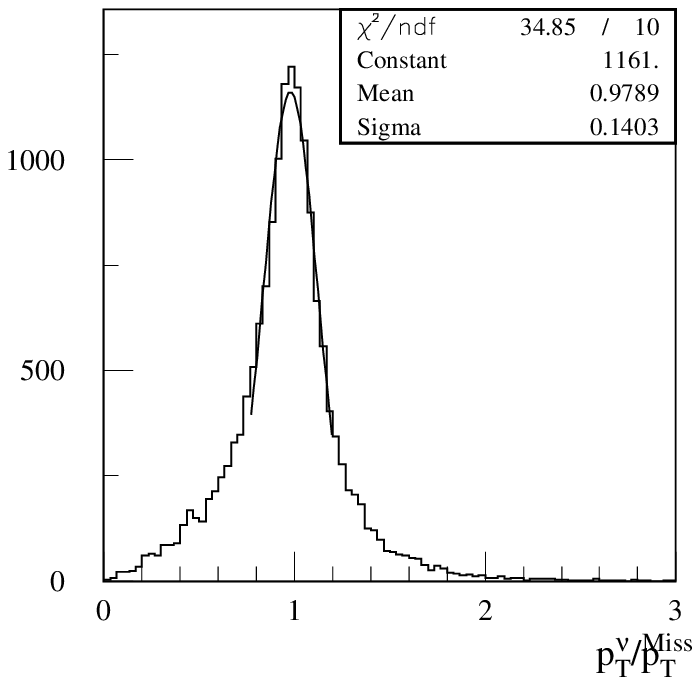,width=6.0cm}
\caption{\label{lepton_ptnu} {\it Ratio of the neutrino transverse momentum at
parton level over the reconstructed neutrino transverse momentum.}}
\end{center}
\end{minipage}
\hfill
\begin{minipage}{0.49\textwidth}
\begin{center}
\epsfig{file=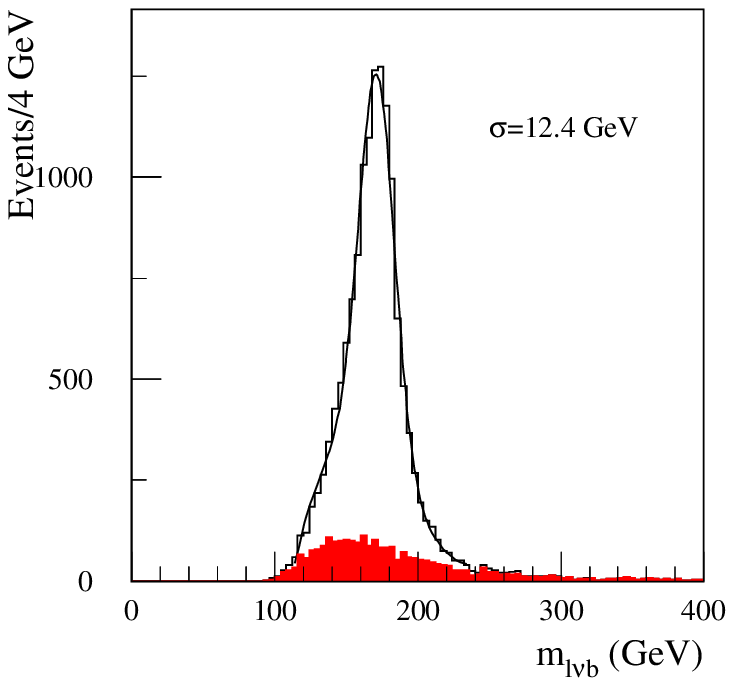,width=6.0cm}
\caption{\label{lepton_mtlep} {\it Reconstructed top mass from the leptonic part.
The combinatorial background contribution is shown (shaded area) together
with the full invariant mass distribution (full line).}}
\end{center}
\end{minipage}
\end{figure}

The leptonic final state cannot be directly reconstructed due to the presence
of the undetected neutrino. Nevertheless, the neutrino four-momentum can be
estimated in two steps. First, the transverse component of the neutrino momentum
can be approximated by the transverse missing energy
(see figure \ref{lepton_ptnu}). The longitudinal component of the neutrino
momentum can then be deduced with a quadratic ambiguity, by constraining the invariant
mass of the lepton-neutrino system to the known W mass value. Finally, the
remaining b-tagged jet is associated to the reconstructed W. In most of the cases,
there are only two b-tagged jets present in the event, and one has already been associated.
In case of additional b-jets, the closest one to the isolated lepton is chosen.
Per event, two leptonic top masses are computed, corresponding to the two
neutrino $p_Z$ solutions. The distribution of the leptonic top mass the
closest to the hadronic mass is represented in figure \ref{lepton_mtlep}. It is
fitted by a third order polynomial plus a Gaussian, leading to a peak width of
12.4 GeV, similar to the hadronic resolution. The event is eventually kept if one
of the two leptonic top masses is within the same top mass window as the
hadronic part.

Therefore, the entire $t \bar{t}$ final state can be reconstructed, with a
twofold ambiguity due to the neutrino reconstruction. After selection and
mass window cuts, the final sample is composed of 18000 signal events and
7000 combinatorial events (the contribution from other physical background
processes is totally negligible). The total signal ($t \bar{t}$ events with
all jets well assigned) efficiency is $0.7 \%$ with a purity of $73 \%$.

\subsubsection{Top mass determination}

The kinematic fit is performed in such a way that the jets and lepton energy,
the jets direction (in terms of $\eta$ and $\phi$) and the three components
of the reconstructed neutrino momentum can vary freely within their
corresponding resolutions. The following kinematic constraints were
employed: \\
\centerline{$\rm m_{jj}=M_W^{PDG}$, $\rm m_{l\nu}=M_W^{PDG}$ and
$\rm m_{jjb}=m_{l\nu b}=M_{top}^{fit}$}

On an event-by-event basis and for both neutrino solutions, a $\rm \chi ^2$ is
minimized \cite{lepton_pallin1}. The output of the fit is the top mass estimator
$\rm M_{top}^{fit}$. The neutrino solution with the lowest $\rm \chi ^2$ is selected.

\begin{figure}
\begin{center}
\includegraphics[scale=0.75]{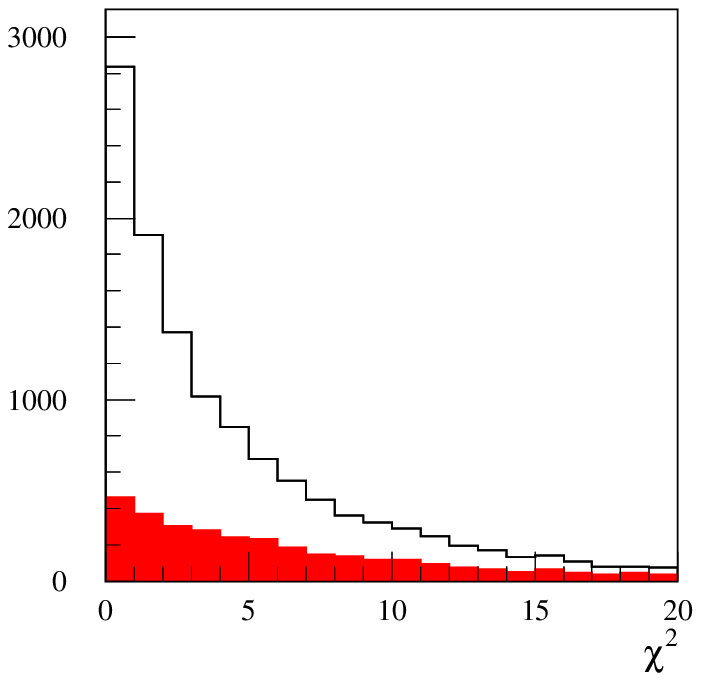}
\includegraphics[scale=0.75]{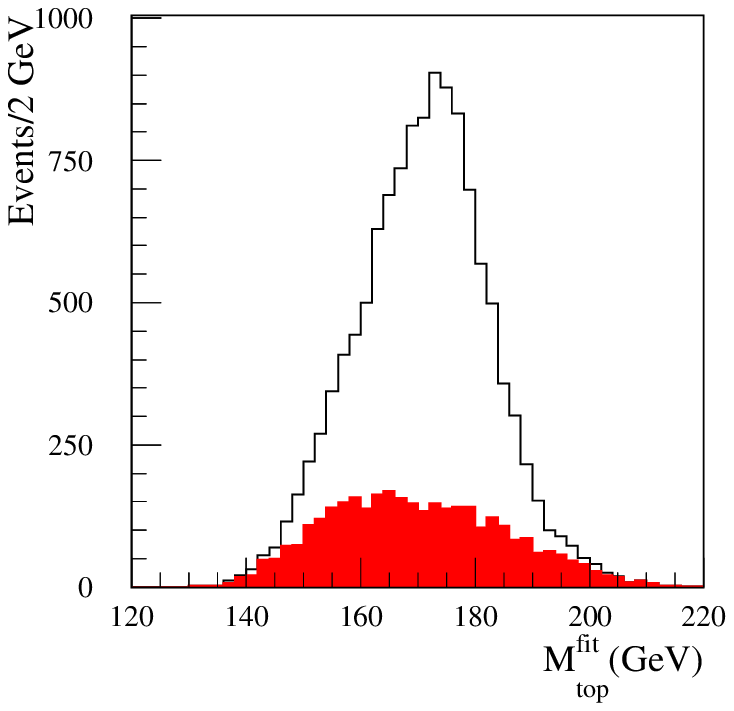}
\caption{\label{lepton_chi2} \it{ Left plot: $\chi ^2$ distribution.
Right plot: top mass after kinematic fit.}}
\end{center}
\end{figure}

The shapes of the $\rm \chi ^2$ distributions are different for signal and
combinatorial background, as is shown in figure \ref{lepton_chi2}:
a cut at $\rm \chi ^2<4$ increases the $\rm t\bar{t}$ purity to more than 80$\%$.
The fitted top mass is also shown in figure \ref{lepton_chi2}.

\begin{figure}
\begin{center}
\includegraphics[scale=0.75]{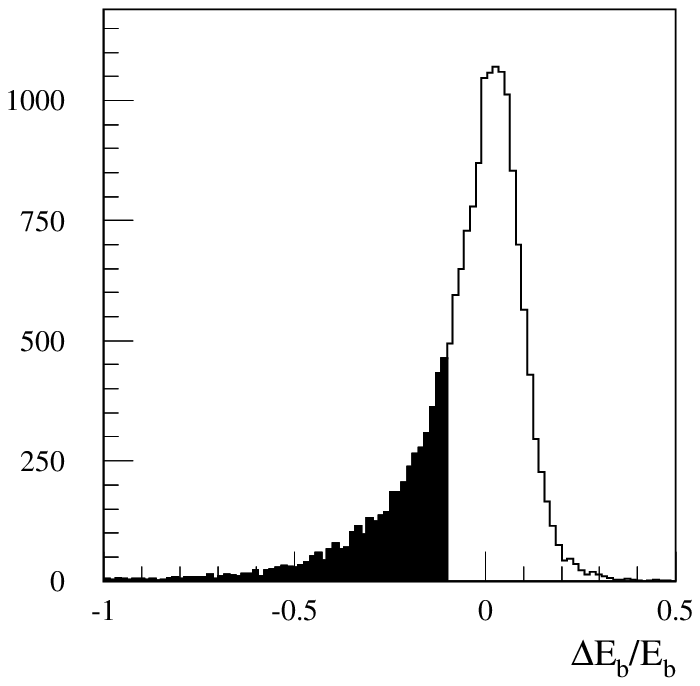}
\includegraphics[scale=0.75]{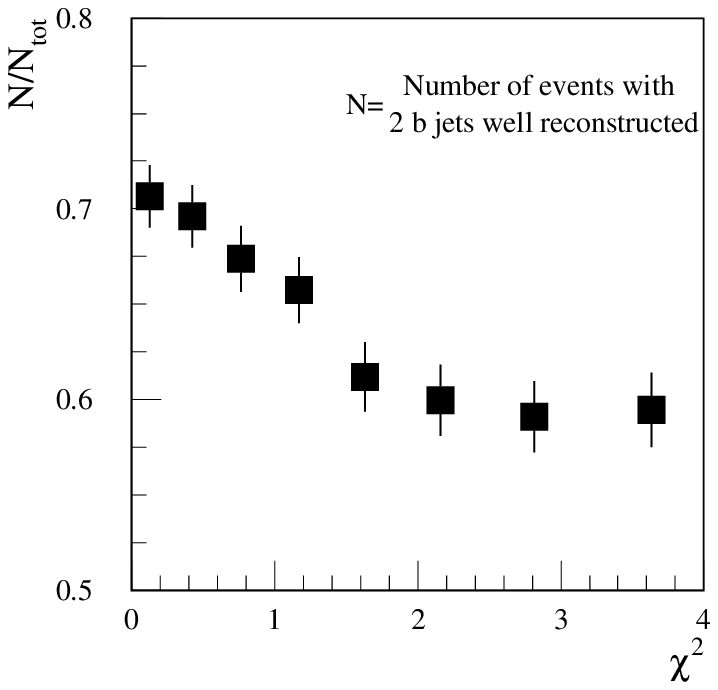}
\caption{\label{lepton_qual} \it{ Left plot: relative difference between
b-jets and b-partons energies. The shaded area defines the ``badly reconstructed''
b-jets. Right plot: probability to have the two b-tagged jets ``well
reconstructed'' as a function of $\chi ^2$.}}
\end{center}
\end{figure}

The constraints which define the $\rm \chi ^2$ are strong on the di-jet and the
lepton-neutrino systems, due to the good knowledge of the W mass, but give a poor
relative constraint on the two b-tagged jets. As a consequence, the $\rm \chi ^2$
is directly related to the quality of the reconstruction of the b-jets, particularly
as the b-jet energy can be underestimated when FSR or leptonic decays occurs.
Furthermore, the quality of the top mass reconstruction can be altered, with a mass
value underestimated in correlation with the effects on the b-jets energy
reconstruction. The left plot of figure \ref{lepton_qual} shows the relative
difference between the b-jets and b-partons energies. The jets which belong
to the tail of this distribution can be defined as badly reconstructed jets.
The probability to have the two b-jets ``well reconstructed'' decreases when
the $\rm \chi ^2$ increases (see figure \ref{lepton_qual}).
The top mass dependence on the $\rm \chi ^2$ is shown in figure \ref{lepton_mtvsx2}.
It can be noticed that the fitted top mass becomes independent on the $\rm \chi ^2$
if only events with ``well reconstructed'' b-jets
are kept. To some extend, the $\rm \chi ^2$ value allows to distinguish between
``well reconstructed'' b-jets from the others.

\begin{figure}
\begin{center}
\includegraphics[scale=0.75]{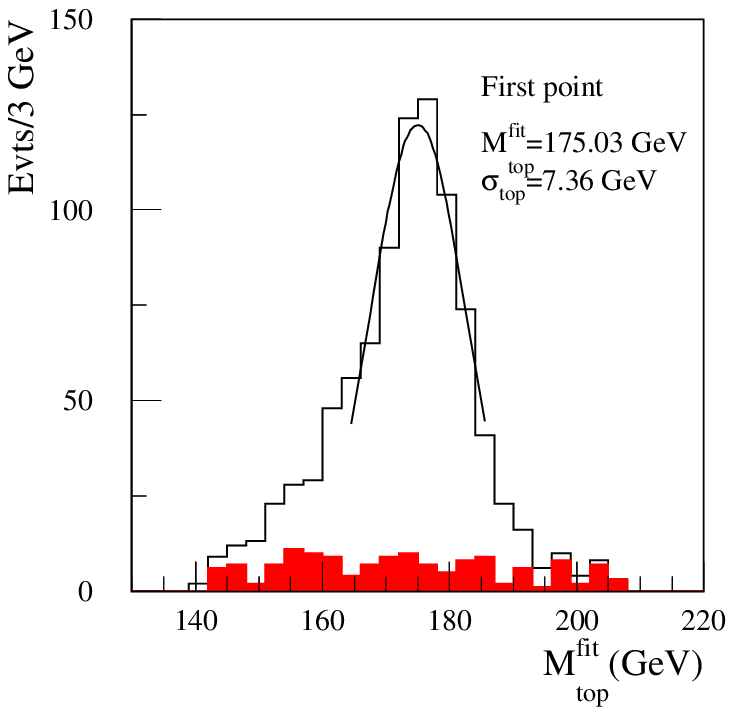}
\includegraphics[scale=0.75]{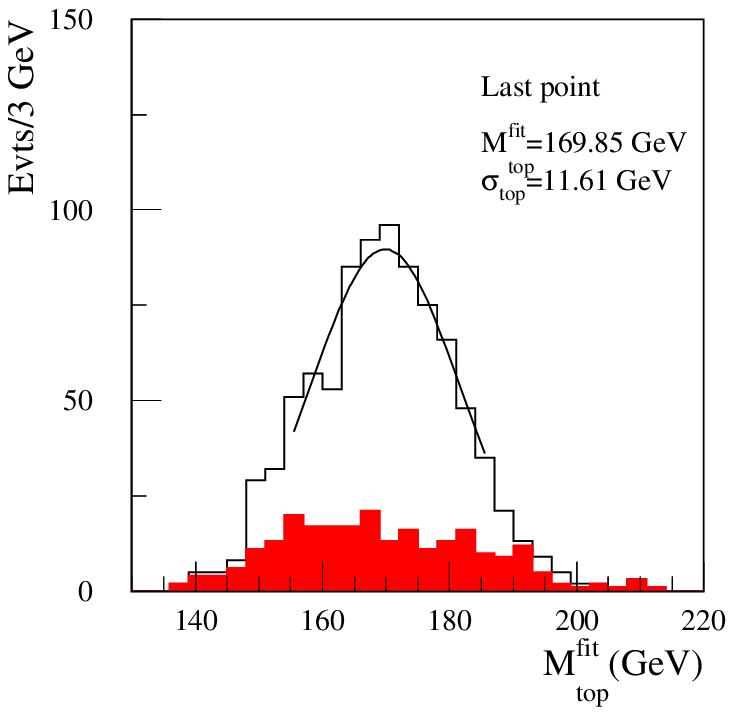}
\caption{\label{lepton_contmt} \it{ Top mass distributions for different values
of $\chi ^2$. One can observe that for the highest $\chi ^2$ the corresponding top
mass is lower, and the mass resolution is higher. The combinatorial background
contributions is shown (shaded area) together with the full invariant mass
distribution (full line).}}
\end{center}
\end{figure}

The top mass is estimated in the following way. Equal samples per slices of
$\rm \chi ^2$ are built. The top mass is computed for each sample, from a
Gaussian fit around $\rm \pm 1.5\sigma$ of the mass peak. Figure
\ref{lepton_contmt} presents two examples, for slices of $\rm\chi ^2$ with
mean values $\rm\chi ^2=0.12\;and\; \chi ^2=3.63$. Finally, the top mass is
determined as $\rm m_{t}=M_{top}^{fit}(\chi ^2 =0)$, from a fit by a linear
function to the distribution (see figure \ref{lepton_mtvsx2}). In one year
of running at low luminosity (per $10fb^{-1}$) the statistical error would be 120 MeV.

\subsubsection{Systematic uncertainties}
\label{sec:systcert}

\begin{figure}[ht!]
\begin{minipage}{0.49\textwidth}
\begin{center}
\epsfig{file=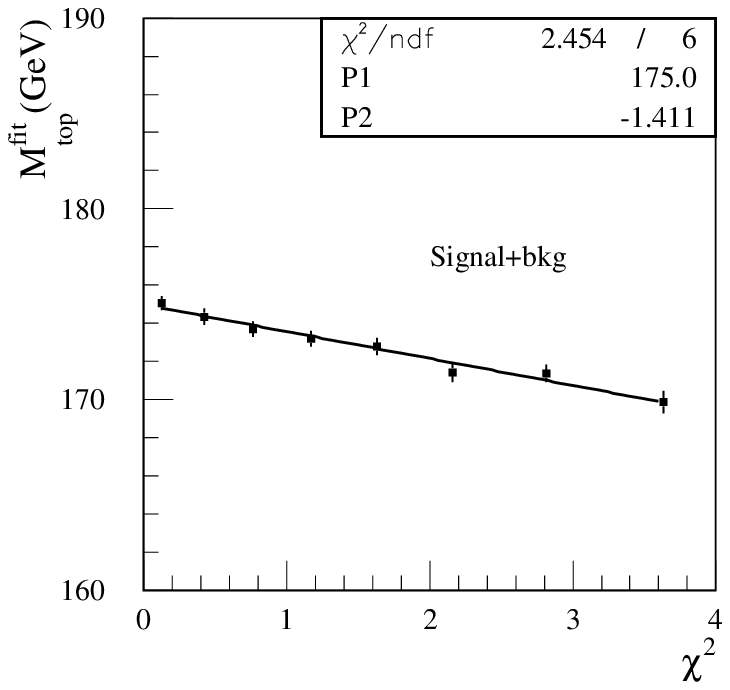,width=6.0cm}
\caption{\label{lepton_mtvsx2} {\it Fitted top mass versus $\chi ^2$.}}
\end{center}
\end{minipage}
\hfill
\begin{minipage}{0.49\textwidth}
\begin{center}
\epsfig{file=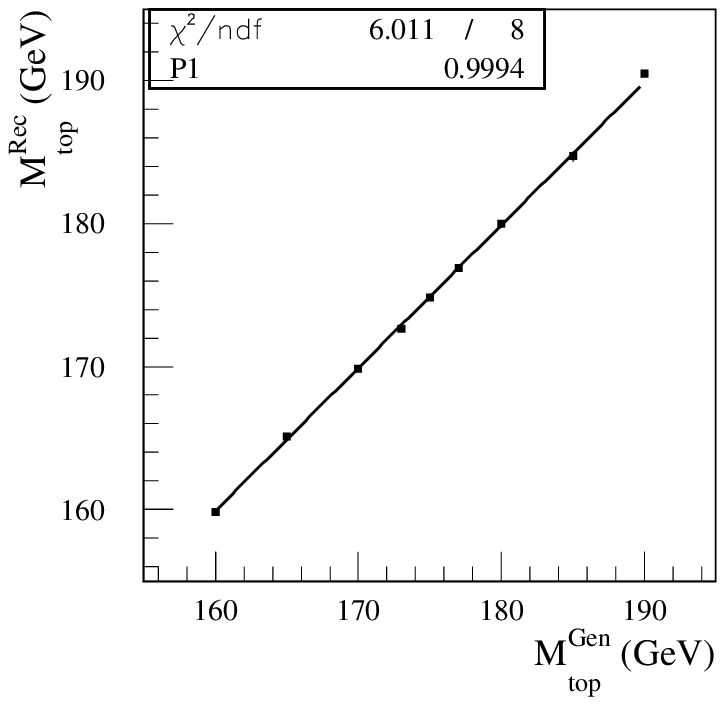,width=6.0cm}
\caption{\label{lepton_mtrvsmtg} {\it Reconstructed top mass versus generated top
mass. The superimposed line is described by the function
$M_{top}^{Rec}=P_1 \times M_{top}^{Gen}$.}}
\end{center}
\end{minipage}
\end{figure}

The study of the systematic uncertainties were handled in the same way as in
the previous section. The results are presented in table \ref{lepton_syst}.
For FSR, taking $20 \%$ of the mass shift obtained between FSR switched on and
off leads to a systematic error on the top mass of 0.5 GeV. However, this estimate
is an upper limit as the mass shift takes into account effects due to a wrong
b-quark jet calibration. This effect could possibly be reduced: if the absolute b-jet scale is
obtained with Z+b events generated with FSR off, the systematic
error would be decreased to 0.1 GeV. This is not surprising, since when
the size of the FSR contribution increases, only the slope in the figure
\ref{lepton_mtvsx2} is modified. This demonstrate that events with large FSR
contributions populate high $ \chi ^2 $ values.

\begin{table}
\begin{center}
\begin{tabular}{l c} \hline
Internal Systematics  & $\rm \delta m_{t}$ (GeV)\\ \hline
light jet energy scale & 0.2 \\
Initial State Radiation & 0.1 \\
Final State Radiation & $\leq$ 0.5 \\
b-quark fragmentation & 0.1 \\
Combinatorial background & 0.1 \\
Total & $\leq$ 0.6 \\ \hline
Statistical error & 0.1 \\ \hline
b jet energy scale & 0.7$\times x\%$ \\ \hline
\end{tabular}
\caption{\label{lepton_syst} \it{Summary of the systematic errors for the top mass
in the inclusive lepton plus jets sample, when the top mass is reconstructed
using a kinematic fit.}}
\end{center}
\end{table}

The linearity of the method has been checked using several samples generated
with different top masses. The same algorithm was used on all samples, with
the same calibration functions for light and b jets. As is shown in figure
\ref{lepton_mtrvsmtg} the estimated top mass depends linearly on the generated
top mass.

In table \ref{lepton_syst}, the errors are divided into three types: i) the
statistical error, ii) internal systematic errors and iii) the systematic
error due to the b-quark jet absolute energy scale which depends on
external inputs. No systematic errors have been accounted for the Monte
Carlo description of the decay as it has been demonstrated to be negligible
\cite{lepton_zjet3}. Assuming the b-quark jet absolute energy scale can be
determined within $1 \%$, the total error on the top mass measurement is of
the order of 0.9 GeV.

In summary:

\centerline{$\rm m_{t}=m\pm 0.1_{\;(stat.)} \pm (0.3-0.6)_
{\;(internal\;syst.)}(\pm 0.7 \times x)_{\;(external\;syst.)}$}

\noindent
where x = b-quark jet miscalibration in $\%$.

\subsection{An additional technique: mass measurement using a continuous jet
algorithm}

The main sources of systematic uncertainty  entering in the top mass measurement
in the inclusive lepton plus jets channel arise from the non-precise knowledge
of the correction factors applied to the jet energy and of some physics processes
parameters.
In order to reduce the impact of the latter effects (mainly final state
radiation) in the top mass determination, an approach based on the continuous
definition of jets has been investigated
\cite{lepton_vadim}.

This approach is based on the following ingredient: a continuous jet definition
and mass estimation.
This idea was first introduced in \cite{lepton_tkachev}. It is based on the
consideration that the discrete nature of the standard jet definition may
cause some problems. Two of these problems are: i) from the mathematical point
of view, the breakdown of continuous distributions gives rise to instabilities
(large statistical fluctuations) and therefore increases the statistical error
on the measurement result, ii) the transition from a continuous energy deposition
in the calorimeter to a fixed structure jet causes the loss of important information.
As an example, let us consider the case of final state radiation.


As a result, the fraction of the
total energy contained in the cone around the quark direction in any standard
jet finding algorithm is subject to large fluctuations which cannot be precisely
predicted from theory. This induces a large contribution to the systematic
errors in the mass measurement.
A continuous jet definition  allows instead to reduce the dependence
of the estimation of the details of the jet shape description in the Monte Carlo
and in the  jet reconstruction procedure.

In addition, this method can be coupled to a jet energy calibration based on the
W mass measurement and on the reconstruction of the $t \bar{t}$ final state with
a constraint kinematic fit, as presented in the previous section.

\subsubsection{Method}

The event selection criteria are similar to the ones used in the previously
described analysis (see section 2.1).
The continuous jet algorithm has been realized in the following way. Initially, a
"fixed cone" jet finding algorithm is used, with a definite cone size.
For the same event, the analysis is then repeated varying the jet cone size. Here,
the cone sizes range from 0.3 to 1.0 with step of 0.1.

In a preliminary stage of the analysis for each cone size a jet energy correction
factor has been defined by calculating the two-jet invariant mass distribution for
non b-tagged jets with  $p_{T}>40$ GeV. The W-peak position has been fitted using
the sum of a Gaussian and Tchebyshev polynomials up to the fourth order.

The same corrections as for the light quark jets have been applied for the b-jets,
even if not being the optimal ones. Two non b-tagged jets with $p_{T}>40$ GeV
have been selected for which the combined invariant mass was close to the mass of
the W peak for a given cone size chosen for the jet reconstruction algorithm.
The requirement \mbox{$|M-M_{W_{true}}|<25~~GeV$} has been used. A b-quark jet
energy rescaling has been applied to the b-tagged jets with $p_{T}>40$ GeV,
according to the correction factor for the given cone size. At least two b-quarks
are required in each event. If there are more, the combination with the highest
$p_{T}$ is selected.

For the given cone size, a constraint fit procedure was then applied to the selected
jets. This kinematic fit uses three constraints $m_{jj}= M_W$, $m_{l\nu}= M_W$ and
$m_{l \nu b} = m_{jjb}$ which, together with the missing momentum measurement, allows
to determine the neutrino momentum unambiguously. Only the energies of the jets are
tuned during the fit.
However, in the method proposed here, a slightly different approach has been exploited
to improve the accuracy of the reconstruction procedure: a robust modification
\cite{lepton_maronna,lepton_andrews} of the fitting functional have been used instead
of the usual $\chi^2$ \cite{lepton_vadim}.

Only the events with a small minimum value of the fitting functional have been
retained. This value is not a $\chi^2$ because of the robust fitting method and
depends on jet and missing momentum error parameterizations. The invariant mass of
the 3-jet system obtained after the constraint fit has then been considered as the
top quark mass estimate for the current event and current jet cone size.

\begin{figure}
\begin{center}
\epsfig{file=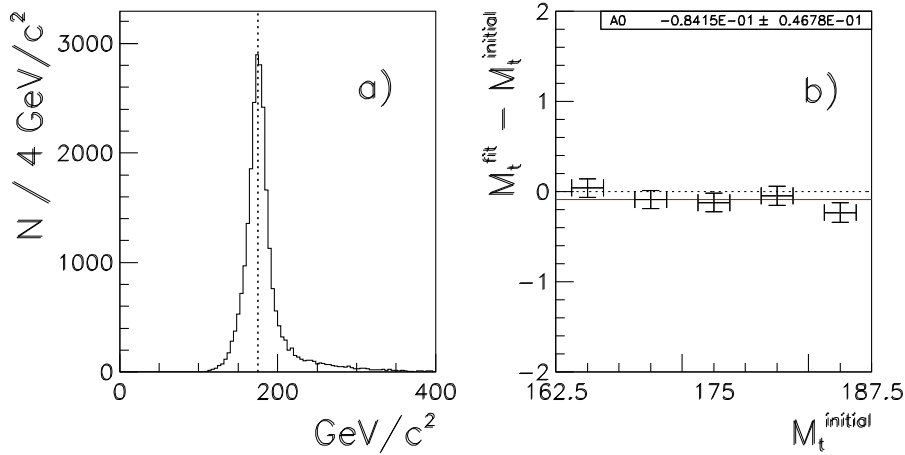,width=12.cm}
\caption{\label{lepton_cont_result} {\it Left plot: Summary distributions of the top
quark mass estimates for all cone sizes. Right plot: Difference between the generated
and reconstructed top quark masses, for different top mass values, after having applied
the mass reconstruction procedure as described in the text. The result of a linear
(constant) fit is also shown.}}
\end{center}
\end{figure}

After having a top quark mass estimate for a certain jet cone size, the cone size
is changed and the whole  procedure is repeated for the given event, starting again
from the jets finding. If the jet finding procedure is unable to find the  required
minimum number of jets with $p_{T}>40$ GeV the algorithm skips to the next event.
All the top quark masses calculated for all the events and all the jet cone sizes
are finally summed up in one single histogram. The  distribution obtained as the
result of the reconstruction procedure is shown in the left plot of
figure \ref{lepton_cont_result}.

\begin{figure}[ht!]
\begin{center}
\epsfig{file=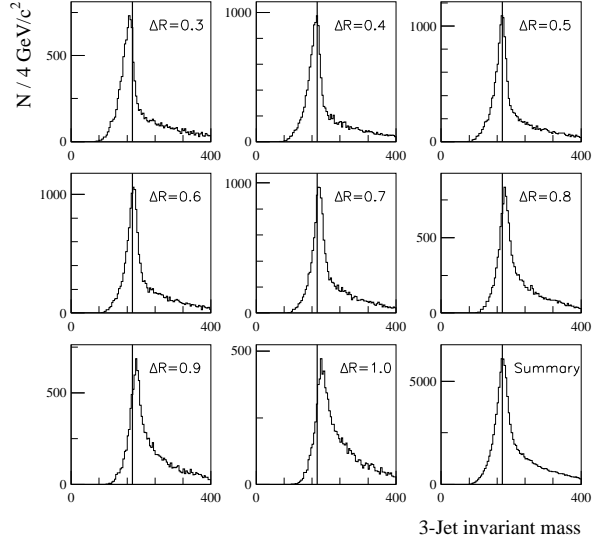,width=9.0cm}
\caption{\label{lepton_contjet} {\it Mass distributions for the three jets
system and for various cone sizes. The vertical line indicates the generated
top mass value. The bottom-right plot represents the sum of all distributions.}}
\end{center}
\end{figure}

For different events, the mass value at the peak position is obtained for  different
cone sizes. The point here is that a position of accumulation should be a more robust
estimator of the jet system invariant mass with respect to the single value which is
obtained by calculation of only one mass for each event. In figure \ref{lepton_contjet}
the 3-jet system invariant masses (the top quark mass estimator) are shown as a function
of the cone size $\Delta R$ as well. The summed distribution for all the $\Delta R$
values is shown at the bottom of figure \ref{lepton_contjet}.

The top quark decay peak is then fitted with a Breit-Wigner function including a
4th order Tchebyshev polynomial to describe the background. The position of the peak
is considered as the final estimate of the top quark mass. The Breit-Wigner shape
has been selected because it gives a much better description of the signal
compared with a Gaussian shape. However, an equally good $\chi^2$ may be obtained
by using a sum of two Gaussian with the same mean for the signal description.
The difference between the results with the two methods has been treated as a
systematic error.

\subsubsection{Results}

\begin{figure}
\begin{center}
\epsfig{file=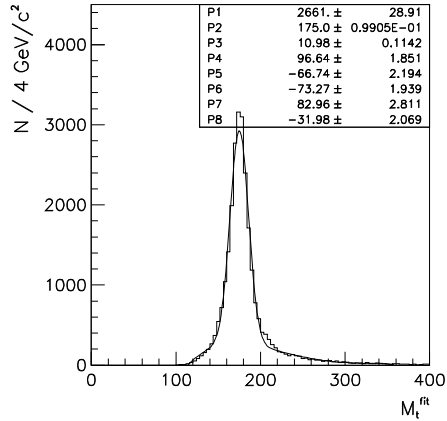,width=6.0cm}
\caption{\label{lepton_cont_fig} {\it Reconstructed top mass. The full fine
represents the result of a fit by a Gaussian plus a 4th order polynomial.}}
\end{center}
\end{figure}

A typical mass distribution obtained by applying the top quark mass reconstruction
procedure described above is shown in figure \ref{lepton_cont_fig} for 400000
$t \bar{t}$ generated events with $m_t^{MC} = 175$ GeV.

A mean statistical error of the top quark mass estimation obtained with the fit of
the mass distribution in figure \ref{lepton_cont_fig} by a Gaussian function plus
a 4th order polynomial background is $\delta m_t \sim 100 $ MeV, for 400000 generated
events.

In order to evaluate the statistical properties of the top quark mass estimation
procedure from $\delta m_t$, some additional steps are needed. Due to the continuous
jet definition method and to the data based b-jet energy correction, the statistical
error given by the fit of the distribution is not correct. The invariant mass distributions
obtained from the same event sample with different cone sizes used by the algorithm are
strongly correlated and cannot be treated as independent. The statistical error of
the b-jet energy correction factors has to be taken into account. Both effects lead to
an underestimate of the statistical error as obtained from the invariant mass
distribution fitting procedure. For the correction, one should rescale statistical
errors from the fitting procedure.
On the other side the mass distributions
obtained with the different jet cone sizes are not the same because even
the number of reconstructed jets in event may be different
for $\Delta R=0.3$ and $\Delta R=1.0$, then the summed distribution has
more information than one obtained with the single $\Delta R$.
So one must rescale a statistical error obtained in the fitting procedure
by a factor in the range $1...\sqrt{8}$.
The determination of such rescaling factor is not
an easy task to be performed analytically. However, one can easily
determine the
corresponding correction in Monte Carlo experiment, with the help of pull
distribution. For each of the top quark masses ($m_t =
165,170,175,180,185$ GeV),
$400000$ events were generated and reconstructed. This procedure has been
repeated five
times. The differences between the generated and reconstructed top masses
are determined for all input top masses and the corresponding pull
distribution is obtained.
The dispersion of the pull distribution
(equal to the scaling factor) is 1.6.
After multiplying $\delta m_t$ by 1.6, taking into account b-tagging
efficiency and rescaling
to an integrated luminosity of 10 fb$^{-1}$,
an estimated statistical error about $\delta m_t
\sim 65$ MeV is found. The final statistical error of the top mass will be less than 100 MeV.

\subsubsection{Systematic errors}

As expected, and as can be seen in figure \ref{lepton_contjet}, the reconstructed
three-jet mass depends on the jet cone size. The final value extracted for the top
mass will, therefore, depend on the choices for the minimum and maximum
cone sizes to use in the analysis.  The analysis, which used a cone size step of
0.1 with minimum and maximum values of 0.3 and 1.0 respectively, was re-done
excluding either the minimum or maximum values and again by shifting the grid
of cone sizes by 0.05.  The largest change observed in the resultant
top mass, namely 250 MeV, has been assigned as the systematic error in the top
mass due to uncertainty in the range of cone sizes to use.

The effect of the $\chi ^2$ cut value on the determination of the top mass has been
checked. The difference between the reconstructed top mass and the generated value
has been plotted versus the $\chi ^2$ cut value. A maximum difference of 200 MeV was
found. This value has been taken as the systematic error due to the $\chi ^2$
dependence of the top quark mass determination.

The description of the background and signal shapes in the fitted invariant mass
distribution has been taken into account. Changing the background description, the
degree of polynomial, the top mass peak shape from a Gaussian to a Breit-Wigner or to
the sum of two Gaussians with common means lead to a systematic error on the top
mass of 90 MeV.

The effects of initial and final state radiation have been computed in the same
way as before. The systematic error due to ISR is found to be negligible and the
error due to FSR is found to be 200 MeV.

Since the light quark jets are calibrated in-situ, a negligible systematic error
in the top mass measurement results from the uncertainty in the light quark jet
energy scale.  For the b-jet
energy scale, a 1\% miscalibration induces a top mass shift of 700 MeV, as
was discussed in section \ref{sec:systcert} for the kinematic fit with fixed cone size.

A study was performed to investigate whether one could reduce the
b-jet energy scale systematic error by calibrating the b-jets
using the same in-situ calibration obtained for the light quark
jets.  Proceeding this way will increase the statistical error,
due to the data based calibration, but also introduce a systematic
shift of the top mass due to any differences of energy losses
between light quark jets and b-jets. These differences are
expected due to physics effects, and also due to detector and
reconstruction effects. The systematic error due to the b-quark
fragmentation parameter $\epsilon_b$ was estimated as before and
was found to be 50 MeV. The b-jet energy scale depends also on the
branching fraction of the semi-leptonic b-hadron decays, due to
the presence of the undetected neutrino. To estimate the influence
of the imprecise knowledge of the semi-leptonic decay fraction of
b-hadrons (which is known with an accuracy of $7\%$), all
semi-leptonic branching ratios were scaled by 1.07 in the Monte
Carlo, resulting in a top mass shift of 60 MeV. Due to the fixed
jet cone reconstruction procedure, the b-jet energy scale depends
on the parton shower evolution and in particular on the $\alpha_s$
evolution parameters. Introducing the b-quark mass corrections
into the default $\alpha_s (p_T)$ evolutions in showers led to a
systematic error of 90 MeV. Combining these uncertainties in
quadrature would give a total systematic error on the top mass due
to the differences of the physics effects between light quark jets
and b-jets of 130 MeV. While this result is very encouraging, it
must be stressed that these comparisons were made using a fast,
parameterized Monte Carlo description of the ATLAS detector.
Further study will be performed with full GEANT-based simulation
of the detector to increase the confidence of this potential
reduction in the systematic error.

The shape of the W signal used to rescale the b-jet energy is another source of
systematic uncertainty. The systematic error was estimated by taking into account
the asymmetric shape of the W signal distribution, changing the background shape
under the W peak and the fitting region. These sources account for a systematic
error on the top mass of 100 MeV.

The various contributions to the top mass systematic error for the continuous
jet analysis are summarized in table \ref{vadim_table}. Adding the various contributions
in quadrature leads to a total systematic error on the top quark mass of the order of
1 GeV, dominated  by the uncertainty in the  external calibration for the b-jets.
Should it be possible to realize the improvement suggested by the study of
b-jet calibration using the in-situ light quark calibration, the total error could
be reduced to of order 400 MeV.

\begin{table}
\begin{center}
\begin{tabular}{l c} \hline
Source  & $\rm \delta m_{t}$ (GeV)\\ \hline
Range of jet cone sizes & 0.25 \\
$\chi ^2$ dependence & 0.2 \\
Signal and background shape & 0.1 \\
ISR and FSR & 0.2 \\ \hline
External b-jet calibration 1\% & 0.7 \\\hline
Internal b-jet calibration &  \\
\ \ \ \ Physics effects & 0.13 \\
\ \ \ \ W signal shape & 0.1 \\ \hline
\end{tabular}
\caption{\label{vadim_table} \it{Summary of the systematic errors for the top mass
in the inclusive lepton plus jets sample, when the top mass is reconstructed
using a continuous jet definition. For more details, see the text.}}
\end{center}
\end{table}

\subsection{Summary}

Three methods to determine the top quark mass in the lepton plus jets channel
have been presented in chapters 2.2, 2.3 and 2.4. In the first method (2.2) the top mass is extracted from the
invariant three jets mass of the hadronic top decay, in the second
method (2.3) the top mass is determined from a kinematical fit of the entire
$t \bar{t}$ decay, and in the third method the top mass is determined from
a kinematic fit and using a continuous jet algorithm.
The main sources of uncertainties arise from final
state radiation and b-quark jet energy scale. It was shown that the
contribution from light quark jets energy scale to the systematics
errors can be reduced to a negligible level using an in situ calibration.
Provided that the b-quark jet absolute energy scale can determined within
$1 \%$, the top quark mass can be measured with a precision at the level
of 1 GeV in one year of LHC running at low luminosity (per $10fb^{-1}$).

%

\subsection{Top mass measurement using large $p_T$ events}
In this section, we present an alternative method which
uses a special sub-sample of the single lepton plus jet events
where the top has high transverse momentum, for example $p_T > 200$ GeV.

\begin{figure}
\begin{minipage}{0.49\textwidth}
\begin{center}
\epsfig{file=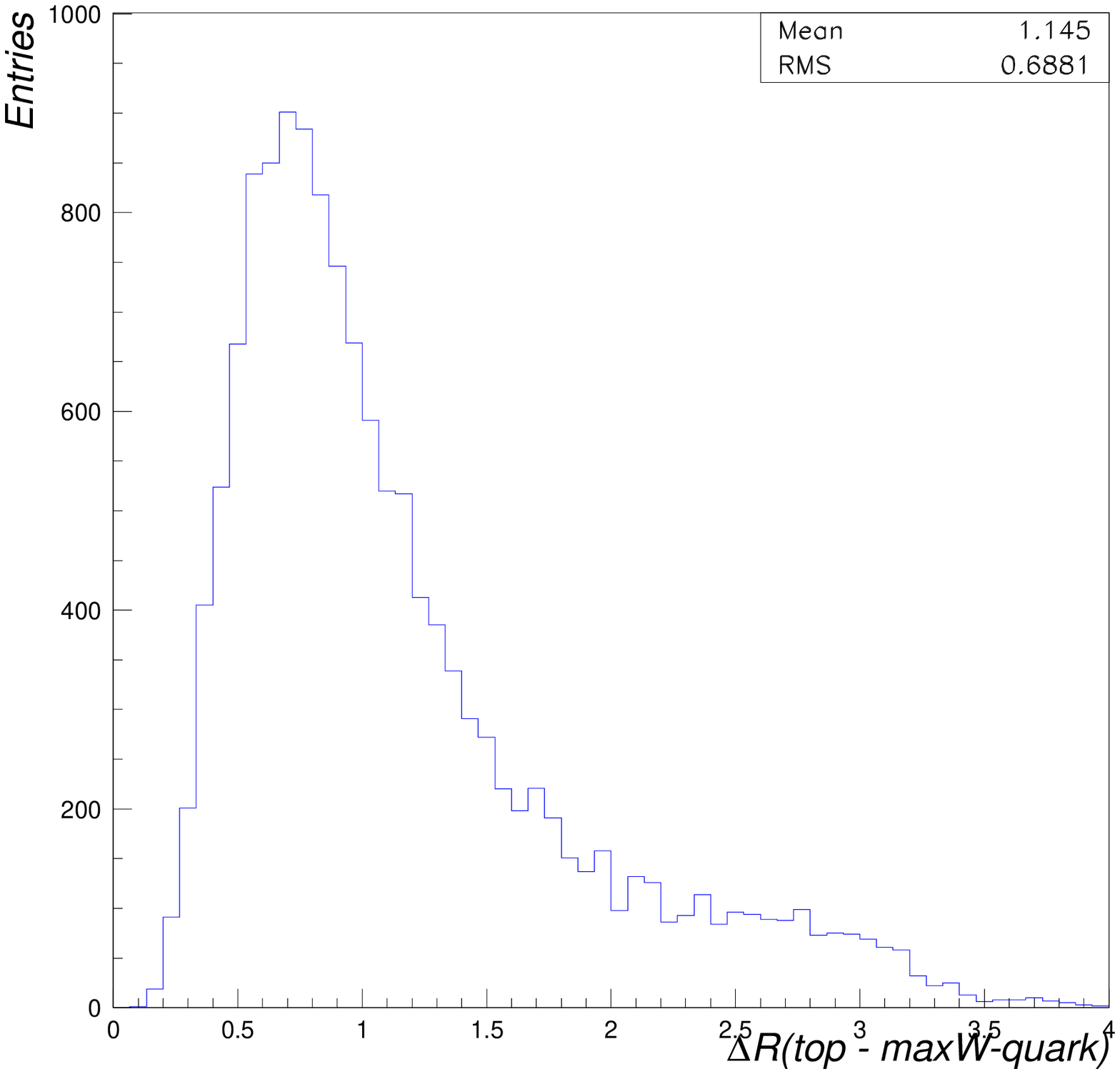,width=6.0cm}
\caption{\label{highpt_dr1} {\it $\Delta R$ distance between the top quark
and the furthest quark from the hadronic W decay.}}
\end{center}
\end{minipage}
\hfill
\begin{minipage}{0.49\textwidth}
\begin{center}
\epsfig{file=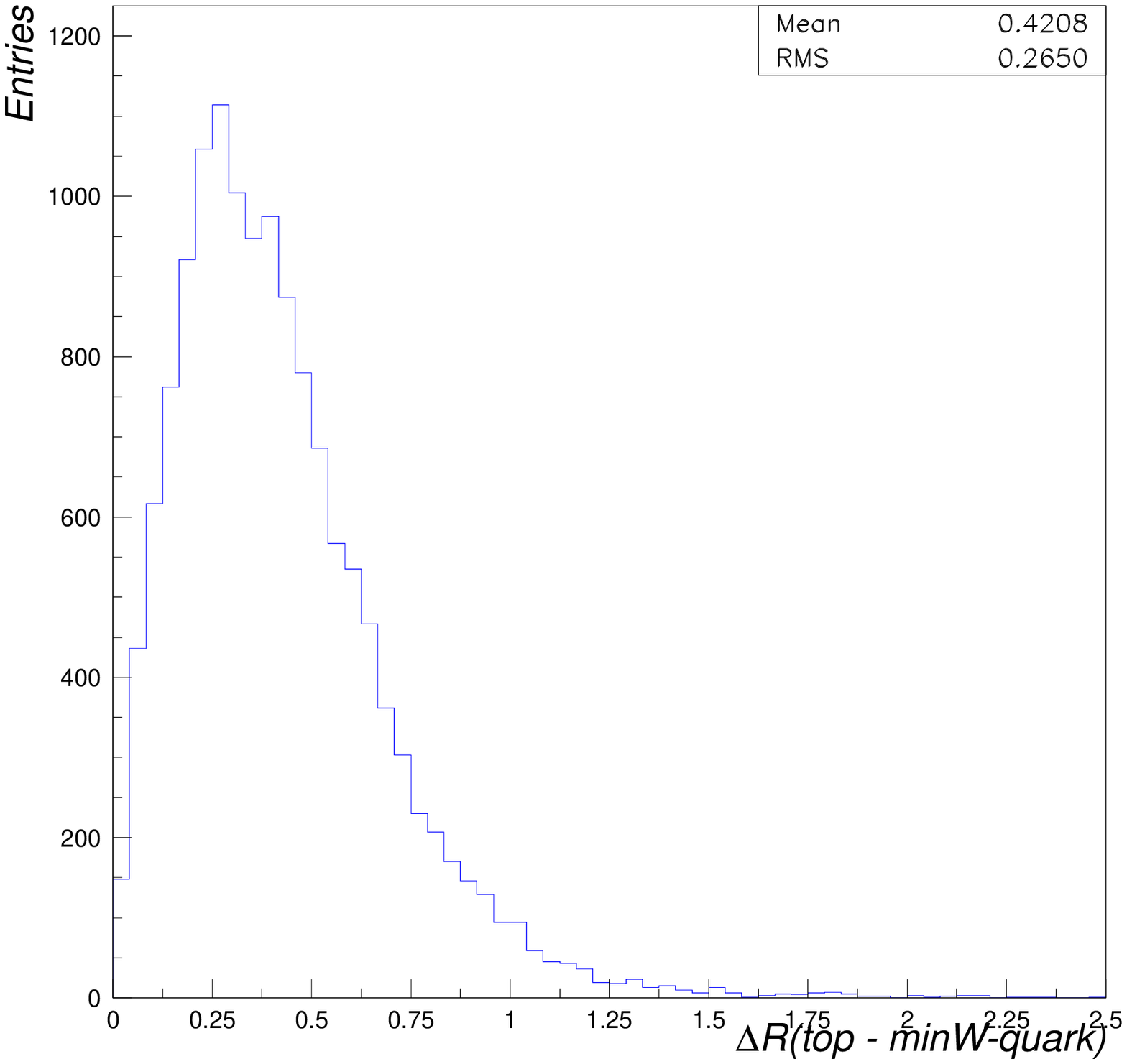,width=6.0cm}
\caption{\label{highpt_dr2} {\it $\Delta R$ distance between the top quark and
the closest quark from the hadronic W decay.}}
\end{center}
\end{minipage}
\end{figure}

\begin{figure}
\begin{minipage}{0.49\textwidth}
\begin{center}
\epsfig{file=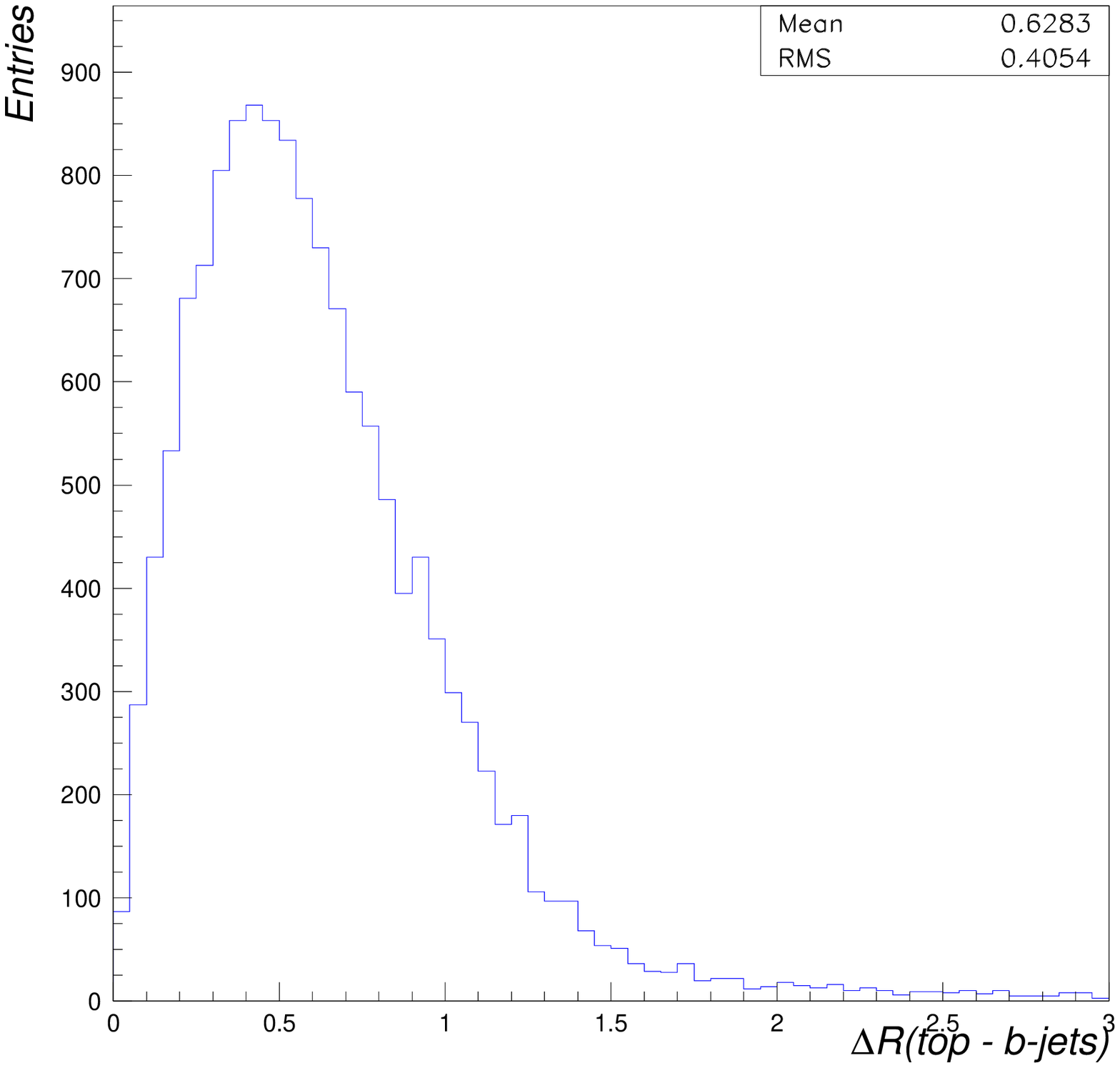,width=6.0cm}
\caption{\label{highpt_dr3} {\it $\Delta R$ distance between the top and the
b-quark at parton level.}}
\end{center}
\end{minipage}
\hfill
\begin{minipage}{0.49\textwidth}
\begin{center}
\epsfig{file=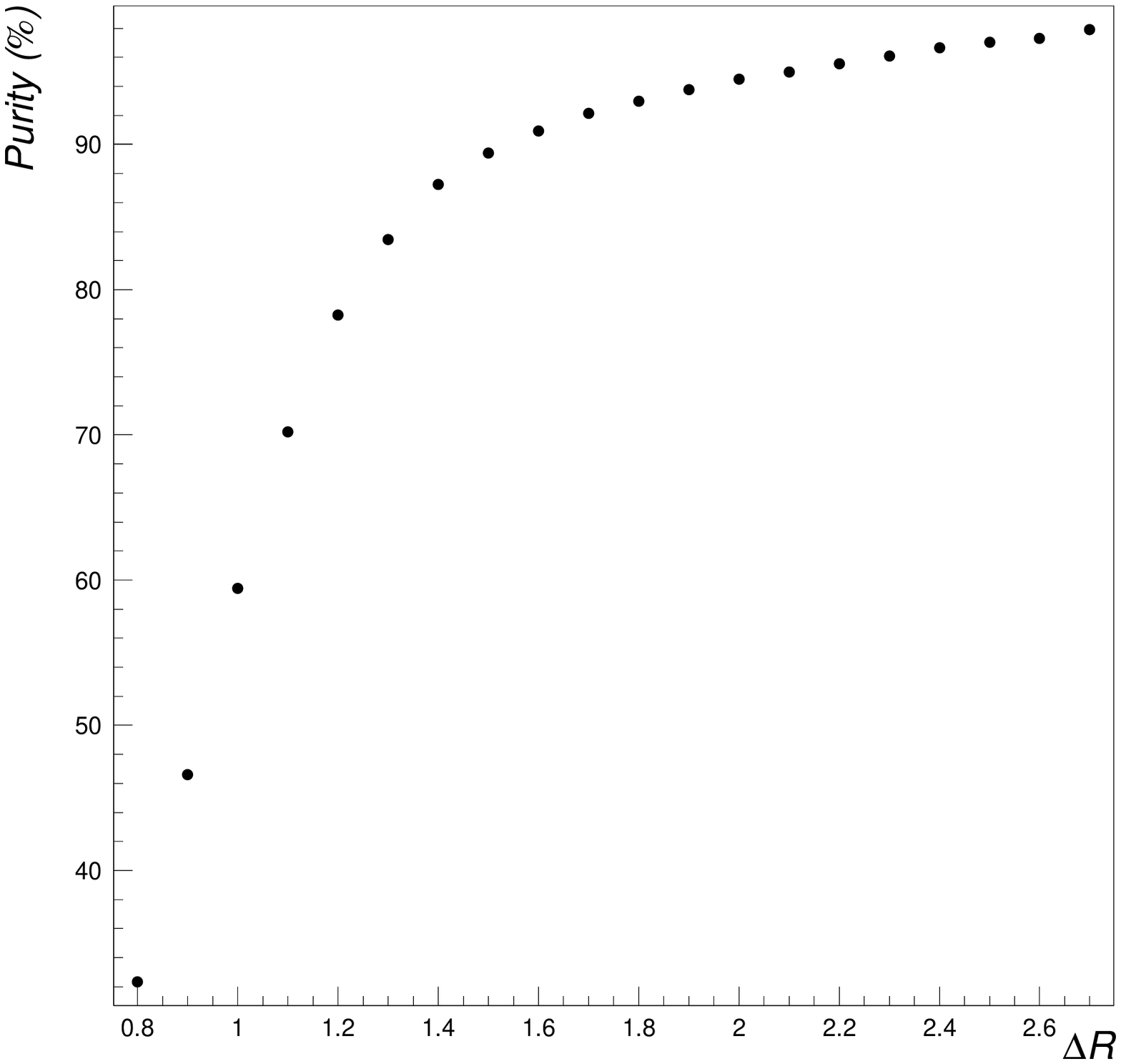,width=6.0cm}
\caption{\label{highpt_seleff} {\it Percentage of the events having all the
three jets from the hadronic top decay within a cone aperture $\Delta R$ from
the top, at parton level.}}
\end{center}
\end{minipage}
\end{figure}

In this topology, the two quarks are produced back-to-back, and the daughters
from the two top decays would appear in distinct hemispheres of the detector:
the {\em ``hadronic''} one from the decay $ t \rightarrow W b \rightarrow jjb$,
and the {\em ``leptonic''} one from the decay
$t \rightarrow W b \rightarrow l \nu b$. Due to the high $p_T$ of the event,
detector systematics as well as backgrounds from other processes are expected
to be very small. The distinct feature of these events which is exploited here,
is the fact that due to the high $p_T$ the three jets from the hadronic top decay
tend to overlap in space, as shown in figures~\ref{highpt_dr1},\ref{highpt_dr2},
and  \ref{highpt_dr3}.

Therefore one could reconstruct the top mass without using the jets as
in the methods described in the previous section, but from summing up the individual
calorimeter towers over a large cone ($\Delta R$ in [0.8-1.8]) around the top
direction. The top direction itself can be determined in two ways: i) as opposite
to the top direction reconstructed in the leptonic decay, where the missing
energy in the event is used to reconstruct the neutrino, and ii) as the direction
of the invariant mass of the three jets in the hadronic top decay.
Figure \ref{highpt_seleff} shows the percentage of the generated events with all
the three jets from the hadronic top decay lying within a distance $\Delta R$
from the top quark direction.

Using directly the calorimeter towers, avoids problems with the jet reconstruction
and energy calibration, one of the major source of systematic errors in the top
mass measurement, and introduces a different set of systematic errors. This method is
therefore very interesting and useful for the final combined top mass determination
by ATLAS.

\subsubsection{Event selection and reconstruction}

Samples of high $p_T$ $t \bar{t}$ events were generated with a cut of
200 GeV in the center of mass of the hard scattering. The cross-section
of this topology corresponds to about $2 \%$ of the total $t \bar{t}$
cross-section. Events are selected to pass the trigger selection and by
requiring one isolated lepton with $p_T > 30 $ GeV and $| \eta | < 2.5$,
the transverse missing energy greater than 30 GeV ($E_T^{miss} > 30$ GeV),
and at least four jets reconstructed using a cone of $\Delta R$ = 0.4 with
$p_T > 40 $ GeV and $| \eta | < 2.5$, of which two must be tagged as b-jets.

The overall efficiency is 9$\%$, resulting to
$\sim$15000 selected events per (10 fb$^{-1}$)~\footnote{
To save computing time, the studies presented here are done using the muon
channel only, and assuming similar efficiencies for the electrons. Several ATLAS
studies have demonstrated that this is a good approximation when working with
high $p_T$ objects as in this analysis.}. Due to the high $p_T$ of the event and
the requirement for two tagged b-jets, the background (mainly W+jet, WW or
QCD events) is reduced into negligible levels and therefore not discussed
further.

For the events passing the preselection cuts described above, the top quark
direction was determined as described. First the hadronic W invariant mass was
reconstructed from the two highest $p_T$ non b-tagged jets. Combinations where
$m_{jj} = M_W \pm 20$ GeV were selected. The two jets were combined with the
closest b-jet to reconstruct the top. Finally, the reconstructed top $p_T$ was
required to be above 235 GeV. After all the cuts $\sim$3600 events remain per
$10^{-1}$ fb, with an overall efficiency of $2 \%$.

Once the top direction is determined, the invariant mass of all the calorimeter
towers ($\Delta\eta\times\Delta\phi = 0.1\times0.1$) around this direction is
evaluated according to the formula:
\begin{displaymath}
m_{clust}^{2}(\Delta R) = (E^{2} - p^{2}) = (\sum_{i=1}^{n(\Delta
  R)}E_{i})^{2}-
(\sum_{i=1}^{n(\Delta R)}\vec{p}_{i})^{2}.
\end{displaymath}
where $E_i$ is the total calorimeter energy in the i-th tower evaluated in
electromagnetic scale, and $\vec{p}_i$ is its three momentum vector. The index
$n(\Delta R)$ runs over all the towers within the selected cone radius. This
invariant mass is directly proportional to the top quark mass:
$m_{clust}=m_{clust}^{top}$.

\begin{figure}[ht!]
\begin{minipage}{0.49\textwidth}
\begin{center}
\epsfig{file=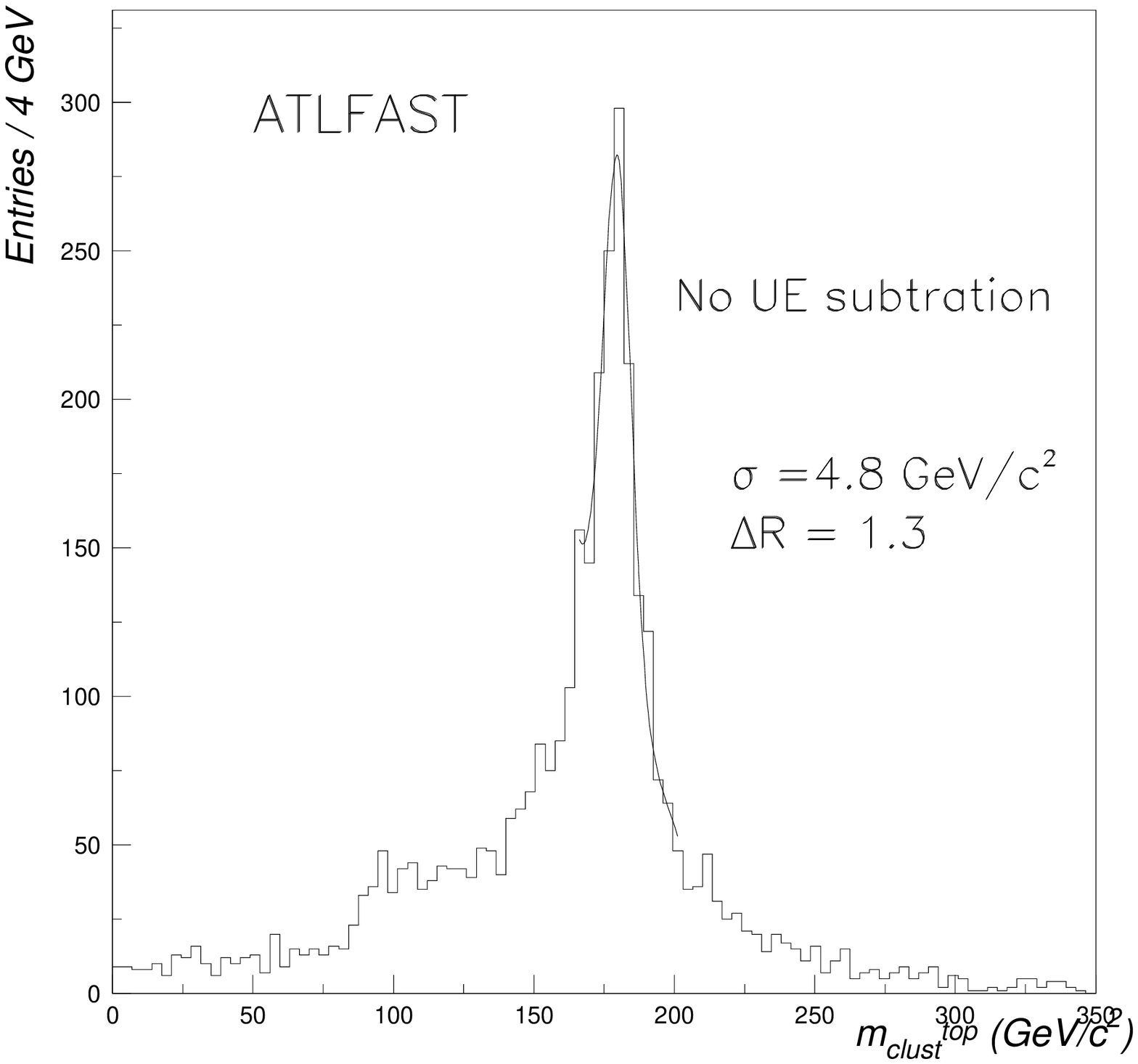,width=6.0cm}
\caption{\label{highpt_mt-cone-ue-on} {\it Reconstructed $m_{clust}^{top}$
spectrum for $\Delta R$=1.3.}}
\end{center}
\end{minipage}
\hfill
\begin{minipage}{0.49\textwidth}
\begin{center}
\epsfig{file=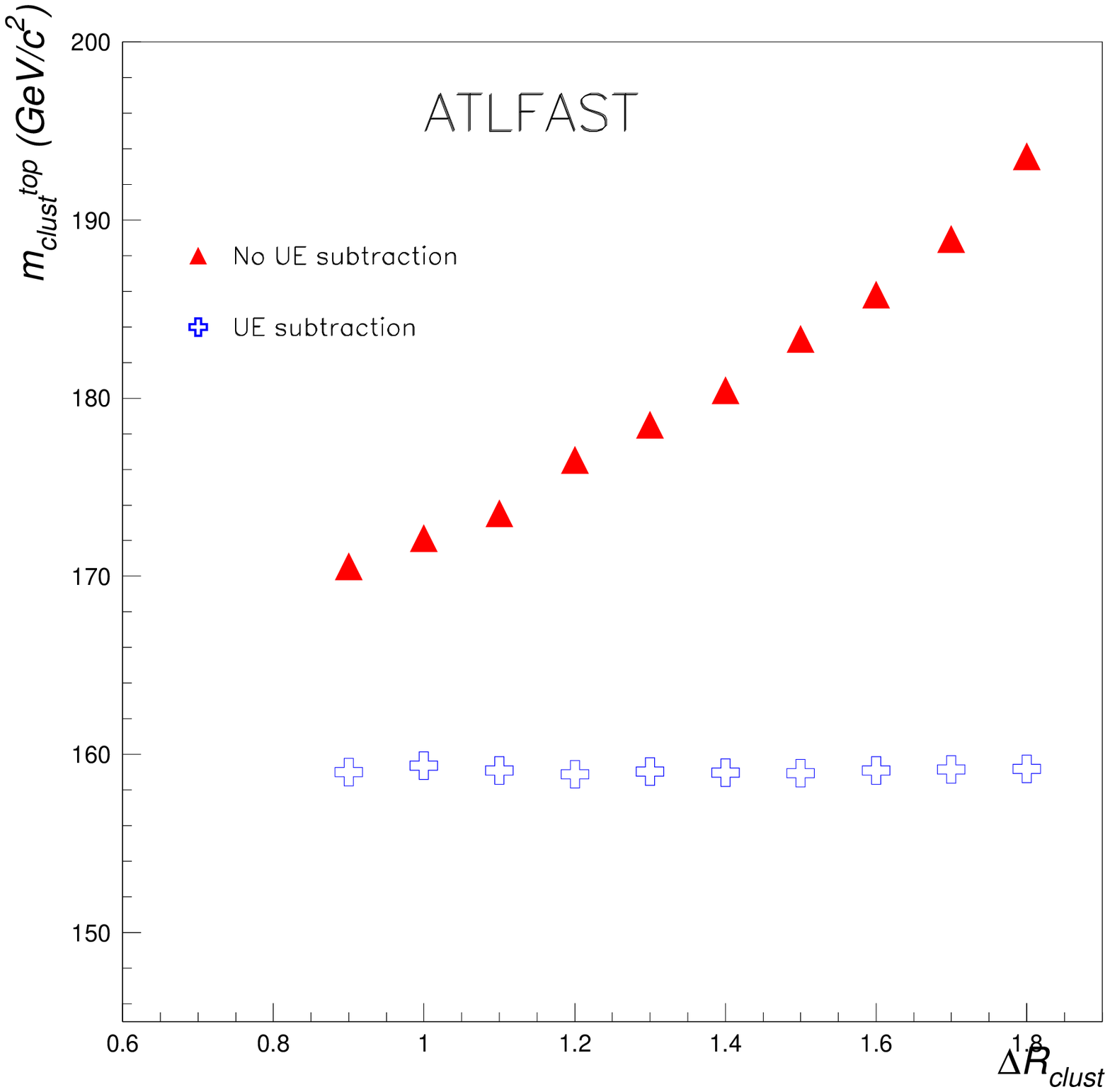,width=6.0cm}
\caption{\label{highpt_mt-vs-dis} {\it Fitted $m_{clust}^{top}$ invariant mass,
as a function of the cone size, before and after the $UE_{tower}$ is subtracted.}}
\end{center}
\end{minipage}
\end{figure}

Figure \ref{highpt_mt-cone-ue-on} shows the reconstructed $m_{clust}^{top}$
invariant mass for a cone size of $\Delta R$=1.3. A clear Gaussian distribution
is observed with the peak value around the nominal top mass. Fitting the peak region
with a Gaussian, we obtain a peak width of 9.6 GeV, comparable to that obtained
with the jet method.

As shown in figure \ref{highpt_seleff}, more than 80$\%$ of the events
where the three jets are at $\Delta R \leq 1.3$ from the top quark direction
are selected.

The invariant mass $m_{clust}^{top}$, is evaluated using various cone sizes
in the range from $\Delta R$=0.8 to 1.8, as there is no reason a priori to select
a given value. On the contrary, the resulting invariant mass has to be independent
from the cone size used. For each $\Delta R$ cone size, a Gaussian is fitted
around the peak position in the invariant mass $m_{clust}^{top}$ distribution.
The fitted peak values for the different cone sizes are shown in
figure \ref{highpt_mt-vs-dis}.
The variation observed can be attributed to the Underlying Event
contribution which is added for each calorimeter tower, resulting in an
increased invariant mass value as the cone size increases. A method to
evaluate the UE contribution in each tower follows.

\subsubsection{The Underlying Event (UE) estimation}

The Underlying Event contribution per calorimeter tower ($UE_{tower}$) was
estimated from the same high $p_T$ top sample. It represents the average
transverse energy $E_T$ deposited per calorimeter tower in each event, once all
the towers related with the high $p_T$ products are excluded.
The values as well as the number of towers used in each case have been computed
for different rapidity regions \cite{highpt_ilias}.
An average over all rapidity and isolation cut range, gives a value of
$UE_{tower}$=447.5 MeV, which is subtracted from the energy of
each tower in the calculation of $m_{clust}$.

In figure \ref{highpt_mt-vs-dis} the invariant mass $m_{clust}^{top}$ is
shown after the the $UE_{tower}$ is subtracted. The resulting values are now
independent of the cone size, with an average value of 159 GeV and with
all values within $\pm$0.15$\%$. Varying the $UE_{tower}$  by $\pm$10$\%$,
a cone size dependence is again observed, raising to a bit less than $\pm$2$\%$,
which demonstrates that the value used is the correct one and gives the precision
required for the target top mass measurement error. In ATLAS, once the real
data become available, the $UE_{tower}$ will be calculated in situ as done
here, but also using other event samples, resulting to an overall error of
about $10 \%$.

\subsubsection{Mass scale calibration}

\begin{figure}
\begin{minipage}{0.49\textwidth}
\begin{center}
\epsfig{file=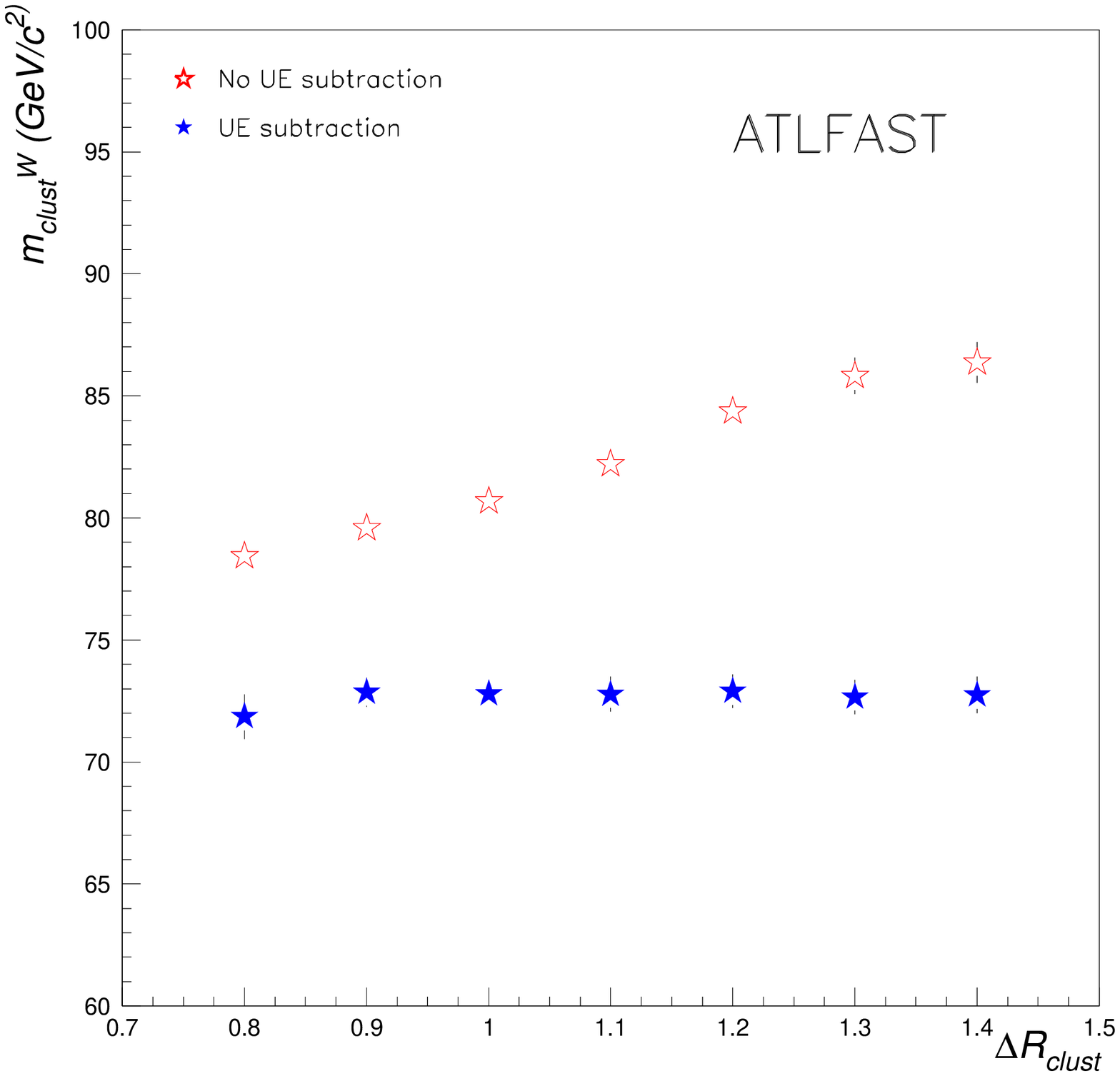,width=6.0cm}
\caption{\label{highpt_mjjinc-vs-dis} {\it The fitted $m_{clust}^{W}$ invariant
mass, before and after UE subtraction.}}
\end{center}
\end{minipage}
\hfill
\begin{minipage}{0.49\textwidth}
\begin{center}
\epsfig{file=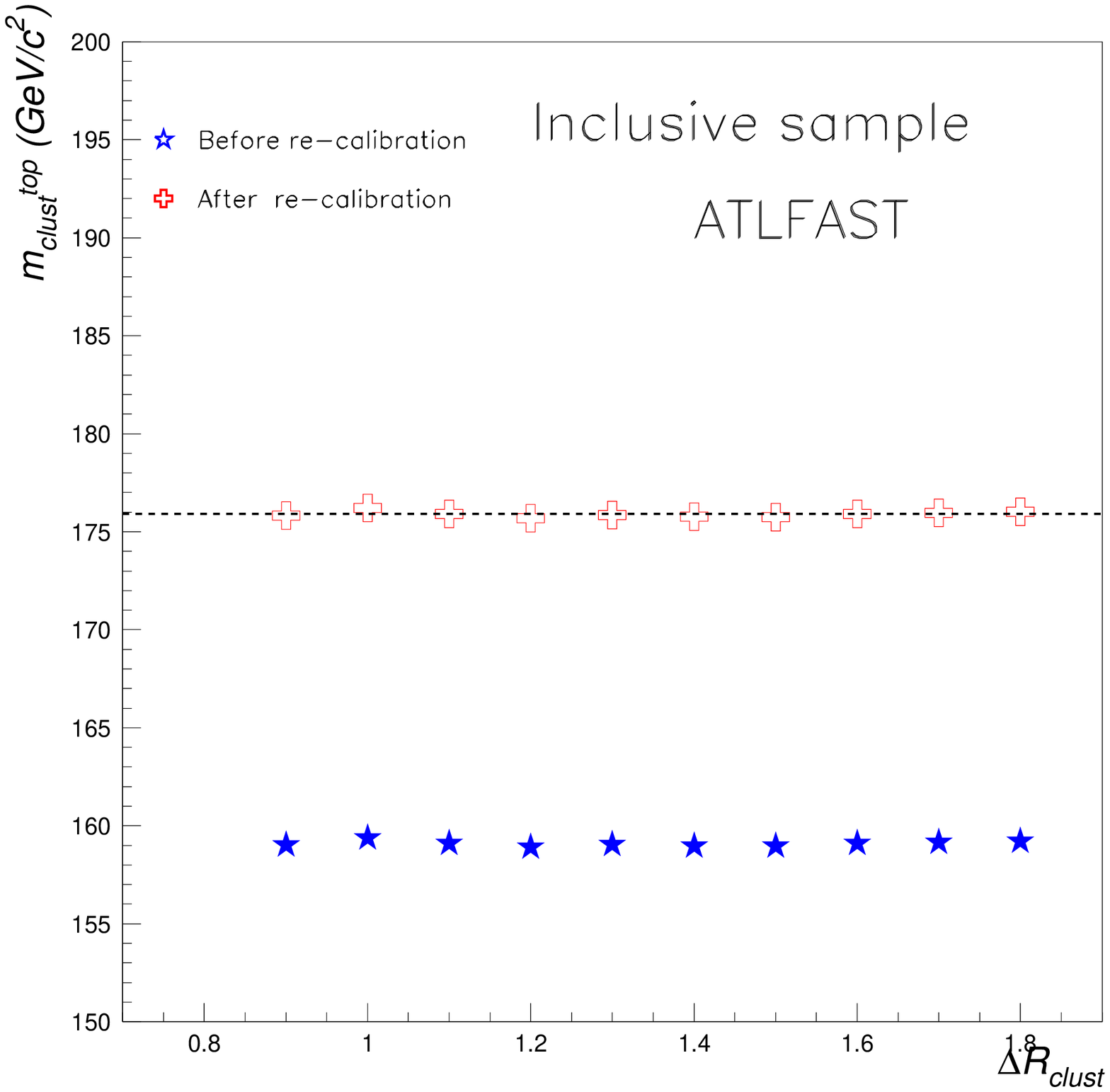,width=6.0cm}
\caption{\label{highpt_mt-resc-winc} {\it The fitted $m_{clust}^{top}$ invariant
mass, after $UE_{tower}$ subtraction and after mass scale is applied. }}
\end{center}
\end{minipage}
\end{figure}

After the $UE_{tower}$ contribution is subtracted, the
reconstructed $m_{clust}^{top}$ invariant mass has become
independent of the cone size,  but the resulting values are now
$\sim$9$\%$ lower than the generated top quark mass. This a priori
is expected as no particular mass scale or absolute energy scale
has been used so far. As a first attempt the $m_{clust}^{top}$
values could be calibrated using the Monte Carlo data. However
doing so the method will be dependent on the exact modelling of
the process and won't be anymore a ``direct'' measurement of the
top mass. The best way is to obtain a mass scale calibration using
the data themselves.

The method studied here is to apply the same reconstruction procedure, with
the same $UE_{tower}$, to other known particles with well measured masses and
extract from there the necessary mass scale calibration factors. In our case
the easiest way is to use the inclusive top sample and apply the same
reconstruction method to the W mass reconstruction. This sample offers high
statistics, and has practically the same event topology as the high $p_T$ sample.
Rescaling the corresponding $m_{clust}^{W}$ invariant
mass values to the nominal mass of the W, we can obtain the mass scale
calibration coefficient $C_{top}$ averaging all the cone sizes $\Delta R$. Finally, to
determine the top mass, the average value of all the $m_{clust}^{top}$ values
after calibration is used.
Fixing the mass scale with the W, and then transferring the results to the top, it
implies that the same calibration is used for the both the light quarks and the b-jets.

Events were generated in the single lepton plus jets topology,
without $p_T$ cut applied to the hard
scattering process. The large statistics available, allow to
apply tight cuts in order to select events where the two jets from the
hadronic W decay are close in space (as in the high $p_T$ sample) and at
the same time far away from the b-jet. Events were selected by requiring an
isolated lepton with $p_T > 20$ GeV and $| \eta |<$2.5, $E_T^{miss}>20$ GeV,
at least four jets (reconstructed in a cone of $\Delta R$=0.4), with $p_T >40$
GeV and $| \eta |<$2.5, of which two are tagged as b-jets. In addition, the
distance between the two highest non b-tagged jets, should be $\Delta R <$1.3 and
the two b-jets of the event should
be at a distance $\Delta R \geq$2.0 away from the reconstructed W direction.

The two highest $p_T$ non b-tagged jets were used to reconstruct the W and
find its direction. Then the $m_{clust}^W$ invariant mass was calculated
around this direction subtracting from each tower the same $UE_{tower}$ as
calculated before. In figure \ref{highpt_mjjinc-vs-dis} the fitted value of
the invariant mass $m_{clust}^W$ are shown. As for the top, the reconstructed
values after the UE is subtracted become independent from the cone size
(within $\pm$0.7$\%$) and about 7.7 GeV lower than the nominal W mass.

Figure \ref{highpt_mt-resc-winc} shows the resulting $m_{clust}^{top}$ values
after the mass scale calibration is applied. The variation between the points
is $\pm$0.22\%. As an example, for $\Delta R$=1.2(1.3) the $m_{clust}^{top}$
invariant mass after the $UE_{tower}$ subtraction was 158.9(159.0) GeV, and
after applying the calibration becomes 175.7(175.8) GeV. Taking the average
of the calibrated $m_{clust}^{top}$ invariant mass for all cone sizes, a value
for $m_t = 175.9$ GeV is obtained, which is within 0.5$\%$ from the generated
top quark mass.

\begin{figure}
\begin{minipage}{0.49\textwidth}
\begin{center}
\epsfig{file=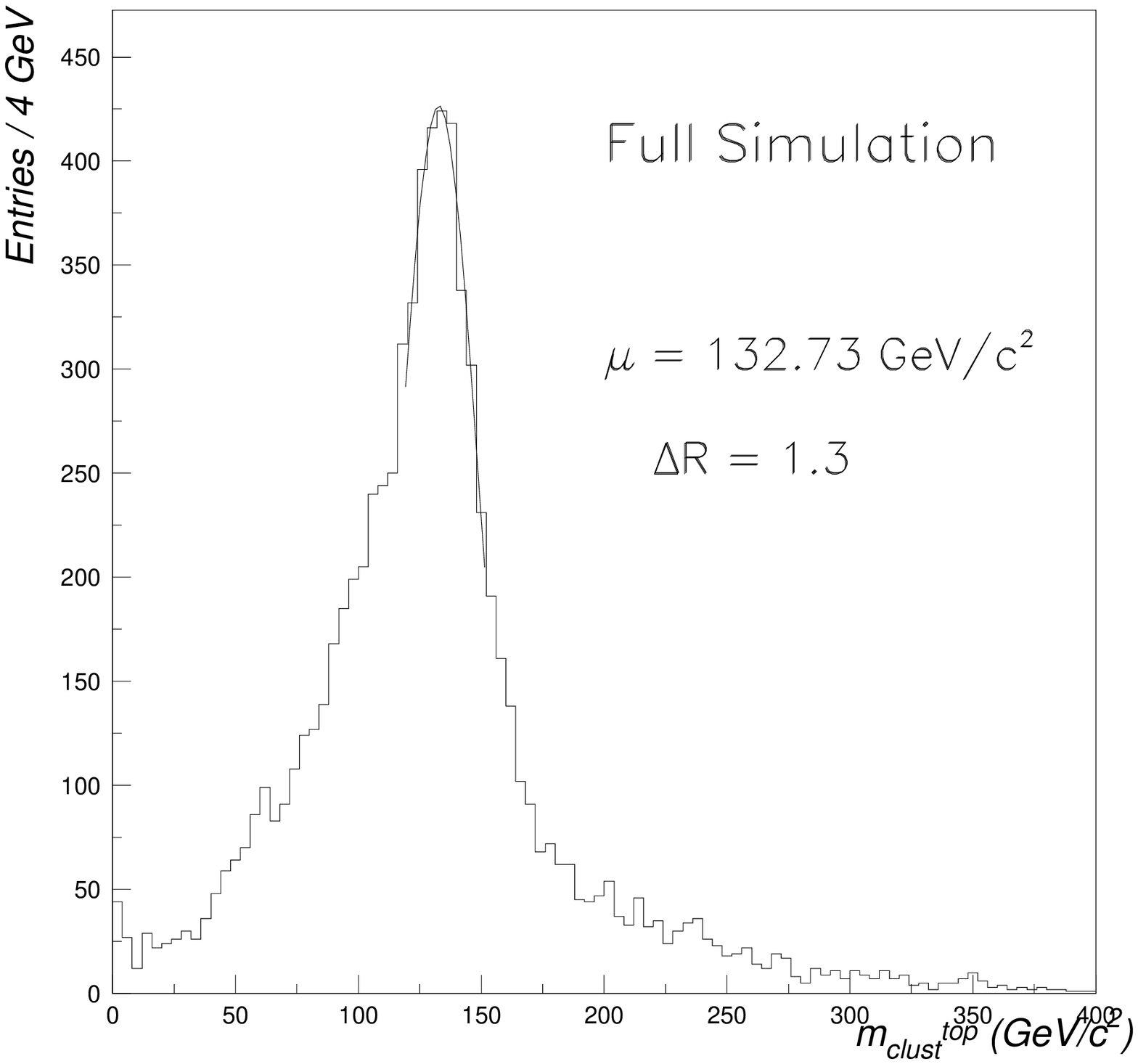,width=6.0cm}
\caption{\label{highpt_mtcone-full-ueon} {\it Reconstructed $m_{clust}^{top}$
spectrum obtained using a cone size of $\Delta R$=1.3 around the top direction.}}
\end{center}
\end{minipage}
\hfill
\begin{minipage}{0.49\textwidth}
\begin{center}
\epsfig{file=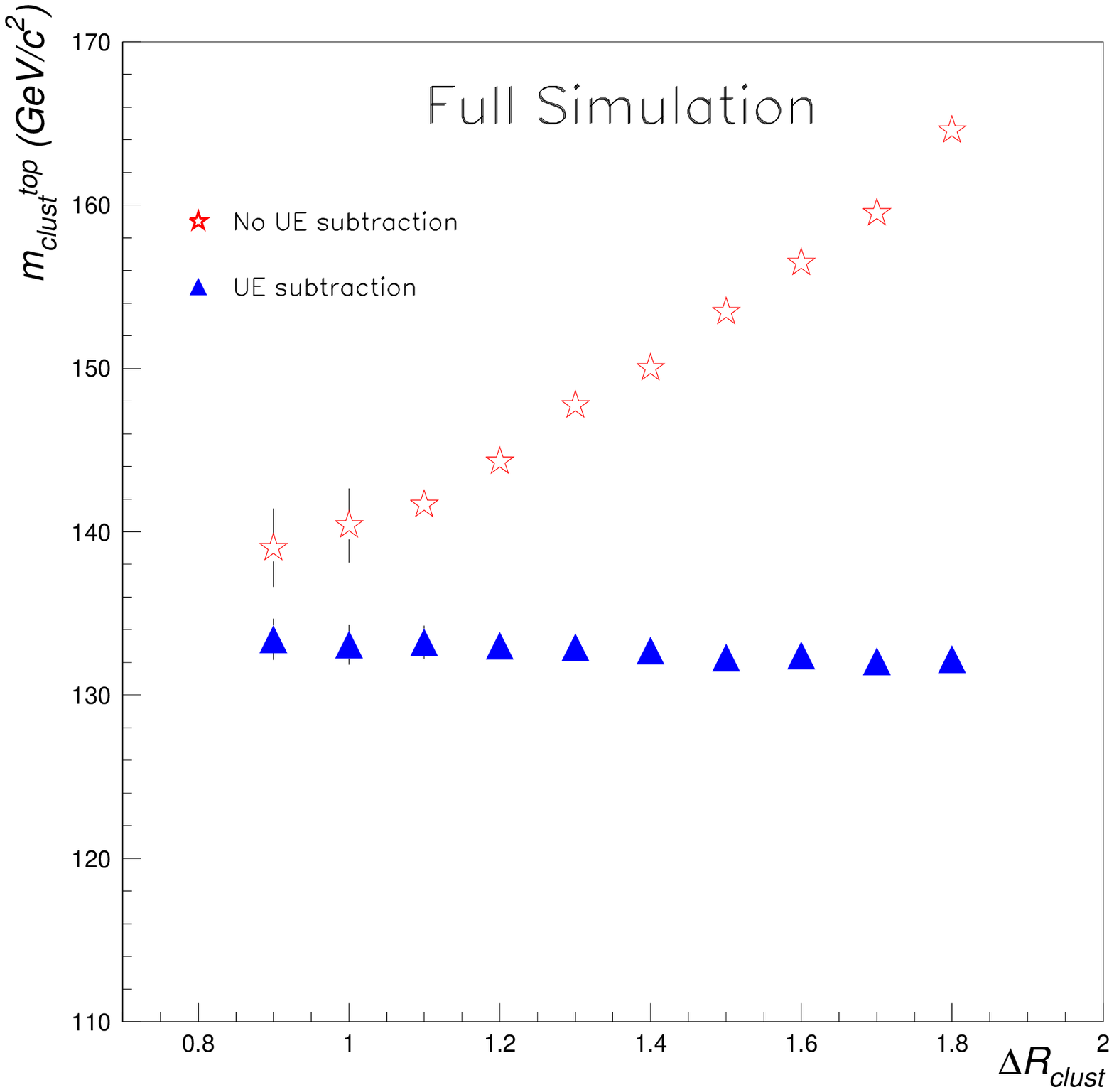,width=6.0cm}
\caption{\label{highpt_mt-vs-rcone-full} {\it The $m_{clust}^{top}$ invariant
mass dependence on the cone size before and after the $UE_{tower}$ is subtracted.}}
\end{center}
\end{minipage}
\end{figure}

\subsubsection{Full simulation results}

The results presented so far were obtained using the fast detector simulation
\cite{atlfast}. The same analysis was repeated with a sample of full (GEANT-based)
simulated events of the ATLAS detector. Figure \ref{highpt_mtcone-full-ueon}
shows the reconstructed $m_{clust}^{top}$ invariant mass spectrum for a cone size
of $\Delta R$=1.3. The variation of the fitted values for different cone sizes is
shown in figure \ref{highpt_mt-vs-rcone-full}.

Although the peak values in this case are lower than those of the fast simulation,
the overall variation for the same cone size range stays about the same. The
difference between the fast and full simulation can be attributed to the
shower shape development which is not included in the fast simulation.

The $UE_{tower}$ was evaluated following the same procedure as before. The average
$UE_{tower}$ is now 42.5 MeV \cite{highpt_ilias}, much lower than the fast simulation. Since
now there are more calorimeter towers contributing but with lower energy in each, compared
to the fast simulation case where the energy for each particle is deposited to a single tower.
This value was used for all the full simulation studies described below.

The $m_{clust}^{top}$ invariant mass after the $UE_{tower}$ subtraction is shown in
figure \ref{highpt_mt-vs-rcone-full} as a function of cone size used. As
expected, it remains basically independent from the cone size, but lower
by $24.6 \%$ from the generated top mass. The variation observed, $\pm$0.9$\%$,
is bigger than with the fast simulation sample, and can be attributed to the poor
quality of the fits due to the lack of statistics. Using only the points up to
$\Delta R$=1.4 the average value is 133 GeV, with a variation of $\pm$0.2$\%$.

\begin{figure}[hb!]
\begin{minipage}{0.49\textwidth}
\begin{center}
\epsfig{file=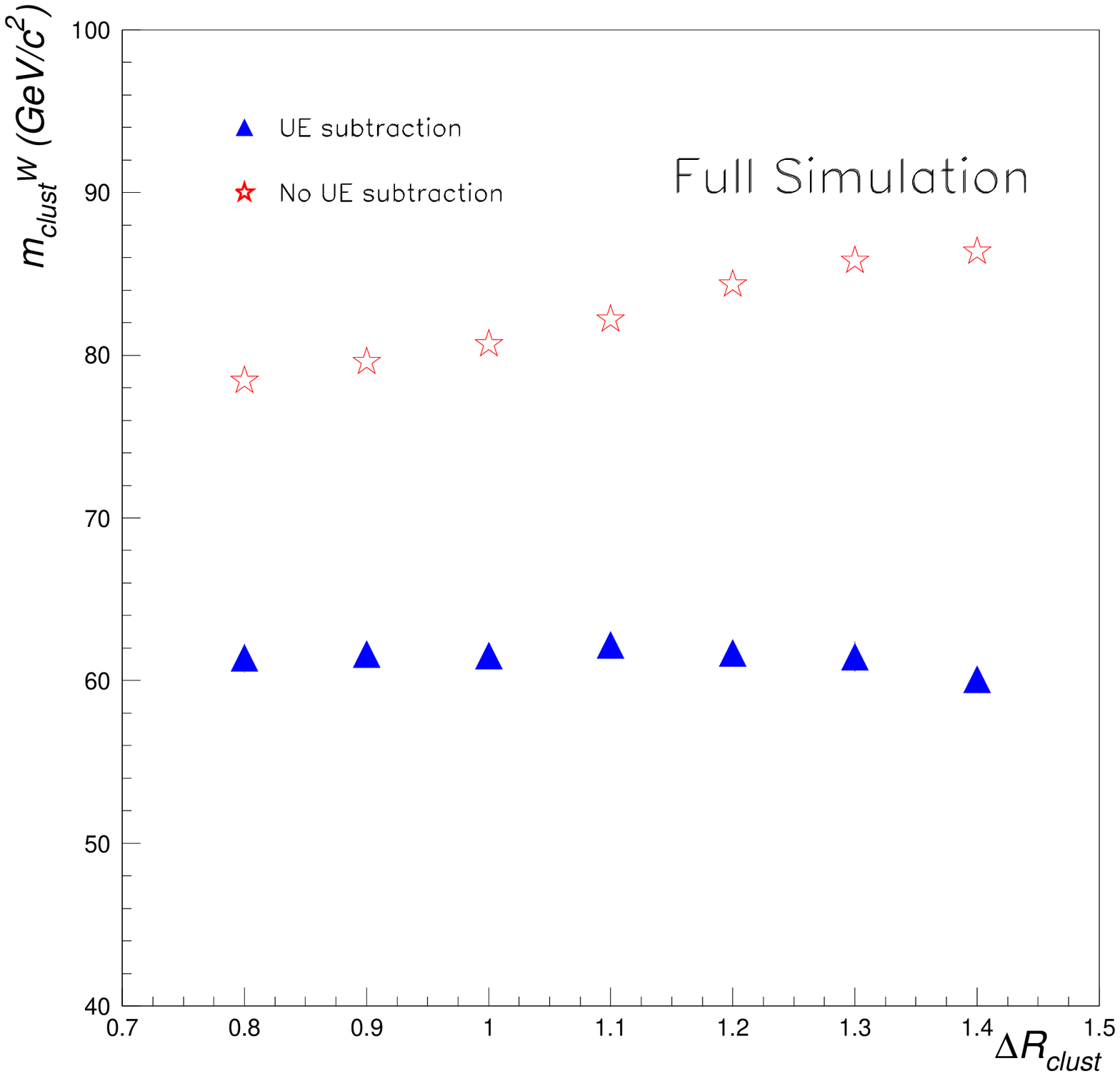,width=6.0cm}
\caption{\label{highpt_mw-vs-rcone-full} {\it The fitted $m_{clust}^W$ invariant
mass, before and after the $UE_{tower}$ is subtracted.}}
\end{center}
\end{minipage}
\hfill
\begin{minipage}{0.49\textwidth}
\begin{center}
\epsfig{file=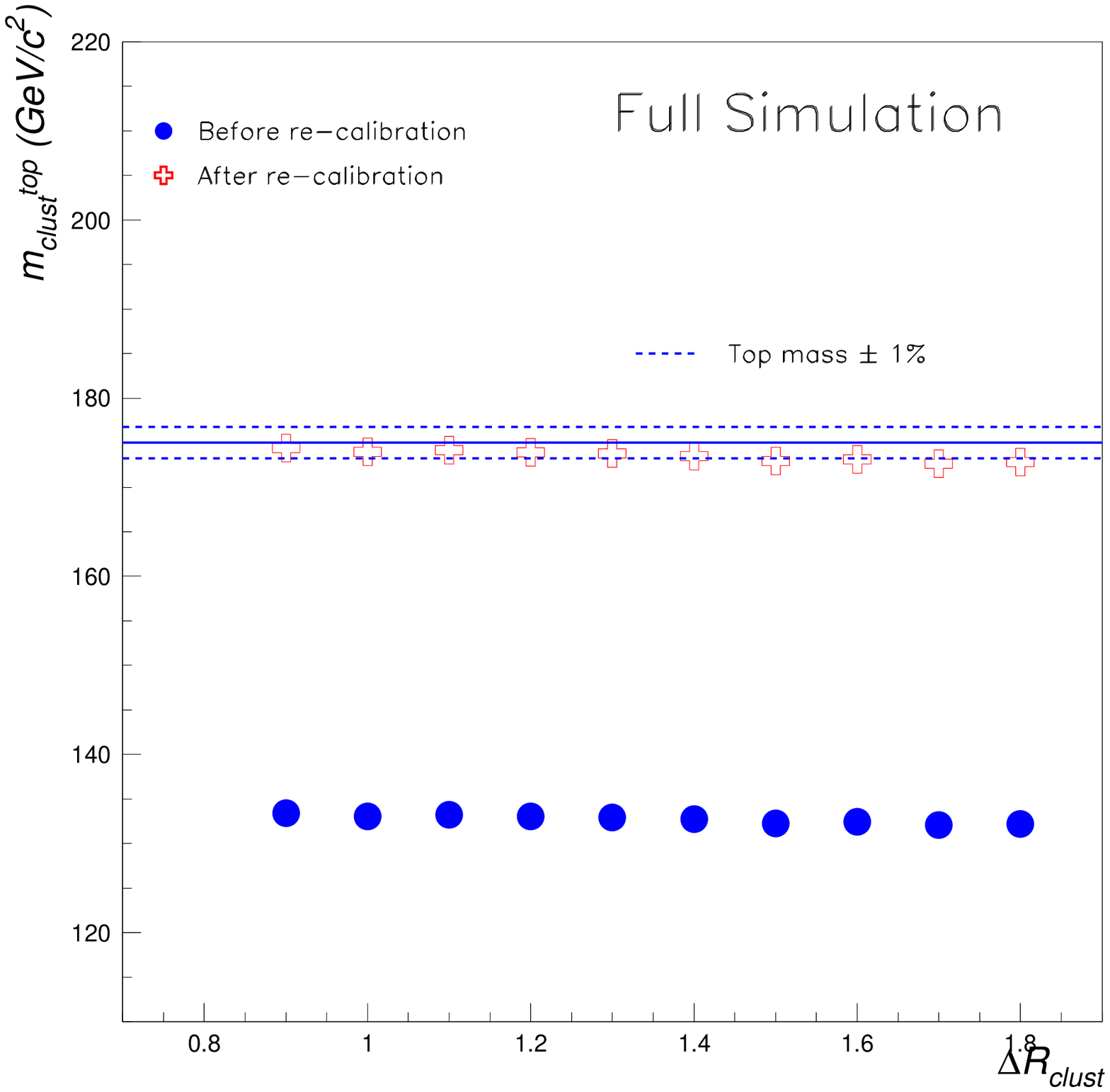,width=6.0cm}
\caption{\label{highpt_mt-resc-winc-full} {\it The fitted $m_{clust}$ invariant
mass before and after rescaling and $UE_{tower}$ subtraction and as a function of
the cone size.}}
\end{center}
\end{minipage}
\end{figure}

The same mass scale calibration procedure was used, with a sample of 30000 fully
simulated inclusive $t \bar{t}$ events, applying the same cuts and reconstruction procedure.
The fitted $m_{clust}^W$ peak values for different values of
$\Delta R$\ are shown in figure \ref{highpt_mw-vs-rcone-full}, before and after
the $UE_{tower}$ is subtracted.
After the UE contribution is subtracted, the resulting $m_{clust}^W$ invariant mass
remains independent from the cone size but 23.5$\%$ below the nominal W mass.
The mass scale is determined as before, and $m_{clust}^{top}$ invariant mass values
after the calibration factors are applied are shown in figure \ref{highpt_mt-resc-winc-full}.
The resulting
values are constant within $\pm 1.4 \%$ from the generated top quark mass, for
the whole range of cone values used. The average $m_t$ is 172.6 GeV,
which is 1.4$\%$ below the generated top mass, and with all points within
$\pm$0.9$\%$. Restricting to the range up to $\Delta R \leq$1.4, where basically
we run out of statistics, the $m_t$ changes to 173.8 GeV, with the points
now having a spread of $\pm$0.3$\%$.

In summary, the full and fast simulation data show the similar results and confirm the
competitiveness of the proposed reconstruction method. However, further studies with larger
statistics samples should be made in particular for high $\Delta R$
values.

\subsubsection{Systematic uncertainties}

Several studies have been performed which cover most of the possible systematic
errors. Studies requiring large statistics and several settings of generator parameters
were performed with the fast detector simulation, while for the cases where the exact
detector response is important samples of full detector simulation were used. In some
cases where large computing effort was required only the sensitivity of the results was
investigated without going to details.
More information on these studies can be found in \cite{highpt_ilias}.

To study the linearity of the reconstructed top mass several samples of high $p_T$ top events
with different input top quark
mass in the generator from 160 GeV to 190 GeV were produced and analyzed
in exactly the same way. For all the samples, the same $UE_{tower}$ and mass scale
calibration factors obtained as explained before were used. In the mass range
170-180 GeV, the reconstructed top mass is in very good agreement with the
generated value. For larger values a bigger error is observed which is somehow
expected as the event environment changes and the UE and mass scale calibrations
are not optimal anymore.

The sensitivity to initial (ISR) and final (FSR) state radiation
was studied in the same way as in other analysis. Samples of high
$p_T$ top events were generated with the ISR or FSR contributions switched
off at the generator level, and the analysis was repeated keeping the same
$UE_{tower}$ estimate and the mass scale calibration factors. Doing so, and for a
cone radius $\Delta R=1.3$ a shift in the reconstructed top mass of 0.7(0.3) GeV is
observed when ISR(FSR) was not present. It has to be pointed out this error is
a very pessimistic approach, as the exact level of the ISR(FSR) contributions in the
events will be measured and known at LHC to about 10\% level and the generators will
be correctly tuned to this. Therefore as in the other analysis, the final error
quoted for the top mass is 20\% of the total mass shift observed, equal to 0.1 GeV.

The sensitivity on the b-quark fragmentation was studied by generating samples
where the $\epsilon_b$ parameter in the Peterson formula was varied within its
error currently at 0.0025 \cite{lepton_peterson}. The top mass was reconstructed
in each sample using the same $UE_{tower}$ and mass scale. The observed top mass
shift among the samples is quoted as the error due to this effect. As an example,
for a cone size of $\Delta R$=1.3 the mass shift was 0.3 GeV. Similar values
obtained for other cone sizes.

\begin{figure}[ht!]
\begin{center}
\begin{minipage}{0.49\textwidth}
\epsfig{file=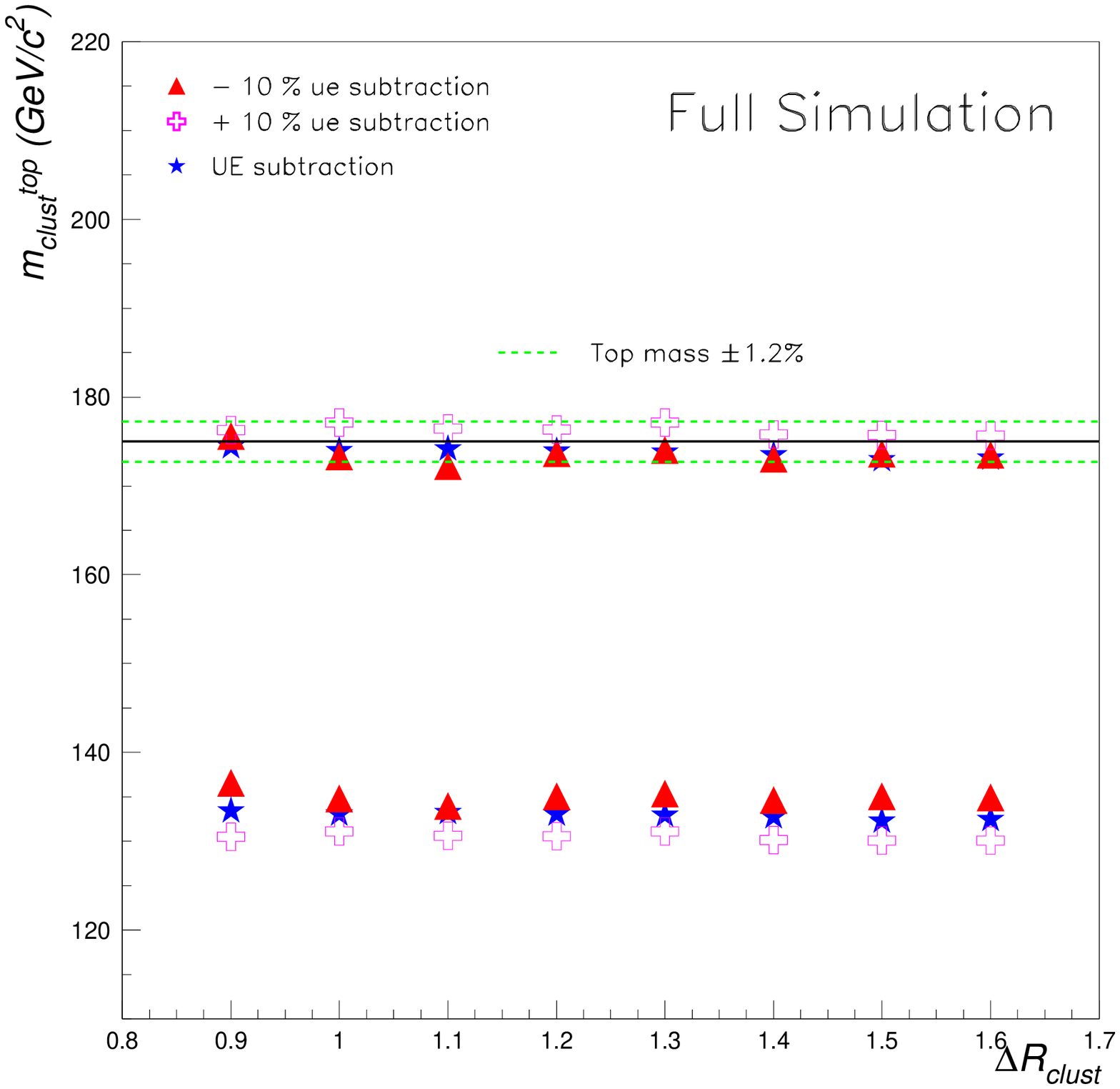,width=6.0cm}
\caption{\label{highpt_uesyst-full} {\it Reconstructed $m_{clust}^{top}$ mass
dependence on the $UE_{tower}$ energy estimate. At each case the mass scaling
factors have been recalculated as explained in the text.}}
\end{minipage}
\hfill
\begin{minipage}{0.49\textwidth}
\epsfig{file=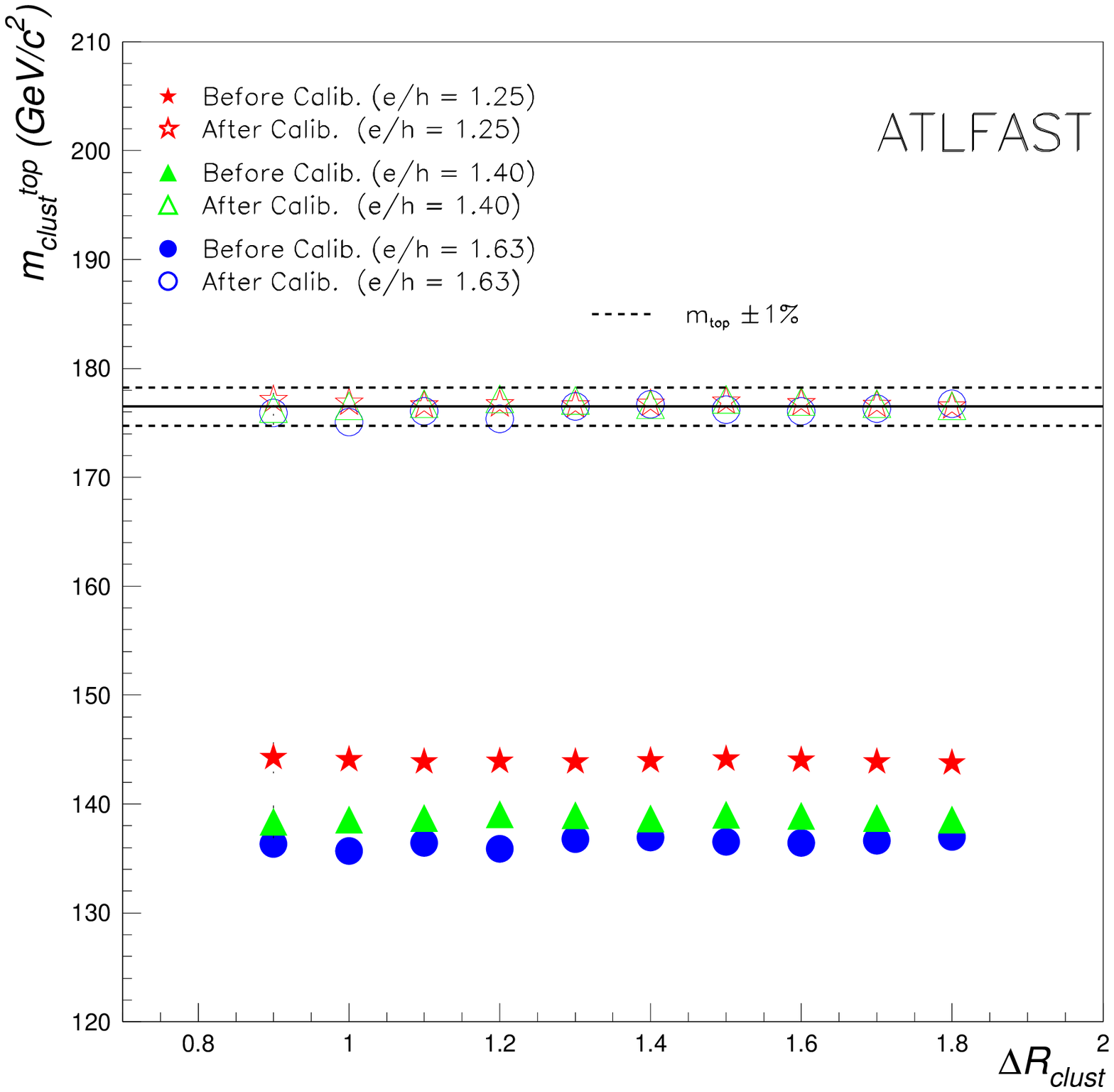,width=6.0cm}
\caption{\label{highpt_mtcalib-lin-compa}{\it Reconstructed $m_{clust}^{top}$
mass values before and after the rescaling from the W reconstruction in the
inclusive sample for three values of $e/h$.}}
\end{minipage}
\end{center}
\end{figure}

The UE energy estimate per tower plays an important role in this top reconstruction
method. To evaluate the sensitivity of the reconstructed mass due to this, the value
calculated (447.5 MeV for the fast and 42.5 MeV for the full simulation data) was
varied by $\pm$10\% and the top reconstruction and the mass scale calibration was
repeated each time. As shown in figure \ref{highpt_uesyst-full} the reconstructed
values stay well within $\pm1 \%$ of the generated value.

To study the contribution of a possible calorimeter mis-calibration in the top
mass measurement, the energy in each tower was varied according to a Gaussian
with different values of sigma from 1 to 5\%, well beyond the expected reach by
ATLAS. The analysis was repeated in each case keeping the $UE_{tower}$ unchanged.
For a cone size of $\Delta R=1.3$ the observed shift in the reconstructed top mass
is 0.6(1.2) GeV when a mis-calibration of 1(5)\% is added (0.7(1.3) GeV in full
simulation). This value however is rather pessimistic, since whatever the cell
mis-calibration would be the $UE_{tower}$ would be computed accordingly, and the
whole error will be incorporated into the mass scale calibration.

Tests with the ATLAS calorimeter prototypes have shown that the
combined $e/h$ is different than 1.0 \cite{highpt_caloeovh}.
Moreover it was demonstrated that in both fast and full detector
simulation programs the $e/h$ effect on the calorimeters is not
correctly treated \cite{highpt_iecalor97}. To study the
sensitivity of the reconstructed top mass due to this, given that
in this method the individual calorimeter towers that combine
information from several calorimeters is used, several fast
simulation samples were generated where the energy deposited by
hadrons has been corrected according to the Groom's formula
\cite{highpt_groom} thus simulating values of $e/h$ between 1.0
and 1.63. The whole analysis was repeated in each case. Changing
$e/h$ to 1.63, a rather extreme value, the $UE_{tower}$ estimate
changes from 447.5 MeV to 417.5 MeV, and the reconstructed top
mass by 0.7\%, as expected from the mass scale calibration
procedure.

Since for the top mass reconstruction the whole calorimeter volume
is used, either to calculate the UE contribution either to
evaluate the top mass, the presence of electronics noise may have
an impact on the results. To study this, a test using the full
simulated data was performed, where the electronics noise was
added as Gaussian noise to the cell energies according to the
expected values for each calorimeter. Repeating the same procedure
as before, the $UE_{tower}$ value changes from 42.5 MeV to 16 MeV,
a surprising low value. Following the same procedure and applying
the mass scale calibration, the average reconstructed top mass, is
175.17 GeV but shows a variation within $\pm$ 1.2\% for the full
range of cone sizes used between 0.9 and 1.6, partially enhanced
due to the lack of statistics. Further studies are needed to fully
understand the effect, once the detector is built.

\subsubsection{Summary}

The method presented here uses a special sub-sample of the single lepton plus jet
events where the top has high transverse momentum. The case with $p_T > 200$ GeV
was studied here, which due to the high $t \bar{t}$ production at LHC offers
good statistics per year at low luminosity ($10fb^{-1}$).

After some pre-selection cuts aiming to efficiently select the
hadronic top decay products with minimal contribution from the
background, the top direction is found using jets similarly to the
inclusive sample. Then, using the unique feature of these events,
where the hadronic top decay products are well collimated in
space, all the calorimeter towers within a cone of radius $\Delta
R_{clust}$ around the reconstructed top direction are summed,
forming a top invariant mass. It was then shown that the
reconstructed invariant mass becomes independent of the cone
radius used (varied between 0.8 and 1.8) once the underlying event
contribution is subtracted. Finally, the mass scale was determined
by applying the same reconstruction method for the $W\rightarrow
jj$ decay in the top events of the inclusive sample. After
recalibration, the results obtained with both the fast and full
detector simulation are comparable with the other methods, and
demonstrate that the top mass can be reconstructed with an
accuracy of 1\%.

Using this method the reconstructed top mass is sensitive to a different set
of systematic errors, a first study of which as was performed. From the results
obtained, the major contribution is in the mass scale calibration procedure, with
all the errors stay below the 1$\%$ level. Table \ref{highpt_syssum} summarizes
results. For some of the studies described above, the final systematic error once
the detector and the data are available will be incorporated in the mass scale
calibration procedure, therefore are not included as individual lines in the table.
\begin{table}[h]
\begin{center}
\begin{tabular}{lcc} \cline{2-3}
                         & $|\Delta m_t|$ (GeV) & $\delta m_t$ (GeV)  \\  \hline
 Initial state radiation & 0.7                     & 0.1              \\
 Final state radiation   & 0.3                     & 0.1              \\
 b-quark fragmentation & 0.3                     & 0.3                \\
 UE estimate ($\pm 10 \%$) & 1.3                 & 1.3                \\
 mass scale calibration  & 0.9                     & 0.9              \\
 \end{tabular}
\caption{\label{highpt_syssum} {\it Top mass shift ($|\Delta m_t|$) and quoted
  systematic error on $m_t$ ($\delta m_t$) due to various sources of
  systematic uncertainties. }}
\end{center}
\end{table}

%
%
\section{Top mass measurement in the dilepton channel}

The dilepton events can provide an indirect measurement of the top quark
mass. The difficulty comes from the fact that, in principle, one cannot
fully reconstruct the top decays due to the presence of undetected neutrinos
in the final state. For the determination of the top mass, previous methods
have exploited the correlation between the top mass and kinematic
quantities, such as the mass $m_{l b}$ of the lepton-b-jet system
\cite{atlastdr}.

Here, in a first step, assuming a value for the top mass, it is proposed to
reconstruct the top decays by solving the set of equations describing the
kinematic constraints of the decays. Then, to determine the top mass, the
solutions obtained for different input top masses will be compared to the
data \cite{dilep_vlada}.

\subsection{Event selection}
The dilepton events are characterized by two high $p_T$ isolated leptons,
large traverse missing energy $E_T^{miss}$ and two jets coming from the
fragmentation of the b-quarks. Taking into account the branching ratio,
about 400000 dilepton events can be expected for integrated luminosity $10fb^{-1}$.

The background is coming mainly from Drell-Yan processes and $Z \rightarrow
\tau \tau$ associated with jets, and from WW+jets and $b \bar{b}$ production.
Events are selected by requiring two opposite sign isolated leptons with
$p_T > 35$ GeV and $p_T > 25$ GeV respectively and $| \eta | < 2.5$,
$E_T^{miss}>40$ GeV, and two jets with $p_T > 25$ GeV. After event selection,
80000 signal events are left, with a signal over background ratio around 10.

\subsection{Method for the final state reconstruction}
For the determination of the momenta of both the neutrino ($\nu$) and
anti-neutrino ($\bar{\nu}$), it is assumed that the masses of the top and
anti-top are known. The reconstruction algorithm is based on solving a set of
equations coming from the kinematic properties of the conservation of momentum
and energy \cite{dilep_vlada}.
The set of equations consists of six equations for six unknown components of momenta of neutrino and antineutrino.
First two equations describe conservation of transversal momentum of the $t\overline{t}$ system, assuming that this momentum is 0.
The other equations constrain invariant masses of both lepton+neutrino systems to the masses of $W^{+}$ and $W^{-}$ bosons, and masses of both lepton+neutrino+jet system to the masses of top and antitop quarks. All $W^{+}$, $W^{-}$, top and antitop masses are assumed to be known.

After some derivations \cite{dilep_vlada}, the following two linear equations
with the unknowns \( p_{x}^{\overline{\nu}}\), \(p_{y}^{\overline{\nu}}\),
\(p_{z}^{\overline{\nu}}\) and \(p_{z}^{\nu}\), are obtained:

\begin{equation} \left(2\frac{{E_{\overline{b}}p_{x}^{l^{-}}}}{{E_{\overline{l}}}}
-2p_{x}^{\overline{b}} \right)p_{x}^{\overline{\nu}} +
\left(2\frac{{E_{\overline{b}}p_{y}^{l^{-}}}}{{E_{\overline{l}}}}-
2p_{y}^{\overline{b}} \right)p_{y}^{\overline{\nu}} +
\left(2\frac{{E_{\overline{b}}p_{z}^{l^{-}}}}{{E_{\overline{l}}}}-
2p_{z}^{\overline{b}} \right)p_{z}^{\overline{\nu}} \end{equation}
\[ + 2E_{\overline{l}}E_{\overline{b}}
 - 2M_{W^{-}} + M_{\overline{t}} - 2(p_{x}^{\overline{l}}p_{x}^{\overline{b}} +
 p_{y}^{\overline{l}}p_{y}^{\overline{b}} +
 p_{z}^{\overline{l}}p_{z}^{\overline{b}} +
\frac{{E_{\overline{b}}M_{W^{-}}}}{{E_{\overline{l}}}}) + m_b^2 = 0 \]

\begin{equation} -2\left( p_{x}^{l} + p_{x}^{\overline{l}} +
p_{x}^{\overline{b}} -k_{1} - \frac{E_{b}p_{x}^{l}}{k_{4}} \right)
p_{x}^{\overline{\nu}} -2\left( p_{y}^{l} + p_{y}^{\overline{l}} +
p_{y}^{\overline{b}} -k_{2} - \frac{E_{b}p_{y}^{l}}{k_{4}} \right)
p_{y}^{\overline{\nu}} \end{equation}
\[ + 2\left( p_{z}^{l} - K_8 + \frac{E_{b}p_{z}^{l}}{k_{4}} \right)
p_{z}^{\nu} + k_{1}^{2} + k_{2}^{2} + k_{4}^{2}+ M_{t} + E_{b}^{2} - k_{8}^{2} \]
\[ - {p_{x}^{\overline{l}}}^{2} - {p_{y}^{\overline{l}}}^{2} -
{p_{x}^{\overline{b}}}^{2} - {p_{y}^{\overline{b}}}^{2} +
2\left(\frac{E_{b}(p_{x}^{l}k_{1}+p_{y}^{l}k_{2})}{k_{4}} +
k_{4}E_{b} + k_{1}p_{x}^{l}+ k_{2}p_{y}^{l}\right) \]
\[ -2\left(\frac{E_{b}M_{W^{+}}}{k_{4}} + p_{x}^{\overline{l}}p_{x}^{\overline{b}}
+ p_{y}^{\overline{l}}p_{y}^{\overline{b}} + M_{W^{+}} \right) = 0 \]

\noindent
Where: $E_{p}$ is the energy of particle $p$, $M_{p}$ is a function of the mass of
particle $p$, $k_i$ is a function of momenta of leptons and b-quarks, $p_{i}^{p}$
represents the i-th component of momentum of particle
$p$, $\overline{l}$ represents either $e^{+}$ or $\mu^{+}$, $l$ represents
either $e^{-}$ or $\mu^{-}$, $\nu$ is either $\nu_{e}$ or $\nu_{\mu}$,
$\overline{\nu}$ is either $\overline{\nu_{e}}$ or $\overline{\nu_{\mu}}$.

Additional derivations lead to one quartic equation with only one unknown,
which is analytically solved. All the derivations were performed using a
software tool for symbolic algebraic manipulations.

The remaining components of both neutrino and anti-neutrino momenta can
be easily computed. Finally, the complete kinematic reconstruction
can be performed.

The reconstruction algorithm can provide no solution or more than one
solution. In the first case, the right-handed sides of the two equations
from the initial set of six equations describing momenta conservation
are varied in the range [-250 GeV:+250 GeV]
starting with 0 until an acceptable solution is found. The solubility
of the system is improved from $88 \%$ to $97.6 \%$ (this means that the
$t \bar{t}$ decay is reconstructed for $97.6 \%$ of the events).
In the second case, the choice of the solution is based on the computing of
weights for known distributions of various kinematic quantities of the
$t \bar{t}$ decay \cite{dilep_vlada}.
The right solution is chosen in $73 \%$ of the events.

Therefore, the reconstruction algorithm exhibits an efficiency of $97.6 \%$
with a purity of $73 \%$.

\begin{figure}[ht!]
\begin{center}
\begin{minipage}{0.49\textwidth}
\epsfig{file=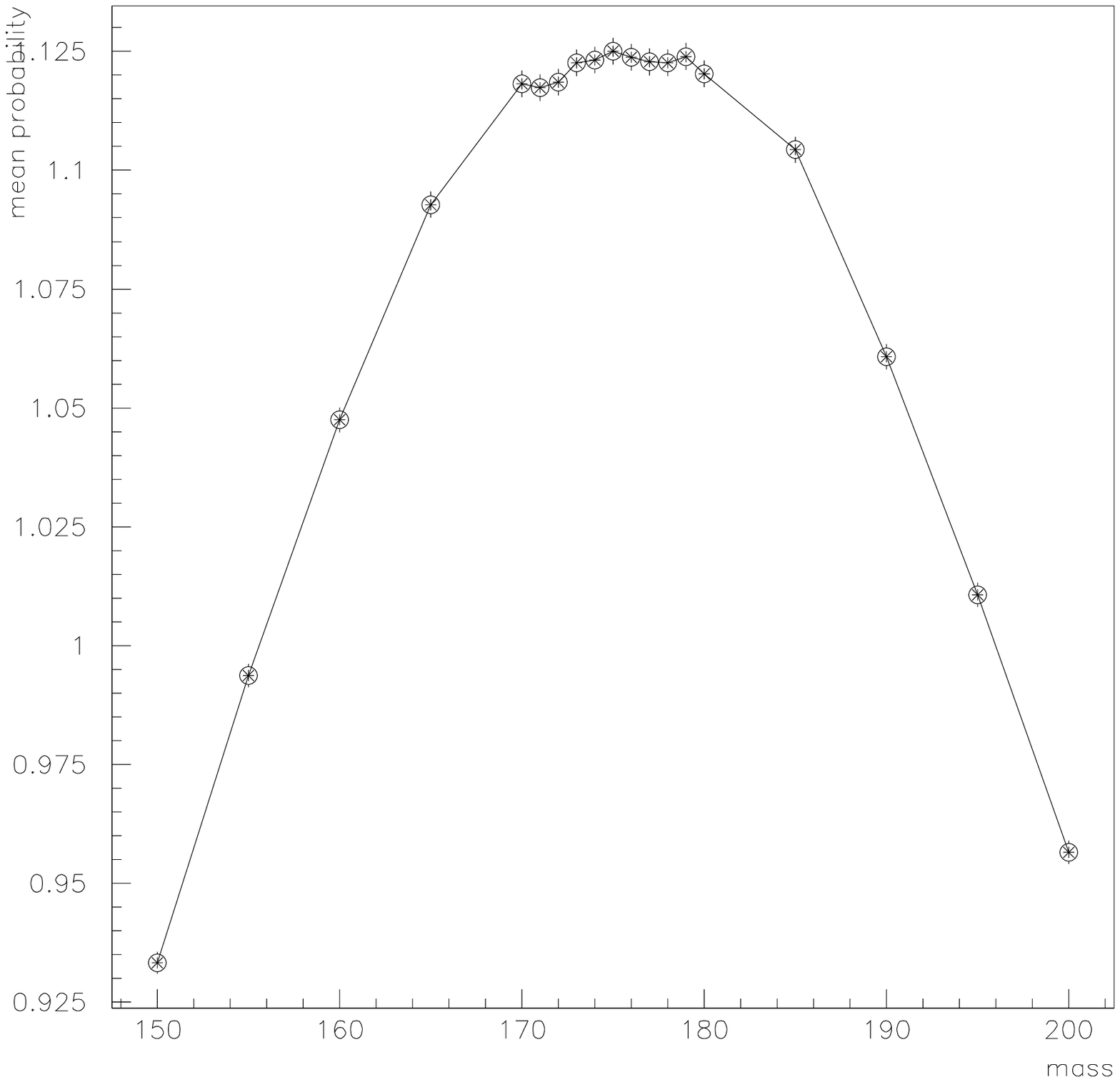,width=6.0cm}
\caption{\label{dilep_fig1} {\it  Mean weight as a function of the input
top mass. The maximum mean value gives the top mass.}}
\end{minipage}
\hfill
\begin{minipage}{0.49\textwidth}
\epsfig{file=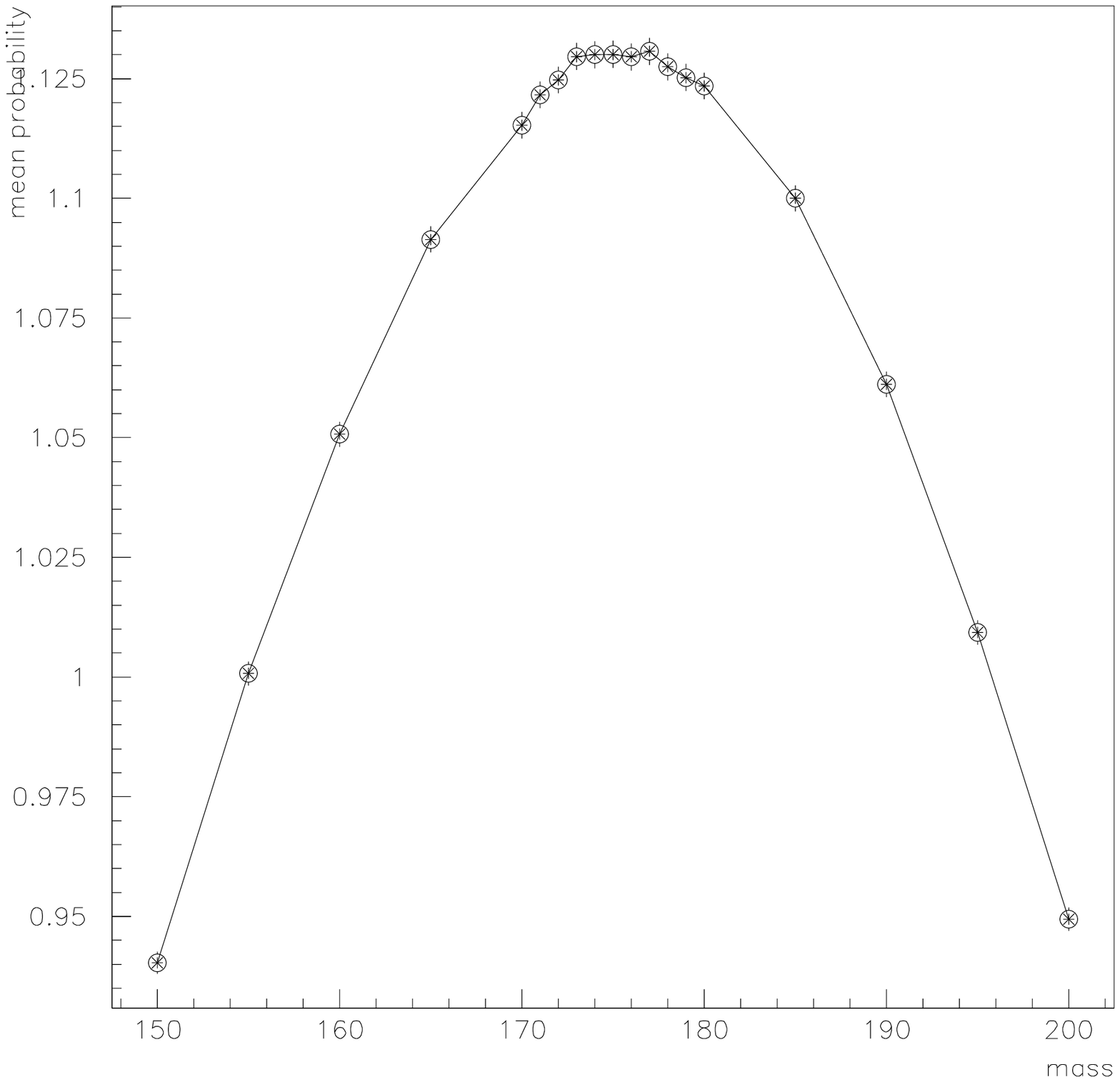,width=6.0cm}
\caption{\label{dilep_fig2}{\it Mean weight as a function of the top
mass, with ISR switched off.}}
\end{minipage}
\end{center}
\end{figure}

\subsection{Top mass determination}

It has just been demonstrated that the entire $t \bar{t}$ decay can be
reconstructed by assuming a value for the top quark mass. For the top mass
determination, the reconstruction algorithm will be fed with various top masses
and the corresponding solutions will be compared to the data.

\subsubsection{Method}
The method was tested using samples of events containing approximately the
same amount of events that will be collected during one year of running at
low luminosity (per $10fb^{-1}$), after selection cuts are applied.

The principle of the determination of the top mass is the following: for
each event, one tries to solve the equations for various input top masses,
and to compute the weight of the best solution for a given top mass value.
If the input top mass value is quite different from the correct value, no
solution may be found, or the solution will have a small weight.

For each input top mass value, a mean weight over the entire set of
events is computed. The top mass is given by the value having the maximum
mean weight.

The weight mean value as a function of the input top mass is represented
on figure \ref{dilep_fig1}. As expected, the curve is peaked
around the generated top mass value of 175 GeV. The maximum mean weight
is obtained by fitting this curve with a quadratic function, leading to a
reconstructed top mass in agreement with the generated value. The error on the
reconstructed value is 0.3 GeV. This error takes into account both statistical
effects and effects due to the reconstruction method itself.

\subsubsection{Systematic uncertainties}

The effect of the systematic uncertainties sources on the top mass determination has
been checked following the methods described in section 2. For initial and final
radiation, top mass shifts were determined by switching ISR and FSR off separately
in the Monte Carlo, and the corresponding error was obtained by taking $20 \%$ of
the mass shifts. The b-quark jets calibration was assumed to be known within 1 $\%$.
For the b-quark fragmentation parameter, a mass shift was computed between the
top mass obtained with the default value $ \varepsilon_b = 0.006$ and with
$ \varepsilon_b = 0.0035$. The error due to the parton distribution function was
estimated by measuring the top mass shift between a sample of events simulated with
the default set and a sample of events simulated with another set.
The values are summarized in table \ref{dilep_syst}.

\begin{table}[h]
\begin{center}
\begin{tabular}{lcc}\hline
source of uncertainty & $\vert\Delta m_t\vert$ (GeV) & $\delta m_t$ (GeV) \\ \hline
Statistics and reconstruction method &               & 0.3                \\
b-jet energy scale                   &      0.6      & 0.6                \\
b-quark fragmentation                &      0.7      & 0.7                \\
Initial state radiation              &      0.4      & 0.1                \\
Final state radiation                &      2.7      & 0.6                \\
Parton distribution function         &      1.2      & 1.2                \\ \hline
\end{tabular}
\caption{ \label{dilep_syst} {\it Top mass shift $\Delta m_t$ and
resulting systematic error on $m_t(\delta m_t)$ due to the various
source of systematic errors, in the dilepton channel.}}
\end{center}
\end{table}

An example, for ISR switched off, of the mean weight as a function of the input
top mass is shown on figure \ref{dilep_fig2}. The values of the
estimated systematic errors listed in table \ref{dilep_syst} are not large. This is
due to a very positive aspect of the reconstruction method. For a given systematic
uncertainty source, the curve of the mean weights versus the input top mass
is modified in two ways compared to the curve obtained with the default sample.
The mean weights are smaller, giving a maximum mean value smaller than
the initial value, and the peak is shifted giving a corresponding shifted top mass.
This second effect only is relevant in the systematic uncertainty studies.

\subsection{Summary}

It was shown that, assuming a mass for the top quark, the final state topology of
dilepton events can be fully reconstructed by solving a set of equations describing
the kinematic constraints of the $t \bar{t}$ decay. The decay reconstruction
algorithm has high efficiency and purity. A step further, for the determination
of the top mass, consists in feeding the reconstruction algorithm with different
input top masses and to compare the solutions with the data.

There is also a possibility to consider more than two jets in
final state. In this case one has to solve the set of equations
for all 2-jets combinations. Surprisingly, this has also no impact
on the estimation of the top mass value, however, it is one of the
subjects to be studied yet.

A preliminary study of the systematics uncertainties shows that the top mass can be extracted with a
reasonable accuracy, at the same level as other techniques. This method can
therefore provide a useful input for the combined ATLAS top mass measurement.

%
\section{Top mass measurement in the six jets channel}

The all jets channel final state topology consists, in the absence of initial
or final state radiation, of six jets (including two $b$-jets), no high $p_T$
leptons, and small transverse missing energy $E_T$. With no energetic neutrinos
in the final state, the all hadronic mode is the most kinematic-ally constrained
of all the $t\bar{t}$ topologies, but it is also the most challenging to measure
due to the large QCD multijet background. Nevertheless, at the Fermilab Tevatron
Collider both the CDF and D{\O} collaborations have shown that it is possible to
isolate a $t\bar{t}$ signal in this channel \cite{alljet_cdf,alljet_d0}.
The CDF collaboration obtained a signal significance over background of better
than three standard deviations \cite{alljet_cdf} by applying simple selection
cuts and relying on high $b$-tagging efficiency. To
compensate for the less efficient $b$-tagging, the D{\O} collaboration developed
a more sophisticated event selection technique based on a neural network
\cite{alljet_d0}.

The potential of the ATLAS detector to study the all hadronic decays of
$t\bar{t}$ pairs has been explored. In the search for an optimal strategy for
signal extraction from background, the kinematic properties of both signal
and background events are investigated, and a kinematic fit of selected
events is performed. Finally, a clean sample is obtained by selecting
events in which both reconstructed top and antitop quarks have a high
traverse momentum ($p_T >$ 200 GeV).  This subsample is then used for
the reconstruction of the top mass. Nevertheless, the top mass is reconstructed
in the inclusive sample as well.

\subsection{Signal selection}
Taking into account the branching ratio, the next-to-leading order
cross-section prediction for the all jets channel is 370 pb. Therefore,
for an integrated luminosity of $10 \ \rm fb^{-1}$, one can expect 3.7
million $t \bar{t}$ pairs with this final state topology.

The main source of background is QCD multijet events, which
arise from $2 \rightarrow 2$ parton processes ($q_{i}q_{j}
\rightarrow q_{i}q_{j}$, $q_{i}g \rightarrow q_{i}g$, $gg
\rightarrow gg$, $q\bar{q} \rightarrow gg$, $q_{i}\bar{q_i}
\rightarrow q_{j}\bar{q_j}$, $gg \rightarrow q\bar{q}$) convoluted
with parton showers. The heavy-flavor ($c\bar{c}$,  $b\bar{b}$,
$t\bar{t}$) content in a QCD multijet sample stems from direct
production (e.g. $q_{i}\bar{q_i} \rightarrow q_{j}\bar{q_j}$, $gg
\rightarrow q\bar{q}$), gluon-splitting (where a final state gluon
branches into a heavy quark pair), and flavor excitation (initial
state gluon splitting). In the analysis that follows, $t\bar{t}$
production was excluded from the QCD background processes. The QCD
background was generated with a $p_T$ cut on the hard scattering
process above 100 GeV, resulting in a production
cross-section of 1.73 $\mu$b.  Processes involving the production
of $W$ and $Z$ bosons (with their subsequent decay into jets) were
not included since their contributions are small compared to the
QCD multijet background.

As the first step in the selection of the all hadronic $t\bar{t}$
topology, events were required to have six or more reconstructed
jets, of which at least two must be tagged as $b$-jets. Jets were
reconstructed using a fixed cone algorithm with $\Delta R$=0.4.
Jets were required to have  $p_T$ greater than 40 GeV, and to
satisfy $\vert\eta\vert<3$ ($\vert\eta\vert<2.5$ for $b$-jet
candidates). The efficiencies for these selection criteria for both
$t\bar{t}$ signal and QCD multijet background are 2.7 $\%$ and
0.011 $\%$ respectively, resulting in a signal over QCD background
of 1/19, indicating that these simple selection cuts can already
reduce the multijet background to manageable level.

\subsection{Signal and background kinematic properties}
Further progress in enhancing the $S/B$ ratio could be
sought using variables that provide discrimination between the
signal and the QCD background. Therefore,  some kinematic variables
sensitive to the energy flow in the event, additional radiation and
event shape (including several variables used in the neural network
analysis of the D{\O} collaboration \cite{alljet_d0}) were examined.
Those variables include:\\
$H_T$: the sum of all jet transverse energies
 in the event ($\sum \limits_{j=1}^{N_{jets}}E_{T_j}$).\\
$H_T^{4j}$: $H_T$ without the transverse energy of the two leading jets.\\
$E_{T}^{b\ jets\ 1,2}$: the transverse energy of the two leading $b$-jets.\\
$\sqrt{ \hat{s}}$: the invariant mass of the jets in the final state.\\
$\it{A}$: the aplanarity, ${3\over2}Q_1$, calculated from the normalized
momentum tensor.\\
$\it{S}$: the sphericity, ${3\over2}(Q_1+Q_2)$, calculated from the
normalized momentum tensor.\\
$\it{C}$: the centrality, $H_T/H_E $, where
$H_E=\sum\limits_{j=1}^{N_{jets}}E_{j}$ is the sum of all the
jet total energies. The centrality characterizes the transverse energy flow.\\
$\Delta R_{jj}^{min}$: the minimal separation between two jets in
$\eta$-$\phi$ space.\\

\begin{figure}
\begin{center}
\begin{tabular}{cc}
\mbox{\epsfig{file=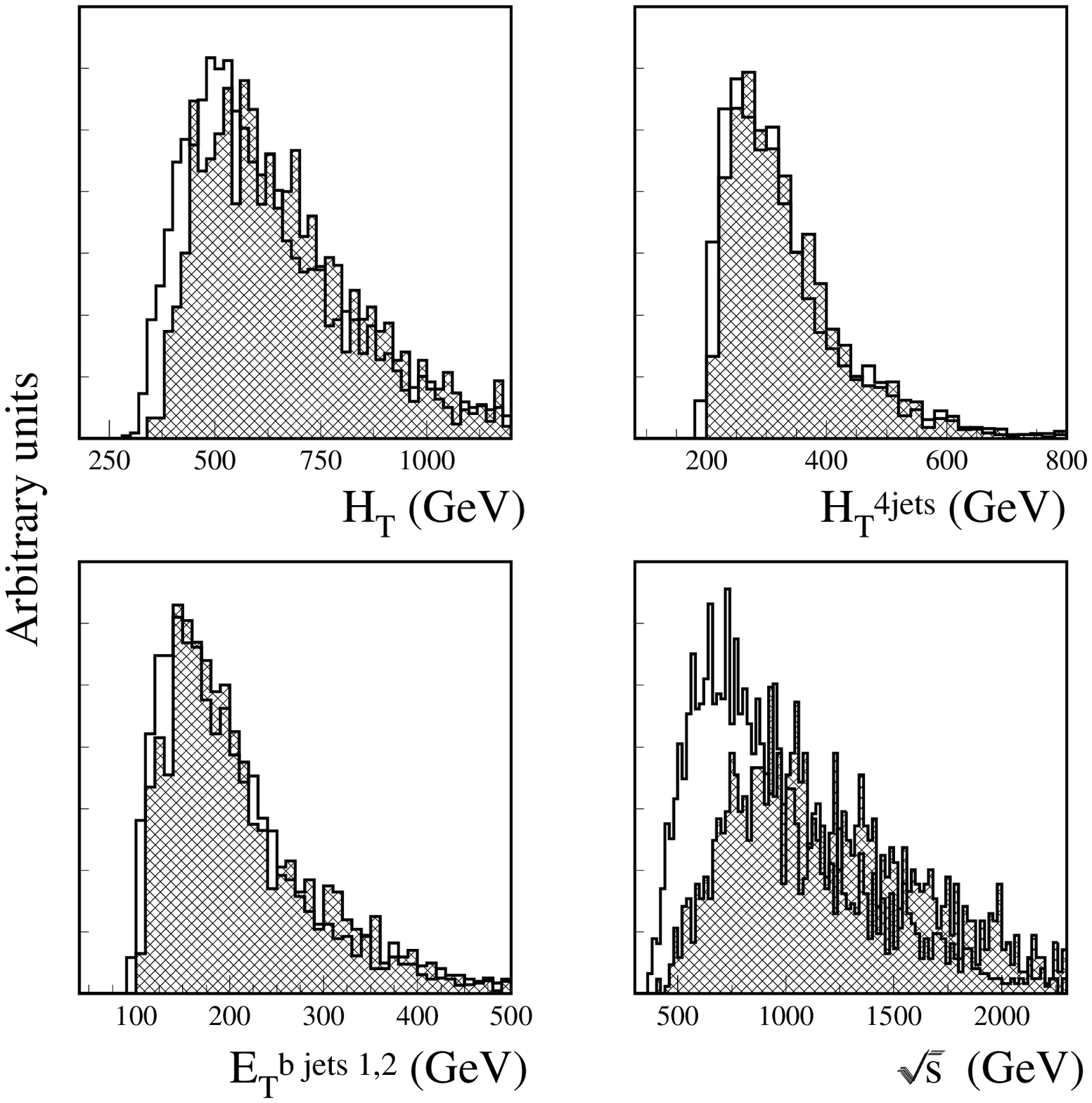,width=6cm}} &
\mbox{\epsfig{file=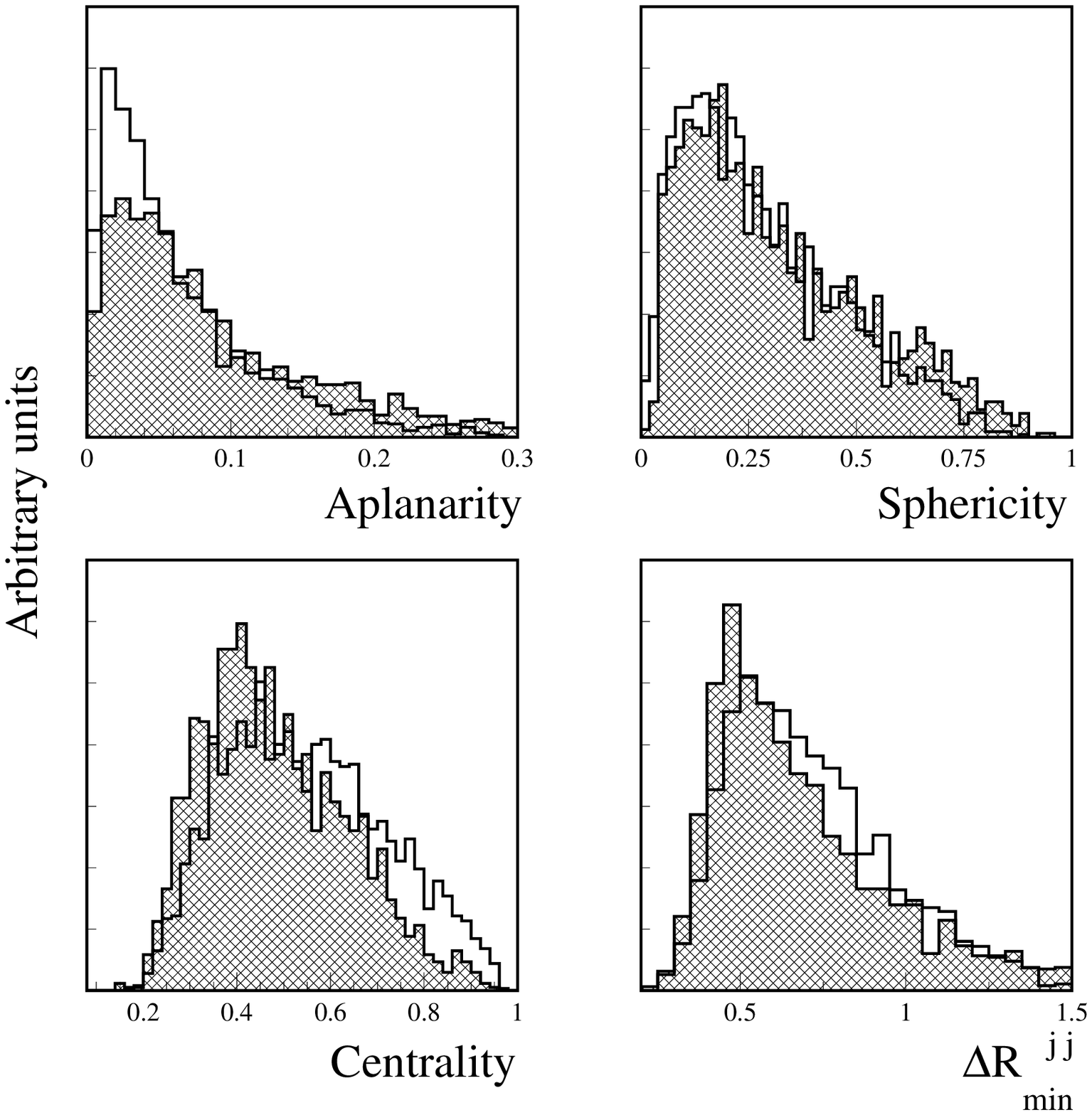,width=6cm}}
\end{tabular}
\caption{ \label{alljet_fig1} {\it Distributions for $t\bar{t}$ signal (hatched)
and QCD multijet background (unhatched). Left: the $H_T$, $H_T^{4\ jets}$,
$E_{T}^{b\ jets\ 1,2}$, and $\sqrt{ \hat{s}}$ distributions. Right: aplanarity,
sphericity, centrality and $\Delta R_{min}^{jj}$
distributions. See the text for more details. }}
\end{center}
\end{figure}

The first four of these variables are related to the energy
deposition in the event, while the others are more related to
the event shape or topology.  The normalized distributions for
these variables, for $p_T^{jet}>$ 40 GeV, are plotted for
$t\bar{t}$ signal and QCD background in figure \ref{alljet_fig1}
(left plot for the first four variables and right plot for the others).
It can be seen that the variables sensitive to the event shape provide a
somewhat better discrimination between the signal and background.
However, it is clear that none of these
variables provides at the LHC the clear discrimination which was
observed at the Tevatron energy \cite{alljet_d0}. Therefore, it would
appear difficult to select a relatively clean signal based on cuts
on these variables, or even the use of a more sensitive cut based
on a multivariate discriminant, where the variables are treated
collectively \cite{alljet_d0,alljet_d00}.

\subsection{Final state reconstruction with a kinematic fit}
The key feature distinguishing top quark events from QCD multijet
background is the fitted mass obtained from the least-squares kinematic
fit of the events to the $t\bar{t}$ decay hypothesis \cite{alljet_ben}.
In order to simplify the analysis, massless jets have been assumed and
the error on the measured jet direction was neglected with respect to
the error on the measured jet energy.

The reconstruction algorithm and fitting procedure proceed in two
steps. First, the two $W \rightarrow jj$ decays are reconstructed by
selecting di-jet combinations from jets not tagged as $b$-jets. This is done
by minimizing a $\chi^2_W$ function \cite{alljet_simic}.

Next, the two $W \rightarrow jj$ candidates are combined with the
$b$-tagged jets to form the top and antitop quark candidates
($jjb$ combination). The energies of the $b$- and $\bar{b}$-jets
are constrained by minimizing a $\chi^2_t$ function \cite{alljet_simic}.
There are two ways to associate the b-tagged jets to the reconstructed
W bosons. The association giving the smallest value of $\chi^2_t$ is
chosen. After the event reconstruction and fitting procedure, additional
qualitative cuts are applied \cite{alljet_simic}.

\begin{table}[h]
\begin{center}
\begin{tabular}{lccc}   \hline
&after selection& after kinematic fit & within the window \\
& cuts& and $\chi^2$ cuts & 130-200 GeV \\ \hline
$t\bar{t}(\%)$ & 2.7 & 0.3 & 0.18\\
QCD $(\%)$ & 0.011 & 0.00017 &  0.000007\\
$S/B$ & 1/19 & 1/2.6 & $\approx 6/1$ \\ \hline
\end{tabular}
\caption{ \label{alljet_table} {\it Efficiency for $t\bar{t}$ signal and
QCD multijet background after applying various level of cuts, and for
$p_T^{jet}$ threshold $>$ 40 GeV.  The last row shows the resulting
signal to background ratio.}}
\end{center}
\end{table}

Table \ref{alljet_table} presents the efficiency and $S/B$ ratio for $t\bar{t}$
signal and QCD multijet background after selection cuts are applied, after
the kinematic fit procedure, and after the additional requirement that
the reconstructed top and antitop quarks masses lie within the
window 130-200 GeV. The kinematic fit and limits on the top
(antitop) mass significantly improve the value of $S/B$ ratio so
that, for top masses within the 130-200 GeV window, the $S/B$
ratio is  $\approx$ 6. This result is obtained with a large
statistical error on the remaining background ($40\%$), and a
more accurate determination would require generation of
significantly larger Monte Carlo background samples.

\subsection{High transverse momentum $t \bar{t}$ events}
The signal over QCD background can be further improved by restricting
the analysis to a sample of high transverse momentum $t \bar{t}$ events
where both reconstructed top and anti-top quarks have $p_T > 200$ GeV. To
study this sample, $t \bar{t}$ signal and QCD background events were generated
with a $p_T$ cut on the hard scattering process above 200 GeV. The
corresponding cross-sections are 53.5 pb for signal and 86.1 for QCD
multijet background.

\subsubsection{Top mass reconstruction}

For the high $p_T$ sample, the same selection cuts were applied as
for the inclusive sample. After the kinematic fit and
the requirement that both reconstructed top and anti-top quarks
have $p_T > 200$ GeV, the selection efficiencies and $S/B$
ratios are given in table \ref{alljet_table2}.

\begin{figure}
\begin{center}
\mbox{\epsfig{file=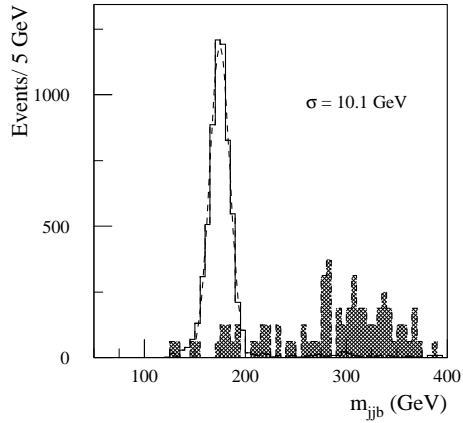,width=6.0cm}}
\caption{ \label{alljet_fig2} {\it Invariant mass distribution of the accepted
 $jjb$ combinations for the high $p_T$ sample, normalized to an
 integrated luminosity of 10 fb$^-1$. The shaded area shows the QCD
 multijet background.} }
\end{center}
\end{figure}

The invariant mass distribution of the accepted  $jjb$
combinations  and for the QCD background (the shaded area) is shown
in figure \ref{alljet_fig2}. Within the window 130-200 GeV the signal
over background ratio is $\approx$ 18.

The distribution fitted by a Gaussian leads to a reconstructed top mass
consistent with the generated value with a peak width of 10.1 GeV.
For an integrated luminosity of 10 fb $^{-1}$, a
sample of 3300 events would be collected with fully reconstructed
top and antitop quarks with $p_T>200$ GeV. This number of events
would lead to a statistical error of $\delta m_{t}$(stat)=$\pm
0.18$ GeV. It can be noted that this clean sample could be used
for the study of differential distributions for both top and anti-top
quarks \cite{alljet_simic}.

\begin{table}[h]
\begin{center}
\begin{tabular}{lcc}   \hline
&after kinematic fit  &  within the window\\
&and $\chi^2$ cuts   & 130-200 GeV \\ \hline
$\varepsilon_{t\bar{t}} (\%)$ &   0.68  &  0.63    \\
$\varepsilon_{QCD }   (\%)$ &  0.00041 & 0.000021 \\
$S/B$&  1/1& $\approx 18/1$ \\ \hline
\end{tabular}
\caption{ \label{alljet_table2} {\it Efficiency for high $p_T$ $t\bar{t}$
signal and QCD multijet background where reconstructed top and antitop
quark both have $p_T>200$ GeV, for different cuts applied and for $p_T^{jet}$
threshold $>40$ GeV. The last row shows the resulting
signal to background ratio.}}
\end{center}
\end{table}

\subsubsection{Systematic uncertainties}
The systematic uncertainties have been treated in a similar way as in the
inclusive lepton plus jets channels. It was assumed that jet energy scale
for both light quark and b-quark jets will be known at the level of $ 1 \%$.
For the b-quark fragmentation parameter $ \varepsilon_b$, a top mass shift was
determined between the top mass obtained with the default parameter
($ \varepsilon_b = 0.006$) and with $ \varepsilon_b = 0.0035$. For initial and
final state radiation, mass shifts were obtained between ISR and FSR switched
on and off separately. The resulting systematic error was taken by considering
$ 20 \%$ of the mass shifts. The results are summarized in table \ref{alljet_syst},
for the high $p_T$ sample.

\begin{table}[h]
\begin{center}
\begin{tabular}{lc}\hline
Systematics & $\delta m_t$ (GeV)           \\ \hline
Light jet energy scale & 0.8               \\
b-jet energy scale & 0.7                   \\
b-quark fragmentation & 0.3                \\
Initial state radiation & 0.4              \\
Final state radiation & 2.8                \\ \hline
\end{tabular}
\caption{ \label{alljet_syst} {\it Systematic error on $m_t(\delta m_t)$ due to
the various source of systematic errors for the high
$p_T\mbox{(top)}$ sample for the all jets channel.}}
\end{center}
\end{table}

The total systematic error is the order of 3.0 GeV for the high $p_T$ sample.
This value is larger than in
the case of the lepton plus jets channel where the top mass is determined in
the same way, as the invariant mass of the three jets coming from the hadronic
top decay (see section 2). Clearly, the sources of systematic uncertainties
have an impact on the resolution of the kinematic fit.

\subsection{Summary}
It has been shown that the top mass can be determined in the all jets channel.
The $t \bar{t}$ signal is extracted from the huge QCD background
($S/B \sim 3 \times 10^{-8}$ at production level) by the use of kinematic
cuts and a kinematic fit which allows to reconstruct the complete final state
topology. The signal over background ratio can be further increased by selecting
high $p_T$ events. Once the entire $t \bar{t}$ decay is reconstructed, the top mass
is determined as the invariant mass of the three jets arising from each top
quark ($t \rightarrow jjb$ and $\bar{t} \rightarrow jj\bar{b}$). It was shown that
a total error on the top mass of the order of 3 GeV can be reached.

%
\section{Top mass measurement in leptonic final states with $J/ \Psi$}

\begin{figure}[h]
\begin{center}
\mbox{\epsfig{file=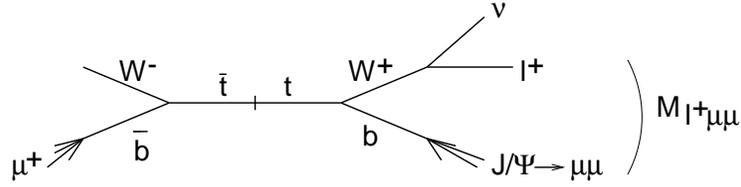}}
\caption{ \label{jpsi_topo} {\it Diagram of the top decay to leptonic
final state with $J/ \Psi$.} }
\end{center}
\end{figure}

Here, one exploits the correlation between the top mass and the invariant
mass $M_{l J/ \Psi}$ of the system made of a $J / \Psi$ from the decay of
a b-hadron and the isolated lepton (e or $\mu$) coming from the associated
W decay (see figure \ref{jpsi_topo}) \cite{jpsi_cms}. In order to uniquely
define the final state topology and therefore to reduce considerably the
combinatorial background, the presence of a muon-in-jet (with the same
sign than the isolated lepton) from the b-quark is required in the other
top quark decay.

The overall branching ratio is $3.2 \times 10^{-5}$. Due to this strong
suppression this method will be only applicable during the high luminosity
phase of the LHC, where 2700 events will be produced per year.

\subsection{Analysis}
\begin{figure}
\begin{minipage}{0.49\textwidth}
\begin{center}
\epsfig{file=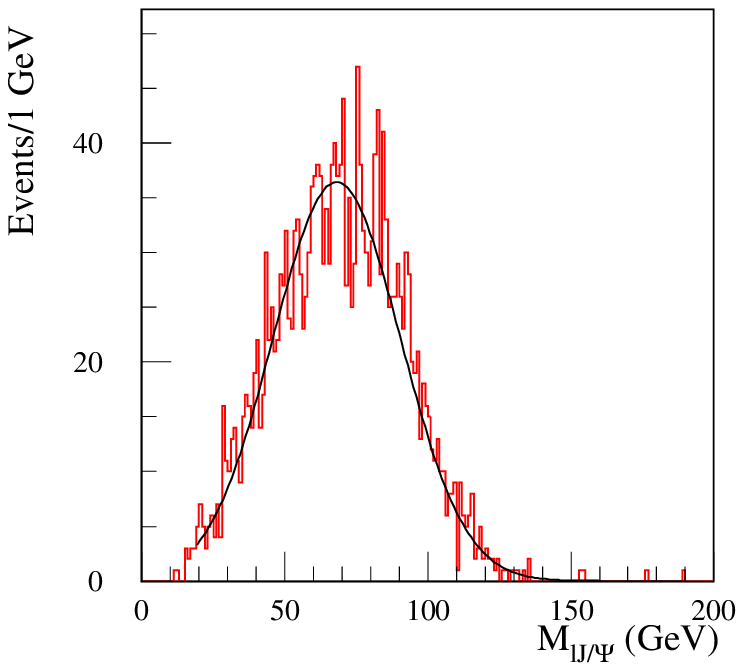,width=6.0cm}
\caption{\label{jpsi_mass} {\it Invariant mass distribution  $M_{l J/ \Psi}$, for
five years of running at high luminosity.}}
\end{center}
\end{minipage}
\hfill
\begin{minipage}{0.49\textwidth}
\begin{center}
\epsfig{file=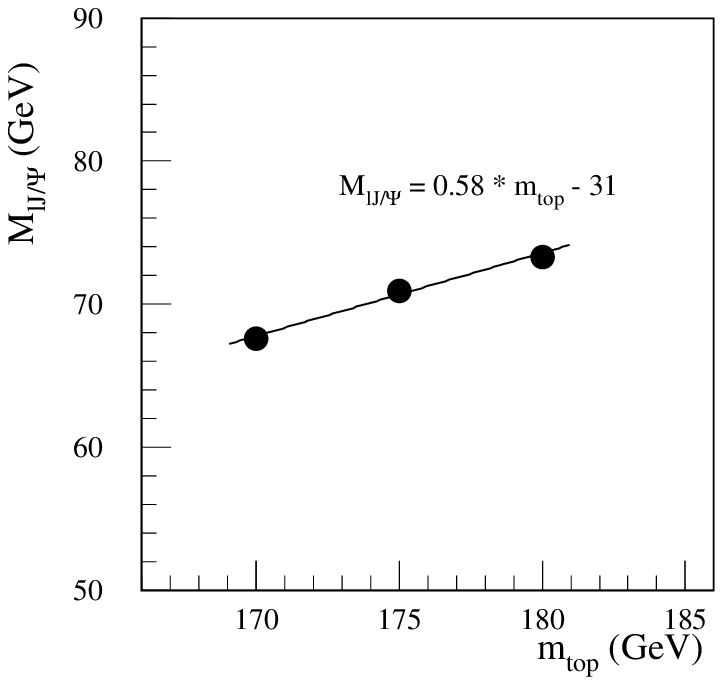,width=6.0cm}
\caption{\label{jpsi_corr} {\it Correlation the top mass and the $l J/ \Psi$
invariant mass $M_{l J/ \Psi}$.}}
\end{center}
\end{minipage}
\end{figure}

Events are selected by requiring one isolated lepton with $p_T > 30 \ \rm GeV$
and $|\eta| < 2.4$, and three non-isolated muons with $p_T > 3 \ \rm GeV$ and
$|\eta| < 2.4$, with the invariant of two of them (with opposite signs) being
compatible with the $J / \Psi$ mass. After selection cuts are applied and for
one year of running at high luminosity, about 430 events are expected. The
$l J/ \Psi$ invariant mass distribution, for five years high luminosity
running, is shown on figure \ref{jpsi_mass}.

The distribution fitted by a Gaussian leads to a mean $l J/ \Psi$
invariant mass of 68.1 GeV with a peak width of 22.4 GeV. The
combinatorial background is small. For five years of running at
high luminosity, the statistical error on $M_{l J/ \Psi}$ would be
approximately 0.5 GeV. To improve the statistics, various
strategies might be considered, such as for example removing the
requirement of the muon-in-bjet and determine the final state
topology by jet charge measurements. Nevertheless, further study
is needed to evaluate the effectiveness of these approaches.

The correlation between the top mass and $l J/ \Psi$ invariant mass is shown
on figure \ref{jpsi_corr}. One expects the uncertainties of the top mass to scale
as a factor $1/0.58 \simeq 1.7$ compared to the estimated errors on $M_{l J/ \Psi}$.
Therefore, for five years of running at high luminosity, the statistical
uncertainty on the top mass would approximately 0.8-0.9 GeV.

\subsection{Background processes}
The major sources of background come from processes involving $b \bar{b}$
production. Potential backgrounds such as W+jets, Z+jets and boson pair
productions are briefly discussed.

The boson pair production processes (WW, WZ and ZZ) have small cross-sections
compared to signal \cite{jpsi_note}. Furthermore, all the final states
topologies are different with respect to the signal (except for one particular
case but with tiny cross-sections \cite{jpsi_note}). These processes can be
neglected.

The cross-sections for W+jets and Z+jets processes are about a factor of
20 higher or similar to the signal cross-section, respectively. Here also
nevertheless, the final states are different with respect to the signal.
These processes can be neglected as well.

The total cross-section for inclusive $b \bar{b}$ production is approximately
$10^6$ larger than the signal cross-section. The final state is simply missing
one isolated lepton compared to the signal. Taking into account the various
branching ratios and for one year of running at high luminosity, of the order
of $3.5 \times 10^9 b \bar{b}$ events with three non-isolated muons can be
expected, compared to 2700 signal events. $10^6 b \bar{b}$ events with three
non-isolated muons have been generated. Only $2.9 \%$ of the events are
reconstructed with three non-isolated muons with $p_T > 3 \ \rm GeV$ and
$|\eta| < 2.4$. Only $0.05 \%$ of the events are reconstructed with
one isolated lepton with $p_T > 30 \ \rm GeV$ and $|\eta| < 2.4$. No events
survive when both of these requirements (corresponding to the signal selection
cuts) are applied. In
addition, a cut on the transverse missing energy could be applied
($p_T^{miss} > 20 \ \rm GeV$) which would reduce the $b \bar{b}$ rate by a factor
of 6 and have no effect on the signal \cite{jpsi_note}. Therefore, and although
more statistic would be helpful, the $b \bar{b}$ background process can
probably be controlled.

The $W b \bar{b}$ process has exactly the same final topology as the signal. The
cross-section, for $W \rightarrow l \nu$, is approximately 85 pb \cite{jpsi_wbb}.
After selection cuts, the reconstruction efficiency if $1.2 \%$ ($16 \%$ for
the signal). Therefore, after selection cuts, the signal over background
ratio would $S/B \approx 55$. This background process can also be controlled.

In conclusion, all the background processes are either negligible or can be
kept under control.

\subsection{Systematic uncertainties}
This technique is insensitive to the jet energy scale which is the main
source of systematic uncertainty in direct top mass measurements. However,
the main limitation to a precise determination of the top mass using this
method relies on how well the Monte Carlo describes the top production and
decay. Particularly, the proper description of the fragmentation of the b
hadrons is a crucial point.

The most relevant sources of systematic uncertainties have been investigated
in the following. Shifts of $M_{l J/ \Psi}$ ($\Delta M_{l J/ \Psi}$) have
been determined, defined as the difference between the value of $M_{l J/ \Psi}$
under nominal conditions and under the condition of the systematic uncertainty
source, as described in the previous sections.
Due to the huge amount of the required Monte Carlo statistics, the
mass shifts have been obtained with a non negligible error \cite{jpsi_note}.
Nevertheless, the quoted numbers should give a realistic estimate of the
impact of the systematic uncertainty sources.

\begin{figure}
\begin{center}
\mbox{\epsfig{file=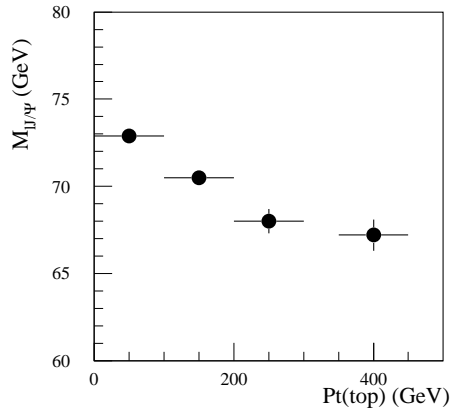,width=6.0cm}}
\caption{ \label{jpsi_pttop} {\it Dependence of the reconstructed
$M_{l J/ \Psi}$ on the top quark transverse momentum.} }
\end{center}
\end{figure}

The mass shifts obtained for initial state radiation, parton distribution function
and the b-quark fragmentation parameter are summarized in table \ref{jpsi_table}.

\begin{table}[h]
\begin{center}
\begin{tabular}{lc}          \hline
Source & Mass shift in GeV \\ \hline
ISR &  0.1   \\
PDF &  0.2   \\
$\epsilon_b \pm 10\%$ & +0.4/0.0  \\
$\epsilon_b \pm 20\%$ & +0.8/0.1
 \\ \hline
\end{tabular}
\caption{ \label{jpsi_table} {\it $M_{l J/ \Psi}$ shifts for the various
systematic uncertainties sources.}}
\end{center}
\end{table}

The $l J/ \Psi$ invariant mass can be determined with a systematic uncertainty
of the order of 0.5 GeV which translates to a systematic error on the top mass
of the order of 1 GeV.

\subsection{Top transverse momentum:}

The stability of $M_{l J/ \Psi}$ as a function of $p_T(top)$ has been
controlled and a strong $p_T(top)$ dependence has been found, as shown
on figure \ref{jpsi_pttop}. This analysis has been repeated at generator
level using the following selection criteria: no cut applied, cuts on
isolated lepton only, cuts on non-isolated muons only, and all cuts
applied. As expected, no dependence is observed when no cuts are applied.
The cuts on the non-isolated muons introduces a small effect whereas the
cuts on the isolated lepton has a strong impact \cite{jpsi_note}. This
effect is a kinematic one. It can be kept under control as long as the
Monte Carlo is well tuned to the data, which is needed in any case for
the determination of the top mass from $M_{l J/ \Psi}$.

\subsection{Summary}
During the high luminosity phase of the LHC, the top quark mass can be
determined in leptonic final states with $J/ \Psi$. This indirect method
relies heavily on the proper Monte Carlo description of the top production
and decay. The top mass can be determined with both a statistical and systematic
uncertainty at the level of 1 GeV.

%
%
%
\section{Conclusion}
The LHC will be an excellent place to study the top quark
properties. The very large sample of top events that will be
accumulated will allow a precision measurement of the top quark
mass. Various methods applied to statistically independent samples
gathering all dominant decay channels of the top quark have been
investigated. The studies have shown that after only one year of
data taking at low luminosity (per $10fb^{-1}$), a total error on
the top mass at the level of 2 GeV can be achieved. In the
inclusive lepton plus jets channels, the error can probably be
further reduced down to 1 GeV. In all channels, the errors are
dominated by systematic uncertainties (the systematic errors are
summarized in table \ref{overall_syst}).

\begin{table}[h]
\begin{center}
\begin{tabular}{|l|c|c|c|c|}\hline
Source of error & Lepton+jets & Lepton+jets & Dilepton & All jets \\
in GeV & inclusive & large clusters & & high pT \\
& sample & sample & & sample \\ \hline
Energy scale & & & & \\
\ \ Light jet energy scale & 0.2 &- &- & 0.8              \\
\ \ b-jet energy scale & 0.7 & - & 0.6 & 0.7                \\
\ \ Mass scale calibration &- & 0.9 &-  &-             \\
\ \ UE estimate  &- & 1.3 &-  &- \\
Physics & & & & \\
\ \ Background & 0.1 & 0.1 & 0.2 & 0.4             \\
\ \ b-quark fragmentation   & 0.1 & 0.3 & 0.7 & 0.3            \\
\ \ Initial state radiation  & 0.1 & 0.1 & 0.1 & 0.4           \\
\ \ Final state radiation   & 0.5 & 0.1 & 0.6 & 2.8           \\
\ \ PDF &- &- & 1.2 &-             \\ \hline
\end{tabular}
\caption{ \label{overall_syst} {\it Summary of the systematics
errors in the top mass measurement, in the lepton plus jets
channel, in the all jets channel and in the dilepton channel.}}
\end{center}
\end{table}

The analyses presented in this paper are differently sensitive to
the various sources of systematic errors. This will allow reliable
cross-checks between the various methods and an efficient
extraction of the combined ATLAS measurement of the top quark
mass.

In the inclusive lepton plus jets channel, errors are dominated by the b-jet energy
scale and the knowledge of FSR. It was shown that these effects can be better
controlled using a continuous jet definition. A possibility has been seen to reduce
the systematic error due to the b-jet scale uncertainty by calibrating the b-jets
with the same calibration as determined for light jets, though this must be further
studied will full detector simulation before conclusions can be reached.
In the lepton plus jets channel, using
a sub-sample of hight $p_T$ top events, the top mass can be reconstructed with a
large calorimeter cluster. In the all jets channels, it was demonstrated that the
$t \bar{t}$ signal can be efficiently extracted from the huge QCD background. In
the dilepton channel, it was presented that despite the two undetected neutrinos the
final state can be fully reconstructed assuming a value for the top mass. Finally,
during the high luminosity phase of the LHC, the top mass can also be precisely
determined in leptonic final states with $J / \Psi$.

\section{Acknowledgments}
This work has been performed within the ATLAS Collaboration and we would
like to thank collaboration members for helpful discussions. We are very
grateful to Marina Cobal and John Parsons for their careful reading of the
paper. We made use of the ATLAS physics simulation and analysis framework
tools which are the fruit of a collaboration-wide effort.


\begin{thebibliography}{99}
\bibitem{lepton_tkachev} F. Tkachev, "Measuring multijet structure of hadronic
energy flow or what is a jet?", HEP-PH/9601308.
\end{thebibliography}
\end{document}